\documentclass[a4paper,12pt]{article}

\usepackage{amsmath}
\usepackage{amssymb}
\usepackage[dvips]{graphicx}
\usepackage[dvips]{psfrag}
\usepackage{bm}

\newcommand {\be}{\begin{equation}}
\newcommand {\ee}{\end{equation}}
\newcommand {\bea}{\begin{eqnarray}}
\newcommand {\eea}{\end{eqnarray}}
\newcommand {\nn}{\nonumber}
\newcommand {\tr}{{\rm tr}}

\newcommand{\cB}{{\cal B}}
\newcommand{\cF}{{\cal F}}
\newcommand{\cI}{{\cal I}}
\newcommand{\cN}{{\cal N}}
\newcommand{\cO}{{\cal O}}

\newcommand{\cV}{{\cal V}}

\newcommand{\ket}{\rangle}
\newcommand{\bra}{\langle}
\newcommand{\vev}[1]{\left\langle #1 \right\rangle}

\newcommand{\rNS}{{\rm NS}}
\newcommand{\rR}{{\rm R}}
\newcommand{\cD}{{\cal D}}

\newcommand {\bA}{{\tt A}}
\newcommand {\bB}{{\tt B}}

\begin{document}
\thispagestyle{empty} \addtocounter{page}{-1}
\begin{flushright}
OIQP-13-07
%
\end{flushright} 
\vspace*{1cm}

\begin{center}
{\large \bf Supersymmetric double-well matrix model as 
two-dimensional type IIA superstring on RR background}\\
\vspace*{2cm}
Tsunehide Kuroki$^*$ and Fumihiko Sugino$^\dagger$\\
\vskip0.7cm
{}$^*${\it Kobayashi-Maskawa Institute for the Origin of Particles and the Universe, }\\
\vspace*{1mm}
{\it Nagoya University, Nagoya 464-8602, Japan}\\
\vspace*{0.2cm}
{\tt kuroki@kmi.nagoya-u.ac.jp}\\
\vskip0.4cm
{}$^\dagger${\it Okayama Institute for Quantum Physics, } \\
\vspace*{1mm}
{\it Kyoyama 1-9-1, Kita-ku, Okayama 700-0015, Japan}\\
\vspace*{0.2cm}
{\tt fumihiko\_sugino@pref.okayama.lg.jp}\\
\end{center}
\vskip2cm
\centerline{\bf Abstract}
\vspace*{0.3cm}
{\small 
In the previous paper, the authors pointed out correspondence of a supersymmetric double-well matrix model 
with two-dimensional type IIA superstring theory on a nontrivial Ramond-Ramond background 
from the viewpoint of symmetries and spectrum. 
In this paper we further investigate the correspondence from dynamical aspects by comparing 
scattering amplitudes in the matrix model and those in the type IIA theory. 
In the latter, cocycle factors are introduced to vertex operators in order to reproduce correct transformation laws 
and target-space statistics. By a perturbative treatment of the Ramond-Ramond background as insertions 
of the corresponding vertex operators, various IIA amplitudes are explicitly computed including 
quantitatively precise numerical factors. We show that several kinds of amplitudes in both sides indeed 
have exactly the same dependence on parameters of the theory. 
Moreover, we have a number of relations among coefficients 
which connect quantities in the type IIA theory and those in the matrix model. 
Consistency of the relations convinces us of the validity of the correspondence.  
}
\vspace*{1.1cm}



\newpage

\section{Introduction}
Although matrix models~\cite{Banks:1996vh,Ishibashi:1996xs,Dijkgraaf:1997vv} have been proposed as 
nonperturbative formulations of superstring/M theory, 
it is still difficult to compute perturbative string amplitudes from these models. 
Regarding two of them~\cite{Banks:1996vh,Dijkgraaf:1997vv}, since the models are formulated 
relying on nonperturbative objects (D-branes), it is not straightforward to see perturbative aspects 
of fundamental strings. 
The remaining one~\cite{Ishibashi:1996xs} is based on the Schild gauge formulation of the type IIB superstring theory. 
It is a fully interacting theory, 
and its analytical treatment to carry out the computation of a perturbative S-matrix has not been found yet. 

In this situation, it will be an interesting direction to make correspondence between 
a supersymmetric matrix model and simpler noncritical superstring theory, in both of which  
perturbative scattering amplitudes are computable and the correspondence is explicitly confirmed.       
In fact, we pointed out in the previous paper~\cite{Kuroki:2012nt} correspondence of a supersymmetric double-well matrix model 
to two-dimensional type IIA superstring theory on a nontrivial Ramond-Ramond (RR) background 
from the viewpoint of symmetries and spectrum. 
In this paper, 
we further investigate dynamical aspects of the correspondence.  
We compute various amplitudes in the type IIA theory, and compare with 
the calculation of matrix-model correlators obtained in~\cite{Kuroki:2012nt}. 
We carefully introduce cocycle factors to vertex operators in the IIA theory in order to 
realize correct transformation laws and target-space statistics. 
In the calculation of amplitudes, the RR background is treated in a perturbative manner 
by insertions of the corresponding RR vertex operators. 
Amplitudes evaluated at the on-shell momenta are often indefinite or divergent, 
for which we find a reasonable regularization scheme preserving mutual locality of physical vertex operators.  
We thus obtain several kinds of regularized amplitudes in the type IIA theory including precise numerical factors, 
which allows direct comparison with the corresponding correlators in the matrix model at the quantitative level. 
As a result, we find that they indeed have exactly the same dependence on parameters of the theory. 
Furthermore, we obtain a number of relations among coefficients that connect quantities in the type IIA 
theory to those in the matrix model. Remarkably, all of them are consistent with each other, 
which provides strong evidence for the validity of the correspondence. 

The rest of this paper is organized as follows. 
In the next section, 
we present some results for amplitudes of the supersymmetric double-well 
matrix model computed in~\cite{Kuroki:2012nt}. 
In section~\ref{sec:2DIIA}, a brief review of two-dimensional type IIA superstring theory is given, 
together with discussion of cocycle factors in detail that is important for precise evaluation of amplitudes. 
We also explain how to take into account an RR background of the 
IIA theory. 
In section~\ref{sec:basic}, we compute   
basic amplitudes among vertex operators in the type IIA theory 
on the trivial background. 
In section~\ref{sec:check}, 
the results of the basic amplitudes are transcribed to 
amplitudes in the type IIA theory on the RR background, that are compared 
with the matrix-model results presented in section~\ref{sec:MM_results}. 
As a result of the comparison, they agree with each other as functions of parameters in the theory, 
and various relations are obtained among coefficients which link 
quantities in the type IIA theory with those in the matrix model. 
The (R$-$, R$+$) vertex operators representing the background contain nonlocal 
vertex operators which violate the Seiberg bound~\cite{Seiberg:1990eb}. 
They do not satisfy the Dirac equation constraint. 
It would be acceptable in the sense that they do not describe on-shell particles
but represent the background. However, since the nonlocal operators turn out not to be BRST-closed, 
we discuss consistency of amplitudes in the presence of them 
in section~\ref{sec:nonlocalRR}. The results obtained so far are summarized and some future directions 
are discussed in section~\ref{sec:discussions}. 
An identity concerning matrix-model amplitudes is proved in appendix~\ref{app:sum_m}. 
We give a brief summary of the worldsheet superconformal symmetry in the type IIA theory 
in appendix~\ref{app:SCFT}, 
and discuss cocycle factors for 0-picture NS vertex operators in appendix~\ref{app:cocycle_T0}. 
Integral formulas needed to evaluate IIA string amplitudes in the text are presented in 
appendix~\ref{app:integrals}. 
Appendix~\ref{app:4pt_BBBB2_pic} is devoted to a note on the picture changing manipulation of 
Friedan-Martinec-Shenker~\cite{Friedan:1985ge} 
in a certain amplitude in the presence of the nonlocal operator.    

\section{Results of the supersymmetric matrix model}
\label{sec:MM_results}
In the previous paper~\cite{Kuroki:2012nt}, we investigated the supersymmetric 
matrix model: 
\be
S = N \tr \left[\frac12 B^2 +iB(\phi^2-\mu^2) +\bar\psi (\phi\psi+\psi\phi)\right],  
\label{S}
\ee
where $B$, $\phi$ are Grassmann even, and $\psi$, $\bar\psi$ are Grassmann-odd 
$N\times N$ Hermitian matrices, respectively. 
The action $S$ is invariant under supersymmetry transformations generated by $Q$ and $\bar{Q}$: 
\be
Q\phi =\psi, \quad Q\psi=0, \quad Q\bar{\psi} =-iB, \quad QB=0, 
\label{QSUSY}
\ee
and 
\be
\bar{Q} \phi = -\bar{\psi}, \quad \bar{Q}\bar{\psi} = 0, \quad 
\bar{Q} \psi = -iB, \quad \bar{Q} B = 0,  
\label{QbarSUSY}
\ee
which lead to the nilpotency: $Q^2=\bar{Q}^2=\{ Q,\bar{Q}\}=0$. 
Various correlation functions were computed, 
and correspondence of the matrix model to two-dimensional type IIA superstring theory 
was pointed out from symmetry properties and spectrum.  
%
Let us present results in the matrix model for later comparison with the type IIA theory. 
We express the connected correlation function among $n$ single-trace operators $\frac{1}{N}\tr\,\cO_i$ 
($i=1,\cdots, n$) as 
\be
\vev{\prod_{i=1}^n \frac{1}{N}\tr\,\cO_i}_C=\sum_{h=0}^\infty \frac{1}{N^{2h+2n-2}}\,
\vev{\prod_{i=1}^n \frac{1}{N}\tr\,\cO_i}_{C, h}, 
\label{vev_MM}
\ee
where $\vev{\,\cdot\,}_{C,h}$ denotes the connected correlator on a handle-$h$ random surface 
with the $N$-dependence factored out. 
When $\mu^2\geq 2$, the planar limit of the matrix model has an infinitely degenerate supersymmetric vacua 
parametrized by filling fractions $(\nu_+, \nu_-)$, 
which represent configurations that $\nu_\pm N$ of the eigenvalues of $\phi$ are around the minimum $x=\pm |\mu |$  
of the double-well potential $\frac12(x^2-\mu^2)^2$.  
$\mu^2=2$ is a critical point at which the matrix model exhibits the third-order phase transition 
between a supersymmetric phase ($\mu^2>2$) and a nonsupersymmetric phase ($\mu^2<2$)~\cite{Kuroki:2010au}. 
In the limit $\mu^2\rightarrow 2+0$ from the supersymmetric phase, the operators of the scalar matrix $\phi$: 
\be
\Phi_{2k+1}=\frac{1}{N}\,\tr\,\phi^{2k+1} + (\mbox{mixing}) \qquad (k=0,1,2,\cdots)
\ee
( ``mixing'' represents lower power operators of $\phi$ introduced in order to remove nonuniversal 
singular terms in $\mu^2\to 2+0$)~\footnote{
The explicit form of the ``mixing'' for the first few $\Phi_{2k+1}$ is presented in eqs. (4.40) 
in~\cite{Kuroki:2012nt}.}
show critical behavior as power of logarithm:  
\bea
& & \left.\vev{\Phi_{2k+1}}_0\right|_{\rm sing.} 
= (\nu_+-\nu_-)\left[\frac{2^{k+2}}{\pi}\frac{(2k+1)!!}{(k+2)!}\,
\omega^{k+2} \ln \omega +(\mbox{less singular})\right], 
\label{vevPhi_MM} \\ 
& & \left.\vev{\Phi_{2k+1}\,\Phi_{2\ell+1}}_{C,0}\right|_{\rm sing.} \nn \\
& & \hspace{3mm}=-(\nu_+-\nu_-)^2\frac{2k+1}{4\pi^2}\,2^{2k+m}\left[\sum_{p=1}^m\frac{(2p+2k-1)!!}{(p+k)!} 
\frac{(2m-2p+2k-1)!!}{(m-p+k+1)!} \right.\nn \\
& & \hspace{52mm} \left. +2\,\frac{(2k-1)!!}{k!}\frac{(2m+2k-1)!!}{(m+k)!}\right] 
\omega^{2k+m+1}(\ln\omega)^2 \nn \\
& & \hspace{7mm} +(\mbox{less singular}),
\label{vevPhiPhi_MM}
\eea
where 
\be
\omega = (\mu^2-2)/4
\label{omega}
\ee 
and $\ell=k+m$. The symbol $|_{\rm sing.}$ means that entire functions of $\omega$ are removed from 
the expression~\footnote{In general, matrix models can be regarded as a sort of lattice models for 
string theory. Then, entire functions are analogous to lattice artifacts that are irrelevant to continuum physics.}. 
Also, for fermionic operators~\footnote{
$\Psi_1$ and $\bar{\Psi}_1$ have no ``mixing'' term. 
The ``mixing'' terms for $\Psi_3$ and $\bar{\Psi}_3$ are given in eqs. (6.29), (6.30) in \cite{Kuroki:2012nt}.
}   
\bea
\Psi_{2k+1}=\frac{1}{N}\,\tr\,\psi^{2k+1} + (\mbox{mixing}), \qquad 
\bar{\Psi}_{2k+1} = \frac{1}{N}\,\tr\,\bar{\psi}^{2k+1} + (\mbox{mixing}), 
\eea
\be
\left.\vev{\Psi_{2k+1}\bar{\Psi}_{2\ell+1}}_{C,0}\right|_{\rm sing.} 
= \delta_{k,\ell}\,v_k\,(\nu_+-\nu_-)^{2k+1}\omega^{2k+1}\ln\omega 
+(\mbox{less singular}).
\label{vevPsiPsi_MM}
\ee
The coefficient $v_k$ has been computed for $k=0,1$ as $v_0=\frac{1}{\pi}$ and $v_1=\frac{6}{\pi}$. 

According to appendix~\ref{app:sum_m}, the sum with respect to $m$ in (\ref{vevPhiPhi_MM}) is reduced 
to a simple expression: 
\bea
\left.\vev{\Phi_{2k+1}\,\Phi_{2\ell+1}}_{C,0}\right|_{\rm sing.} & = & 
-\frac{(\nu_+-\nu_-)^2}{2\pi^2}\frac{1}{k+\ell+1}\frac{(2k+1)!}{(k!)^2}
\frac{(2\ell+1)!}{(\ell!)^2}\,\omega^{k+\ell+1}(\ln \omega)^2 \nn \\
 & & +(\mbox{less singular}).
\label{vevPhiPhi_MM2}
\eea
 
Some genus-one amplitudes are presented in appendix A in~\cite{Kuroki:2012nt}. Among them, 
\be
\vev{\frac{1}{N}\tr\,B}_1=0
\label{torus_vevB_MM}
\ee
means that the torus free energy is a constant independent of $\mu^2$. 
It is reasonable to expect that the constant vanishes. 
Actually, from the result of eq.~(3.42) in~\cite{Kuroki:2010au}, 
the partition function in the sector of the filling fraction $(\nu_+,\nu_-)$ becomes 
\be
Z_{(\nu_+, \nu_-)}=(-1)^{\nu_-N}. 
\label{Z_MM}
\ee
It is valid in all order in $1/N$ expansion, indicating that 
the free energy defined by $-\ln |Z_{(\nu_+,\nu_-)}|$ is zero at each topology, and that  
the expectation is correct. 
(The sign factor $(-1)^{\nu_-N}$ could not be seen from the conventional string perturbation theory.)

\section{Two-dimensional type IIA superstring} 
\label{sec:2DIIA}
\setcounter{equation}{0}
In this section, we explain the two-dimensional type IIA superstring 
theory, which is discussed in \cite{Kutasov:1990ua,Kutasov:1991pv,Murthy:2003es,Ita:2005ne,Grassi:2005kc}. 
Then, we mention correspondence of physical vertex operators in the type IIA theory with operators 
in the matrix model~\cite{Kuroki:2012nt}. 

The target space is $(\varphi, x)\in\mbox{(Liouville direction)} \times \mbox{($S^1$ with self-dual radius)}$, 
and the holomorphic energy-momentum tensor on the string worldsheet is given by 
\bea
T& = & T_{\rm m} + T_{\rm gh}, \\
T_{\rm m} & = & -\frac12 (\partial x)^2 -\frac12 \psi_x\partial\psi_x -\frac12(\partial \varphi)^2 
+\frac{Q}{2}\partial^2\varphi -\frac12\psi_\ell\partial \psi_\ell, \nn \\
T_{\rm gh} & = & -2b\partial c-\partial b c -\frac32\beta \partial \gamma -\frac12 \partial \beta \gamma
\label{EMtensor}
\eea
with $Q=2$. 
Here, $\psi_x$ and $\psi_\ell$ are superpartners of $x$ and $\varphi$, respectively and 
$b, c$ ($\beta, \gamma$) represent conformal (superconformal) ghosts. 
OPEs for the fields are 
\bea
x(z)x(w) \sim -\ln(z-w), & & \varphi(z)\varphi(w) \sim -\ln(z-w), \nn \\
\psi_x(z)\psi_x(w) \sim \frac{1}{z-w}, & & \psi_\ell(z)\psi_\ell(w) \sim \frac{1}{z-w}, \nn \\
c(z)b(w)\sim \frac{1}{z-w}, & & \gamma(z)\beta(w)\sim \frac{1}{z-w},
\eea
and the others are regular. 

In order to treat the Ramond sector in the RNS formalism, it is convenient to bosonize 
$\psi_x$, $\psi_\ell$, $\beta$ and $\gamma$ as 
\bea
\Psi\equiv \psi_\ell+ i\psi_x=\sqrt{2} e^{-iH}, & &  \Psi^\dagger \equiv \psi_\ell- i\psi_x=\sqrt{2} e^{iH}, \\
\gamma = e^\phi\,\eta, & & \beta = \partial \xi\,e^{-\phi}  
\eea
with 
\be
H(z)H(w)\sim -\ln (z-w), \qquad \phi(z)\phi(w)\sim -\ln(z-w), \qquad \eta(z)\xi(w)\sim \frac{1}{z-w}.
\ee
Some properties of worldsheet superconformal generators are summarized in appendix~\ref{app:SCFT}. 

Vertex operators are constructed by combining the NS ``tachyon'' vertex operator 
(in the $(-1)$ picture): 
\be
T_k(z) = e^{-\phi+ikx+p_\ell\varphi}(z), \qquad 
\bar{T}_{\bar{k}}(\bar{z}) = e^{-\bar{\phi}+i\bar{k}\bar{x}+p_\ell\bar{\varphi}}(\bar{z}) 
\ee
and the R vertex operator (in the $(-\frac12)$ picture): 
\be
V_{k,\,\epsilon}(z) = e^{-\frac12\phi+\frac{i}{2}\epsilon H+ikx +p_\ell\varphi}(z), \qquad 
\bar{V}_{\bar{k},\,\bar{\epsilon}}(\bar{z})= 
e^{-\frac12\bar{\phi}+\frac{i}{2}\bar{\epsilon} \bar{H}+i\bar{k}\bar{x} +p_\ell\bar{\varphi}}(\bar{z}) 
\label{Rvertex}
\ee
with $\epsilon, \bar{\epsilon}=\pm 1$. 
Here the local scale invariance on the worldsheet imposes $p_\ell=1\pm k$ for $T_k$, $V_{k,\,\epsilon}$ and 
$p_\ell=1\pm\bar k$ for $\bar T_{\bar k}$, $\bar V_{\bar{k},\,\bar{\epsilon}}$. 
We consider a branch of $p_\ell = 1-|k|$, $p_\ell = 1-|\bar k|$  
satisfying the locality bound ($p_\ell \leq Q/2$)~\cite{Seiberg:1990eb} for a while. 
Target-space supercurrents are 
\be
q_+(z)\equiv V_{-1,\,-1}(z)=e^{-\frac12\phi-\frac{i}{2}H-ix}(z), \qquad 
\bar{q}_-(\bar{z}) \equiv \bar{V}_{+1,\,+1}(\bar{z}) = e^{-\frac12\bar{\phi}+\frac{i}{2}\bar{H}+i\bar{x}}(\bar{z}). 
\label{supercurrents}
\ee
As discussed in \cite{Kutasov:1990ua,Kutasov:1991pv,Murthy:2003es,Ita:2005ne,Grassi:2005kc}, 
physical vertex operators should satisfy 
locality with the supercurrents, mutual locality, superconformal invariance (including the Dirac equation 
constraint) and the level matching condition. Two consistent sets of physical vertex operators are found 
in ref.~\cite{Ita:2005ne}, which are called as ``momentum background'' and ``winding background''. 
As considered in~\cite{Kuroki:2012nt}, we focus on the ``winding background''. It is given by 
\bea
\mbox{(NS, NS)}: \quad &  T_k\,\bar{T}_{-k}  & \quad (k\in {\bf Z}+\frac12), \nn \\
\mbox{(R$+$, R$-$)}: \quad &  
V_{k, \,+1}\,\bar{V}_{-k, \,-1} & \quad (k\in {\bf Z}+\frac12), \nn \\
\mbox{(R$-$, R$+$)}: \quad &  
V_{k, \,-1}\,\bar{V}_{\bar{k},\, +1} & \quad(k,\,\bar{k}\in {\bf Z}, \, |k|=|\bar{k}|), \nn \\
 \mbox{(NS, R$-$)}: \quad &  
T_{k}\,\bar{V}_{k,\,-1}  & \quad (k\in {\bf Z}+\frac12), \nn \\
\mbox{(R$+$, NS)}: \quad &  
V_{k, \,+1}\, \bar{T}_k\,  & \quad (k\in {\bf Z}+\frac12) 
\label{spectrum_before}
\eea  
before imposing the Dirac equation constraint. 
It requires the states corresponding to the Ramond vertex operators 
to be annihilated by the zero-mode of worldsheet superconformal generator $T_{{\rm m}, F}$: 
$G_0^{\rm m}=\frac{1}{\sqrt{2}}(G^+_0+G^-_0)$ (see appendix~\ref{app:SCFT}). Consequently, 
\be
k=\epsilon |k|, \qquad \bar{k}=\bar{\epsilon}|\bar{k}|. 
\ee
Then, imposing the Dirac equation constraint amounts to 
\bea
\mbox{(NS, NS)}: \quad &  T_k\,\bar{T}_{-k}  & \quad (k\in {\bf Z}+\frac12), \nn \\
\mbox{(R$+$, R$-$)}: \quad &  
V_{k, \,+1}\,\bar{V}_{-k, \,-1} & \quad (k=\frac12,\,\frac32, \cdots), \nn \\
\mbox{(R$-$, R$+$)}: \quad &  
V_{-k, \,-1}\,\bar{V}_{k,\, +1} & \quad(k=0,1,2,\cdots), \nn \\
 \mbox{(NS, R$-$)}: \quad &  
T_{-k}\,\bar{V}_{-k,\,-1}  & \quad (k=\frac12,\,\frac32, \cdots), \nn \\
\mbox{(R$+$, NS)}: \quad &  
V_{k, \,+1}\, \bar{T}_k\,  & \quad (k=\frac12,\,\frac32, \cdots) . 
\label{spectrum_after}
\eea  
Note that the spectrum (\ref{spectrum_after}) is invariant under  
not only the $\cN=1$ but also the $\cN=2$ superconformal symmetry. 
In particular, each of $G^+_0$ and $G^-_0$ annihilates the Ramond states in (\ref{spectrum_after}).   

\subsection{Cocycle factors}
\label{sec:cocycle}
Here, we introduce cocycle factors to realize correct transformation laws and target-space statistics 
for vertex operators~\footnote{
The argument in this subsection also holds for vertex operators of the nonlocal branch $p_\ell=1+|k|$.}. 

Vertex operators without cocycle factors have the following two problems. 
First, the OPE between $q_+$ and $T_k$ ($k\in {\bf Z}+\frac12$): 
\be
q_+(z) T_k(w) = \frac{1}{(z-w)^{\frac12+k}}\, :q_+(z)T_k(w): 
\label{OPE_q+T}
\ee
implies that the radial ordering should be defined as 
\be
R(q_+(z) \,T_k(w))=\begin{cases}  q_+(z) \,T_k(w) & (|z|>|w|) \\ 
(-1)^{-\frac12-k} \,T_k(w) \,q_+(z) & (|z|<|w|). \end{cases}
\label{R_q+NS}
\ee   
The factor $(-1)^{-\frac12-k}$ ensures continuity at $|z|=|w|$. 
Then, the target-space supercharge $Q_+=\oint\frac{dz}{2\pi i}\,q_+(z)$ acts on $T_k$ in a manner 
\bea
Q_+ T_k(w) -(-1)^{-\frac12 -k} T_k(w) Q_+ & =  & \oint_w\frac{dz}{2\pi i}\,R(q_+(z) \,T_k(w)) \nn \\
 & = & \oint_w\frac{dz}{2\pi i}\,\frac{1}{(z-w)^{\frac12+k}}\, :q_+(z)T_k(w):, 
\label{Q+_Tk}
\eea
so that the transformation law can be given by the contour integral. 
The $k$-dependent sign factor 
is due to the fact that the supercurrents (\ref{supercurrents}) carry $x, \bar{x}$-momenta, 
which is peculiar to noncritical superstring theory~\cite{Kutasov:1990ua,Kutasov:1991pv,Murthy:2003es}. 
On the other hand, target-space statistics suggests that $Q_+$ should act on an (NS, NS) field $T_k\,\bar{T}_{-k}$ 
in the form of a commutator. 
It will become consistent with (\ref{Q+_Tk}), if we make $\bar{T}_{-k}$ and $q_+$ noncommuting such as 
\be
\bar{T}_{-k}(\bar{w})\,q_+(z)  = (-1)^{-\frac12-k} q_+(z)\,\bar{T}_{-k}(\bar{w})
\label{Tbar_q+}
\ee
by introducing cocycle factors. 
If this condition is met, it automatically follows that $Q_+$ transformation of $V_{k,+1}\bar T_k$ 
is given by an anticommutator in accordance with the target-space statistics. (See the first formula in 
(\ref{R_q+_other}).) 

For other vertex operators, we have the radial orderings: 
\bea
R(q_+(z)\,V_{k,\,\epsilon}(w)) & = & 
\begin{cases} q_+(z)\,V_{k,\,\epsilon}(w) & (|z|>|w|) \\ 
(-1)^{-\frac14-\frac{\epsilon}{4}-k}\,V_{k,\,\epsilon}(w)\,q_+(z) & (|z|<|w|), \end{cases}  \nn \\
R(\bar{q}_-(\bar{z}) \,\bar{T}_{\bar{k}}(\bar{w})) & = & 
\begin{cases} \bar{q}_-(\bar{z}) \,\bar{T}_{\bar{k}}(\bar{w}) & (|z|>|w|) \\ 
(-1)^{\frac12-\bar{k}} \,\bar{T}_{\bar{k}}(\bar{w}) \,\bar{q}_-(\bar{z}) & (|z|<|w|), \end{cases}
\nn \\
R(\bar{q}_-(\bar{z})\,\bar{V}_{\bar{k},\,\bar{\epsilon}}(\bar{w})) & = & 
\begin{cases} \bar{q}_-(\bar{z})\,\bar{V}_{\bar{k},\,\bar{\epsilon}}(\bar{w}) & (|z|>|w|) \\ 
(-1)^{\frac14-\frac{\bar{\epsilon}}{4}-\bar{k}}\,\bar{V}_{\bar{k},\,\bar{\epsilon}}(\bar{w})\,\bar{q}_-(\bar{z}) & (|z|<|w|). \end{cases} 
\label{R_q+_other}
\eea

Second, the OPE between an (NS, NS) field $T_k\,\bar{T}_{-k}$ and an (NS, R$-$) field $T_k \,\bar{V}_{k,\,-1}$ 
($k,k'\in {\bf Z}+\frac12$):  
\bea
 & & T_k(z) \bar{T}_{-k}(\bar{z})\,T_{k'}(w)\bar{V}_{k',\,-1}(\bar{w}) \nn \\
 & & \hspace{3mm}= r^{-\frac32-2p_\ell p'_\ell}\,e^{i\theta (-\frac12+2kk')}\,
:T_k(z)\,T_{k'}(w):\,:\bar{T}_{-k}(\bar{z})\,\bar{V}_{k',\,-1}(\bar{w}):
\label{OPE_TTbar_TVbar}
\eea
with $z-w=r\,e^{i\theta}, \bar{z}-\bar{w}=r\,e^{-i\theta}$ leads to the radial ordering 
\bea
& & R(T_k(z) \bar{T}_{-k}(\bar{z})\,T_{k'}(w)\bar{V}_{k',\,-1}(\bar{w})) \nn \\
& & \hspace{3mm}= \begin{cases} T_k(z) \bar{T}_{-k}(\bar{z})\,T_{k'}(w)\bar{V}_{k',\,-1}(\bar{w}) & (|z|>|w|) \\
(-1)^{-\frac12+2kk'}\,T_{k'}(w)\bar{V}_{k',\,-1}(\bar{w})\,T_k(z) \bar{T}_{-k}(\bar{z}) & (|z|<|w|).
\end{cases}
\eea
It is consistent with the target-space statistics when $-\frac12 +2kk'$ is even, but not otherwise. 
Similar is the situation for the radial ordering of other fields. 

Let us introduce cocycle factors to resolve these problems. We put the hat ($\,\hat{\,}\,$) on vertex 
operators with cocycle factors. 
For the target-space supercurrents, 
\be
\hat{q}_+(z) = e^{\pi\beta (\frac12 p_{\bar{\phi}} -i\frac12 p_{\bar{h}} -ip_{\bar{x}})}\,q_+(z), \qquad 
\hat{\bar{q}}_-(\bar{w}) = e^{-\pi\beta (\frac12 p_\phi +i\frac12 p_h+ip_x)}\,\bar{q}_-(\bar{w}), 
\label{cocycle_q}
\ee
where $\beta$ is a constant to be determined. $p_{\phi}$, $p_h$ and $p_x$ ($p_{\bar{\phi}}$, $p_{\bar{h}}$ and 
$p_{\bar{x}}$) are momentum modes of holomorphic part (anti-holomorphic part) of free bosons: 
\bea
\phi(z)=\phi_0-ip_\phi\,\ln z +\cdots, & & \bar{\phi}(\bar{z})=\bar{\phi}_0-ip_{\bar{\phi}}\,\ln\bar{z}+\cdots, 
\nn \\
H(z)=h_0-ip_h\,\ln z +\cdots, & & \bar{H}(\bar{z})=\bar{h}_0-ip_{\bar{h}}\,\ln\bar{z}+\cdots, 
\nn \\
x(z)=x_0-ip_x\,\ln z +\cdots, & & \bar{x}(\bar{z})=\bar{x}_0-ip_{\bar{x}}\,\ln\bar{z}+\cdots 
\eea
with $\cdots$ representing oscillator modes.
{}From the commutation relations 
\be
[\phi_0, p_\phi] = [\bar{\phi}_0, p_{\bar{\phi}}]=i,\,\mbox{etc}, 
\ee
we see noncommuting properties
\bea
& & q_+(z)\,e^{-\pi\beta (\frac12 p_\phi +i\frac12 p_h+ip_x)} 
= e^{-i\pi\beta}\,e^{-\pi\beta (\frac12 p_\phi +i\frac12 p_h+ip_x)} \,q_+(z), \\
& & e^{\pi\beta (\frac12 p_{\bar{\phi}} -i\frac12 p_{\bar{h}} -ip_{\bar{x}})}\,\bar{q}_-(\bar{w}) 
= e^{-i\pi\beta}\,\bar{q}_-(\bar{w})\,e^{\pi\beta (\frac12 p_{\bar{\phi}} -i\frac12 p_{\bar{h}} -ip_{\bar{x}})},
\eea 
which lead to 
\be
\hat{q}_+(z)\,\hat{\bar{q}}_-(\bar{w}) = e^{-i2\pi\beta}\,\hat{\bar{q}}_-(\bar{w})\,\hat{q}_+(z).
\label{q+_qbar-}
\ee
The target-space statistics requires 
\be
\beta\in {\bf Z}+\frac12. 
\label{beta}
\ee
Hereafter we assume that $\beta$ satisfies this condition. 
Modified target-space supercharges are given by 
\be
\hat{Q}_+=\oint\frac{dz}{2\pi i}\,\hat{q}_+(z), \qquad 
\hat{\bar{Q}}_-=\oint\frac{d\bar{z}}{2\pi i}\,\hat{\bar{q}}_-(\bar{z}). 
\ee

For vertex operators, we introduce cocycle factors as 
\bea
\hat{T}_k(z) = e^{\pi\beta(p_{\bar{\phi}}+ik p_{\bar{x}})}\,T_k(z), &  &
\hat{\bar{T}}_{\bar{k}}(\bar{z}) = e^{-\pi\beta(p_\phi+i\bar{k}p_x)}\,\bar{T}_{\bar{k}}(\bar{z}), \nn \\
\hat{V}_{k,\,\epsilon}(z) = e^{\pi\beta(\frac12p_{\bar{\phi}}+i\frac{\epsilon}{2}p_{\bar{h}}+ikp_{\bar{x}})}\,
V_{k,\,\epsilon}(z), & & 
\hat{\bar{V}}_{\bar{k},\,\bar{\epsilon}}(\bar{z}) = e^{-\pi\beta(\frac12p_\phi+i\frac{\bar{\epsilon}}{2}p_h 
+i\bar{k}p_x)}\,\bar{V}_{\bar{k},\,\bar{\epsilon}}(\bar{z}). 
\label{cocycle_TV}
\eea
It turns out to be a choice resolving the above problems. 

In the first problem, since 
\bea
\bar{T}_{\bar{k}}(\bar{w})\,e^{\pi\beta(\frac12p_{\bar{\phi}}-i\frac12p_{\bar{h}}-ip_{\bar{x}})} 
 & = & e^{i\pi\beta(-\frac12+\bar{k})}\,e^{\pi\beta(\frac12p_{\bar{\phi}}-i\frac12p_{\bar{h}}-ip_{\bar{x}})}\,
\bar{T}_{\bar{k}}(\bar{w}), \nn \\
e^{-\pi\beta(p_\phi+i\bar{k}p_x)}\,q_+(z) & = & e^{i\pi\beta(-\frac12+\bar{k})}\,q_+(z)\,
e^{-\pi\beta(p_\phi+i\bar{k}p_x)}
\eea
hold, we have 
\be
\hat{\bar{T}}_{\bar{k}}(\bar{w})\,\hat{q}_+(z) 
= e^{i2\pi\beta(-\frac12+\bar{k})}\,\hat{q}_+(z)\,\hat{\bar{T}}_{\bar{k}}(\bar{w}) 
\label{Tbar_q+_hat}
\ee
that realizes (\ref{Tbar_q+}) for (\ref{beta}). Then, $\hat{Q}_+$ 
acts on a hatted (NS, NS) field in the form of a commutator: 
\be
\left[\hat{Q}_+, \,\hat{T}_k(w)\hat{\bar{T}}_{-k}(\bar{w})\right] 
= \oint_w\frac{dz}{2\pi i}\,R(\hat{q}_+(z)\hat{T}_k(w))\,\hat{\bar{T}}_{-k}(\bar{w})  
\label{Q+NSNS}
\ee 
with 
\be
R(\hat{q}_+(z)\hat{T}_k(w)) = e^{\pi\beta (\frac12 p_{\bar{\phi}} -i\frac12 p_{\bar{h}} -ip_{\bar{x}})}\,
e^{\pi\beta(p_{\bar{\phi}}+ik p_{\bar{x}})}\,R(q_+(z)T_k(w)). 
\label{R_q+_Tk} 
\ee
In (\ref{R_q+_Tk}), 
the cocycle factors do not include any modes in the holomorphic sector and can be treated as constants 
in the radial ordering. 

For other cases, owing to the noncommuting properties
\bea
\hat{\bar{V}}_{\bar{k}, \,\bar{\epsilon}}(\bar{w})\,\hat{q}_+(z) & = & 
e^{i2\pi\beta(-\frac14+\frac{\bar{\epsilon}}{4}+\bar{k})}\,
\hat{q}_+(z)\,\hat{\bar{V}}_{\bar{k}, \,\bar{\epsilon}}(\bar{w}), \nn \\
\hat{\bar{q}}_-(\bar{z})\,\hat{T}_k(w) & = & e^{i2\pi\beta(-\frac12-k)}\,\hat{T}_k(w)\,
\hat{\bar{q}}_-(\bar{z}), \nn \\
\hat{\bar{q}}_-(\bar{z})\,\hat{V}_{k,\,\epsilon}(w) & = & e^{i2\pi\beta(-\frac14-\frac{\epsilon}{4}-k)}\,
\hat{V}_{k,\,\epsilon}(w)\,\hat{\bar{q}}_-(\bar{z}), 
\eea
the modified supercharges consistently act on hatted fields as 
\bea
\left[\hat{Q}_+, \,\hat{V}_{k,\,\epsilon}(w)\hat{\bar{V}}_{-k,\,-\epsilon}(\bar{w})\right]
& = & \oint_w\frac{dz}{2\pi i}\,R(\hat{q}_+(z)\hat{V}_{k,\,\epsilon}(w))\,\hat{\bar{V}}_{-k,\,-\epsilon}(\bar{w}), 
\nn \\
\left\{\hat{Q}_+,\,\hat{T}_k(w) \hat{\bar{V}}_{k,\,-1}(\bar{w})\right\}
& = & \oint_w\frac{dz}{2\pi i}\,R(\hat{q}_+(z)\hat{T}_k(w))\,\hat{\bar{V}}_{k,\,-1}(\bar{w}), 
\nn \\
\left\{\hat{Q}_+,\,\hat{V}_{k,\,+1}(w)\hat{\bar{T}}_k(\bar{w})\right\} 
& = &  \oint_w\frac{dz}{2\pi i}\,R(\hat{q}_+(z)\hat{V}_{k,\,+1}(w))\,\hat{\bar{T}}_{k}(\bar{w}),
\nn \\
\left[\hat{\bar{Q}}_-,\,\hat{T}_k(w)\hat{\bar{T}}_{-k}(\bar{w})\right] 
& = & (-1)^{-\frac12-k}\,\hat{T}_k(w)\,\oint_{\bar{w}}\frac{d\bar{z}}{2\pi i}\,R(\hat{\bar{q}}_-(\bar{z})
\hat{\bar{T}}_{-k}(\bar{w})), 
\nn \\   
\left[\hat{\bar{Q}}_-,\,\hat{V}_{k,\,\epsilon}(w)\hat{\bar{V}}_{-k,\,-\epsilon}(\bar{w})\right]
& = & (-1)^{-\frac14-\frac14\epsilon-k}\,\hat{V}_{k,\,\epsilon}(w)\,
\oint_{\bar{w}}\frac{d\bar{z}}{2\pi i}\,R(\hat{\bar{q}}_-(\bar{z})
\hat{\bar{V}}_{-k,\,-\epsilon}(\bar{w})), 
\nn \\
\{\hat{\bar{Q}}_-,\,\hat{T}_k(w) \hat{\bar{V}}_{k,\,-1}(\bar{w})\}
& = & (-1)^{-\frac12-k}\,\hat{T}_k(w)\,\oint_{\bar{w}}\frac{d\bar{z}}{2\pi i}\,R(\hat{\bar{q}}_-(\bar{z})
\hat{\bar{V}}_{k,\,-1}(\bar{w})), 
\nn \\
\left\{\hat{\bar{Q}}_-,\,\hat{V}_{k,\,+1}(w)\hat{\bar{T}}_k(\bar{w})\right\} 
& = & (-1)^{-\frac12-k}\,\hat{V}_{k,\,+1}(w)\,\oint_{\bar{w}}\frac{d\bar{z}}{2\pi i}\,R(\hat{\bar{q}}_-(\bar{z})
\hat{\bar{T}}_k(\bar{w})). 
\eea

In the second problem, due to the cocycle factors, we have the following noncommuting relations: 
\bea
\hat{\bar{T}}_{\bar{k}}(\bar{z})\,\hat{T}_{k'}(w) & = & e^{i2\pi\beta(-1-\bar{k}k')}\,\hat{T}_{k'}(w)\,
\hat{\bar{T}}_{\bar{k}}(\bar{z}), 
\label{Tbar_T_hat} \\
\hat{\bar{T}}_{\bar{k}}(\bar{z})\,\hat{V}_{k',\,\epsilon}(w) & = & e^{i2\pi\beta(-\frac12-\bar{k}k')}\,
\hat{V}_{k',\,\epsilon}(w)\,\hat{\bar{T}}_{\bar{k}}(\bar{z}), 
\label{Tbar_V_hat} \\
\hat{\bar{V}}_{\bar{k},\,\bar{\epsilon}}(\bar{z})\,\hat{T}_{k'}(w) & = & e^{i2\pi\beta(-\frac12-\bar{k}k')}
\,\hat{T}_{k'}(w)\,\hat{\bar{V}}_{\bar{k},\,\bar{\epsilon}}(\bar{z}), 
\label{Vbar_T_hat} \\
\hat{\bar{V}}_{\bar{k},\,\bar\epsilon}(\bar{z})\,\hat{V}_{k',\,\epsilon}(w) & = & 
e^{i2\pi\beta(-\frac14-\frac14\epsilon\bar\epsilon-\bar{k}k')}\,
\hat{V}_{k',\,\epsilon}(w)\,\hat{\bar{V}}_{\bar{k},\,\bar\epsilon}(\bar{z}). 
\label{Vbar_V_hat} 
\eea
Let us see the OPE considered in \eqref{OPE_TTbar_TVbar} in the presence of the cocycle factors. 
By using (\ref{Tbar_T_hat}), 
\bea
\hat{T}_k(z)\hat{\bar{T}}_{-k}(\bar{z})\,\hat{T}_{k'}(w)\hat{\bar{V}}_{k',\,-1}(\bar{w}) 
& = & e^{i2\pi\beta(-1+kk')}\,r^{-\frac32-2p_\ell p'_\ell}\,e^{i\theta (-\frac12+2kk')}\nn \\
& & \times :\hat{T}_k(z)\,\hat{T}_{k'}(w):\,:\hat{\bar{T}}_{-k}(\bar{z})\,\hat{\bar{V}}_{k',\,-1}(\bar{w}):
\label{OPE_TTbar_TVbar_hat}
\eea
is obtained when $|z|>|w|$. 
In the normal ordering of holomorphic fields $:\hat{T}_k(z)\,\hat{T}_{k'}(w):$\,, the cocycle factors 
can be treated as constants. It is similar for 
$:\hat{\bar{T}}_{-k}(\bar{z})\,\hat{\bar{V}}_{k',\,-1}(\bar{w}):$\,. 
On the other hand, when $|z|<|w|$, we use (\ref{Vbar_T_hat}) to have 
\bea
\hat{T}_{k'}(w)\hat{\bar{V}}_{k',\,-1}(\bar{w}) \,\hat{T}_k(z)\hat{\bar{T}}_{-k}(\bar{z}) 
& = & e^{i\pi(-\frac12+2kk')}\,e^{i2\pi\beta(-\frac12-kk')}\,r^{-\frac32-2p_\ell p'_\ell}\,
e^{i\theta (-\frac12+2kk')}\nn \\
& & \times 
:\hat{T}_k(z)\,\hat{T}_{k'}(w):\,:\hat{\bar{T}}_{-k}(\bar{z})\,\hat{\bar{V}}_{k',\,-1}(\bar{w}):\,.
 \label{OPE_TVbar_TTbar_hat}
\eea
Note that (\ref{OPE_TTbar_TVbar_hat}) has the same form as (\ref{OPE_TVbar_TTbar_hat}) since 
\be
e^{i2\pi\beta(-1+kk')}=e^{i\pi(-\frac12+2kk')}\,e^{i2\pi\beta(-\frac12-kk')}
\ee
holds for (\ref{beta}) and $k, k'\in {\bf Z}+\frac12$. Thus, the radial ordering of the hatted fields 
becomes consistent with the target-space statistics:  
\bea
& & R(\hat{T}_k(z) \hat{\bar{T}}_{-k}(\bar{z})\,\hat{T}_{k'}(w)\hat{\bar{V}}_{k',\,-1}(\bar{w})) \nn \\
& & \hspace{3mm}
= \begin{cases} \hat{T}_k(z) \hat{\bar{T}}_{-k}(\bar{z})\,\hat{T}_{k'}(w)\hat{\bar{V}}_{k',\,-1}(\bar{w}) & (|z|>|w|) \\
\hat{T}_{k'}(w)\hat{\bar{V}}_{k',\,-1}(\bar{w})\,\hat{T}_k(z) \hat{\bar{T}}_{-k}(\bar{z}) & (|z|<|w|).
\end{cases}
\eea
Similarly, we can show that the radial ordering of all other hatted vertex operators is consistent. 
Cocycle factors for 0-picture NS fields are discussed in appendix~\ref{app:cocycle_T0}.

\subsection{Correspondence to matrix model operators}
\label{sec:correspondence to MM}
In order to make correspondence to the matrix model, we first note that the vertex operators 
\be
\hat{V}_{\frac12, \,+1}\,\hat{\bar{V}}_{-\frac12, \,-1}, \qquad 
\hat{T}_{-\frac12}\,\hat{\bar{V}}_{-\frac12, \,-1}, \qquad 
\hat{V}_{\frac12,\,+1}\,\hat{\bar{T}}_{\frac12}, \qquad  \hat{T}_{-\frac12}\,\hat{\bar{T}}_{\frac12}
\ee 
form a quartet under $\hat{Q}_+$ and $\hat{\bar{Q}}_-$:
\bea
& 
\left[\hat{Q}_+,  \hat{V}_{\frac12, \,+1}\,\hat{\bar{V}}_{-\frac12, \,-1} \right]  
= \hat{T}_{-\frac12}\,\hat{\bar{V}}_{-\frac12, \,-1}, 
& \left\{ \hat{Q}_+, \hat{T}_{-\frac12}\,\hat{\bar{V}}_{-\frac12, \,-1}\right\}=0, 
\nn \\
 & 
\left\{ \hat{Q}_+, \hat{V}_{\frac12, \,+1}\, \hat{\bar{T}}_{\frac12}\right\} 
= \hat{T}_{-\frac12}\,\hat{\bar{T}}_{\frac12}, 
& \left[\hat{Q}_+, \hat{T}_{-\frac12}\,\hat{\bar{T}}_{\frac12}\right]=0, 
\label{quartet_+} 
\\
& 
\left[\hat{\bar{Q}}_-, \hat{V}_{\frac12,\,+1}\,\hat{\bar{V}}_{-\frac12, \,-1}\right] 
= -\hat{V}_{\frac12,\,+1}\, \hat{\bar{T}}_{\frac12}, 
& \left\{ \hat{\bar{Q}}_-, \hat{V}_{\frac12,\,+1}\, \hat{\bar{T}}_{\frac12}\right\}= 0, 
\nn \\
 & 
\left\{ \hat{\bar{Q}}_-, \hat{T}_{-\frac12}\,\hat{\bar{V}}_{-\frac12,\,-1}\right\} 
= \hat{T}_{-\frac12}\,\hat{\bar{T}}_{\frac12}, 
 & \left[\hat{\bar{Q}}_-, \hat{T}_{-\frac12}\,\hat{\bar{T}}_{\frac12}\right]=0,  
\label{quartet_-}
\eea
which are isomorphic to (\ref{QSUSY}) and (\ref{QbarSUSY}) in the matrix model under identification 
between supercharges in both sides. 
It leads to the correspondence of single-trace operators in the matrix model to integrated 
vertex operators in the type IIA theory: 
\bea
\Phi_1= \frac{1}{N}\tr\,\phi & \Leftrightarrow &  
\cV_\phi(0)\equiv g_s^2\int d^2z\, \hat{V}_{\frac12, \,+1}(z)\,\hat{\bar{V}}_{-\frac12, \,-1}(\bar{z}) , \nn \\
\Psi_1 = \frac{1}{N}\tr\,\psi & \Leftrightarrow &  
\cV_\psi(0)\equiv g_s^2\int d^2z\, \hat{T}_{-\frac12}(z)\,\hat{\bar{V}}_{-\frac12, \,-1}(\bar{z}), \nn \\
\bar{\Psi}_1=\frac{1}{N}\tr\,\bar{\psi} & \Leftrightarrow &  
\cV_{\bar{\psi}}(0) \equiv g_s^2\int d^2z\, \hat{V}_{\frac12, \,+1}(z)\, \hat{\bar{T}}_{\frac12}(\bar{z}), \nn \\
\frac{1}{N}\tr\,(-iB) & \Leftrightarrow &  
\cV_B(0)\equiv g_s^2\int d^2z\, \hat{T}_{-\frac12}(z)\,\hat{\bar{T}}_{\frac12}(\bar{z}),   
\label{mat_IIA_op1}
\eea 
where the bare string coupling $g_s$ is put in the r.h.s. to count the number of external lines 
of amplitudes in the IIA theory. 
Furthermore, it can be naturally extended as 
\bea
\Phi_{2k+1}=\frac{1}{N}\tr\,\phi^{2k+1}+\mbox{(mixing)} & \Leftrightarrow & 
\cV_\phi(k)\equiv g_s^2\int d^2z\, \hat{V}_{k+\frac12, \,+1}(z)\,\hat{\bar{V}}_{-k-\frac12, \,-1}(\bar{z}) , \nn \\
\Psi_{2k+1}=\frac{1}{N}\tr\,\psi^{2k+1}+\mbox{(mixing)} & \Leftrightarrow &  
\cV_\psi(k) \equiv g_s^2\int d^2z\, \hat{T}_{-k-\frac12}(z)\,\hat{\bar{V}}_{-k-\frac12, \,-1}(\bar{z}), \nn \\
\bar{\Psi}_{2k+1} = \frac{1}{N}\tr\,\bar{\psi}^{2k+1}+\mbox{(mixing)} & \Leftrightarrow &  
\cV_{\bar{\psi}}(k)\equiv g_s^2\int d^2z\, \hat{V}_{k+\frac12, \,+1}(z)\, \hat{\bar{T}}_{k+\frac12}(\bar{z})  
\label{mat_IIA_op2}
\eea
for higher $k(=1,2,\cdots)$. 
As discussed in~\cite{Kuroki:2012nt},   
if we regard $\psi$ ($\bar{\psi}$) as a target space fermion in the (NS, R) sector (the (R, NS) sector) 
in the corresponding type IIA theory,  
$\phi$ and $B$ are interpreted as an operator in the (R, R) sector and that in the (NS, NS) sector, respectively.  
Then, $(\nu_+-\nu_-)$ represents the RR charge (up to a proportional constant). 
The correspondence (\ref{mat_IIA_op1}) and (\ref{mat_IIA_op2}) are consistent with this interpretation. 

The (R$-$, R$+$) vertex operators behave as singlets under $\hat{Q}_+$ and $\hat{\bar{Q}}_-$. 
Actually, 
\be
\left[\hat{Q}_+, \,\hat{V}_{-k, \,-1}(z)\,\hat{\bar{V}}_{k, \,+1}(\bar{z})\right] 
= \left[\hat{\bar{Q}}_-, \,\hat{V}_{-k, \,-1}(z)\,\hat{\bar{V}}_{k, \,+1}(\bar{z})\right]=0  
\ee
for $k=0,1,2,\cdots$ is shown by taking the OPEs. 
The vertex operators can be expressed as $\hat{Q}_+$- and $\hat{\bar{Q}}_-$-exact forms: 
\bea
\hat{V}_{-k, \,-1}(z) & = & 
\left[\hat{Q}_+, \,\frac{1}{k!}:(\partial^k \hat{q}_+^{-1})\hat{V}_{-k,\,-1}(z):\right], \nn \\
\hat{\bar{V}}_{k,\,+1}(\bar{z}) & = & 
\left[\hat{\bar{Q}}_-, \,\frac{1}{k!}:(\bar{\partial}^k \hat{\bar{q}}_-^{-1})\hat{\bar{V}}_{k,\,+1}(\bar{z}):\right],  
\label{R-R+exactform}
\eea
where the homotopy operators 
\bea
& & \hat{q}_+^{-1}(z) = e^{\pi\beta(-\frac12p_{\bar{\phi}}+i\frac12p_{\bar{h}}+ip_{\bar{x}})}\, 
q_+^{-1}(z), \qquad 
q_+^{-1}(z)\equiv e^{\frac12\phi+\frac{i}{2}H+ix}(z), \nn \\
& & \hat{\bar{q}}_-^{-1}(\bar{z})= e^{-\pi\beta(-\frac12p_{\phi}-i\frac12p_{h}-ip_{x})}\, 
\bar{q}_-^{-1}(\bar{z}), \qquad 
\bar{q}_-^{-1}(\bar{z})\equiv e^{\frac12\bar{\phi}-\frac{i}{2}\bar{H}-i\bar{x}}(\bar{z}) 
\eea
give the inverses of the supercurrents in the sense that 
\be
\oint_z \frac{dw}{2\pi i}\,\hat{q}_+(w)\,\hat{q}_+^{-1}(z) = 1, \qquad 
\oint_{\bar{z}}\frac{d\bar{w}}{2\pi i}\,\hat{\bar{q}}_-(\bar{w})\,\hat{\bar{q}}_-^{-1}(\bar{z})=1.
\ee  
However, $:(\partial^k \hat{q}_+^{-1})\hat{V}_{-k,\,-1}(z):$ and 
$:(\bar{\partial}^k \hat{\bar{q}}_-^{-1})\hat{\bar{V}}_{k,\,+1}(\bar{z}):$ 
appearing in the r.h.s. of (\ref{R-R+exactform}) are not physical operators, because 
they belong to the NS sector but have integer $x$- and $\bar{x}$-momenta~\footnote{
Although they are in the 0-picture, 
the same conclusion holds after the picture is changed to $-1$, 
because the values of $x$- and $\varphi$-momenta are intact in the picture changing operation.}. 
Thus, we conclude that the (R$-$, R$+$) vertex operators are singlets under the target-space supersymmetry. 

In the correspondence (\ref{mat_IIA_op1}) and (\ref{mat_IIA_op2}), the (R$-$, R$+$) vertex operators seem to have 
no counterparts in the matrix model. 
The result of the amplitudes (\ref{vevPhi_MM}) and (\ref{vevPsiPsi_MM}) implies that correlators of operators 
with nonzero Ramond charges do not vanish from the viewpoint of the correspondence. 
Hence it is anticipated that the matrix model represents the type IIA theory on a nontrivial 
background of (R$-$, R$+$) operators.

\subsection{Type IIA theory on RR background} 
\label{sec:IIA_RR}
The worldsheet action of the type IIA theory consists of the free CFT part and the Liouville-like 
interaction part: 
\bea
S_{{\rm IIA}} & = & S_{{\rm CFT}} + S_{{\rm int}}, \\
S_{{\rm CFT}} & = & \frac{1}{2\pi}\int d^2z\left[\partial \varphi_{{\rm tot}}\bar{\partial}\varphi_{{\rm tot}}
+\frac{Q}{4}\sqrt{\hat{g}}\hat{R}\varphi_{{\rm tot}} 
+\partial x_{{\rm tot}}\bar{\partial}x_{{\rm tot}} +\partial H_{{\rm tot}}\bar{\partial}H_{{\rm tot}}\right] 
+(\mbox{ghosts}), \nn\\
& & \label{SCFT} \\
S_{{\rm int}} & = & \mu_1 \cV^{(0,0)}_B(0)\equiv 
\mu_1 \int d^2z\,\hat{T}^{(0)}_{-\frac12}(z)\hat{\bar{T}}^{(0)}_{\frac12}(\bar{z}), 
\label{Sint}
\eea  
where $d^2z=d({\rm Re}\,z)\,d({\rm Im}\,z)$, each boson with the suffix ``tot'' represents the sum of its 
holomorphic and anti-holomorphic parts, and 
the 0-picture NS fields $\hat{T}^{(0)}_{-\frac12}(z)$ and $\hat{\bar{T}}^{(0)}_{\frac12}(\bar{z})$ 
do not have the $\epsilon=-1$ and $\bar{\epsilon}=+1$ parts in (\ref{cocycle_T0}): 
\bea
\hat{T}^{(0)}_{-\frac12}(z) & = & \hat{T}^{(0)}_{-\frac12, \,+1}(z) = 
e^{\pi\beta(ip_{\bar{h}}-i\frac12p_{\bar{x}})}\,\frac{i}{\sqrt{2}}\,e^{iH-i\frac12x+\frac12\varphi}(z), \nn \\
\hat{\bar{T}}^{(0)}_{\frac12}(\bar{z}) & = & \hat{\bar{T}}^{(0)}_{\frac12, \,-1}(\bar{z}) = 
e^{-\pi\beta(-ip_{h}+i\frac12p_{x})}\,\frac{i}{\sqrt{2}}\,
e^{-i\bar{H}+i\frac12\bar{x} +\frac12\bar{\varphi}}(\bar{z}). 
\eea
Here and in what follows, superscripts indicating the picture numbers are put on vertex operators except 
that they have the natural pictures ($(-1)$ for NS fields and $(-\frac12)$ for R fields). 
The form of $S_{{\rm int}}$ corresponds to the term $N\tr (-i\mu^2 B)$ in the matrix model action 
via (\ref{mat_IIA_op1}) (up to a choice of the picture under identification of $1/N$ and $g_s$). 
The Liouville coupling $\mu_1$ is related to $\mu^2$ in the matrix model, 
which is clarified in section~\ref{sec:check}.   

In the trivial background, the genus-zero amplitude with insertion of integrated vertex operators 
$\cV_i=\int d^2z\,\hat{V}_i(z,\bar{z})$ 
reads 
$\frac{1}{{\rm Vol}({\rm CKG}(S^2))} \vev{\prod_i\cV_i}$
with  
\be
\vev{\prod_i\cV_i} = 
\int \cD_{\hat{g}}x_{{\rm tot}}\cD_{\hat{g}}\varphi_{{\rm tot}}\cD_{\hat{g}}H_{{\rm tot}}\cD_{\hat{g}}(\mbox{ghosts})\,
e^{-S_{{\rm IIA}}}\,\prod_i\cV_i. 
\label{vev_IIA}
\ee
Dividing by the conformal Killing group of the sphere is equivalent to fixing the positions of 
three vertex operators with $c\bar{c}$ inserted at each of the fixed positions: 
\be
\frac{1}{\mbox{Vol(CKG($S^2$))}} \vev{\prod_i\cV_i}
=\vev{\prod_{i=1}^3c\bar{c}\hat{V}_i(z_i,\bar{z}_i)\, \prod_{j\ge 4}\cV_j}.
\ee 
We take a usual choice of $(z_1, z_2, z_3)=(\infty, 1, 0)$.  
As the amplitude on a nontrivial (R$-$, R$+$) background, we consider~\footnote{
Similar treatment of an RR flux background is discussed in ref.~\cite{Takayanagi:2004ge}.} 
\bea
\left\bra\!\!\!\vev{\prod_i\cV_i}\!\!\!\right\ket & \equiv & \vev{\left(\prod_i \cV_i\right)e^{W_{{\rm RR}}}}
\nn \\
 & = & \vev{\left(\prod_i \cV_i\right) \left(1+W_{{\rm RR}} + \frac{1}{2!}(W_{{\rm RR}})^2 + \cdots\right)},  
\label{vev_IIA_RR}
\eea
where the background is incorporated as a linear combination of vertex operators in the (R$-$, R$+$) sector 
with numerical coefficients $a_k$: 
\bea
W_{{\rm RR}} &= & q_{{\rm RR}}\sum_{k\in {\bf Z}}a_k\,\mu_1^{k+1}\cV^{{\rm RR}}_k, 
\label{WRR0} \\
\cV^{{\rm RR}}_k & \equiv & \begin{cases} 
\int d^2z\,\hat{V}_{k,\,-1}(z)\,\hat{\bar{V}}_{-k,\,+1}(\bar{z}) & (k=0,-1,-2,\cdots) \\
 \int d^2z\,\hat{V}^{({\rm nonlocal})}_{-k,\,-1}(z)\,\hat{\bar{V}}^{({\rm nonlocal})}_{k,\,+1}(\bar{z}) & (k=1,2,\cdots). 
\end{cases}
\label{WRR}
\eea
$q_{{\rm RR}}$ is an RR charge related to $(\nu_+-\nu_-)$ in the matrix model. 
$\cV^{{\rm RR}}_k$ ($k=0,-1,-2,\cdots$) are the (R$-$, R$+$) vertex operators in (\ref{spectrum_after}).
On the other hand, for $\cV^{{\rm RR}}_k$ ($k=1,2,\cdots$), 
we choose (R$-$, R$+$) vertex operators of the nonlocal branch ($p_\ell = 1+|k|$). 
Since the nonlocal vertex operators are invariant under $\hat{Q}_+$ and $\hat{\bar{Q}}_-$ 
as well as the local ones, 
$W_{{\rm RR}}$ consists of the maximal set of (R$-$, R$+$) vertex operators preserving 
the target-space supersymmetry~\footnote{
Ref.~\cite{Ita:2005ne} considers different nonlocal RR vertex operators with $\epsilon, \bar{\epsilon}$ flipped 
in addition to $p_{\ell}=1+|k|$. 
They satisfy the Dirac equation constraint, but do not preserve the target-space supersymmetry.}. 
Notice that we do not regard the (R$-$, R$+$) operators as particles in asymptotic states, 
but as a background. 
From this point of view it will be natural to include nonlocal operators in addition to the local ones 
in $W_{{\rm RR}}$~\footnote{
According to ref.~\cite{Jevicki:1993zg}, we find some similarity in the $c=1$ Liouville theory. 
Nonlocal $W_\infty$ operators (forbidden by the Seiberg bound) there deform the linear dilaton background to  
a two-dimensional black-hole one~\cite{Witten:1991yr} with an infinitesimally small mass.}. 
Actually, as we will see later, the inclusion of nonlocal operators is crucial to match the matrix-model amplitudes with 
the type IIA ones.  
It should be pointed out that the nonlocal operators do not satisfy the Dirac equation constraint. 
More precisely, they are invariant under a half of the worldsheet supersymmetry transformations
($G_0^-, \bar{G}_0^+$) but not under the other half ($G^+_0, \bar{G}_0^-$). 
This point will be discussed in some detail in section~\ref{sec:nonlocalRR}. 
It seems somewhat similar to a boundary operator, and tempts us to interpret it as a certain brane-like object, 
which however preserves linear combinations of 
holomorphic and anti-holomorphic generators (for example, $G_0^++i\bar{G}_0^-$ and 
$G_0^-+i\bar{G}_0^+$ for an A-brane configuration~\cite{Ooguri:1996ck}). 
Such a brane could exist without breaking the target-space supersymmetry in our case, since 
the supersymmetry does not induce translations in the target space. 
It would be interesting to proceed analysis from the viewpoint of this interpretation. 

The treatment of the background (\ref{vev_IIA_RR}) is a perturbation from the trivial background, and 
valid for small $|q_{{\rm RR}}|$. 
We exactly compute the path-integral with respect to the constant mode of the Liouville coordinate $\varphi_{{\rm tot}}$ 
as performed in refs.~\cite{Gupta:1989fu,Goulian:1990qr} to obtain 
\be
\vev{\prod_i\cV_i}= 2\Gamma(-s)\mu_1^s\,\frac{1}{V_L} \vev{\left(\prod_i\cV_i\right) 
\cV^{(0,0)}_B(0)^s}_{\rm CFT},
\label{amp_CFT}
\ee
where $s=-2\sum_ip_{\ell\, i} +Q\,\chi(S^2)$ ($\chi(M)$ denotes the Euler number of the manifold $M$), 
$V_L$ is the volume of the Liouville direction, and 
the suffix ``CFT'' means the correlator computed under the free CFT action $S_{\rm CFT}$ in \eqref{Sint}. 
Calculation of the amplitude (\ref{amp_CFT}) can be explicitly carried out only when $s$ is a nonnegative integer. 
Then, according to \cite{Di Francesco:1991ud}, the divergent factor $\Gamma(-s)$ is regularized as 
\be 
\Gamma(-s)\to \frac{(-1)^s}{s!}\,\ln \frac{1}{\mu_1}. 
\label{Liouville_V}
\ee 

As usual in computations in the RNS formalism, the total picture in each of holomorphic and anti-holomorphic 
sectors should be adjusted to $2h-2$ for a handle-$h$ Riemann surface. 
Although cocycle factors in (\ref{Sint}) and (\ref{WRR}) might seem to induce nonlocal interactions, 
we will see in the following that they merely give phase factors to amplitudes reflecting target-space 
statistics. 
 
\section{Basic amplitudes}
\label{sec:basic}
\setcounter{equation}{0}
As a preparation to obtain IIA amplitudes on the RR background, we compute some basic 
CFT amplitudes in the form (\ref{amp_CFT}). We will consider various amplitudes which contain 
RR fields and are relevant to comparison with the matrix model results. 
The nonlocal branch ($p_{\ell}=1+|k|$) as well as the local one ($p_{\ell}=1-|k|$) are considered 
for (R$-$, R$+$) vertex operators. Amplitudes which consist only of 
``tachyons" are briefly mentioned at the end of this section. 

Before computation, we notice that in the spectrum \eqref{spectrum_after} the target-space bosons 
coming from the (NS, NS) and (R$\pm$, R$\mp$) sectors are ``winding-like" $\bar k=-k$, while 
the target-space fermions from the (NS, R$-$) and (R$+$, NS) sectors are ``momentum-like" $\bar k=k$. 
Then it immediately follows that momentum/winding in the $x$-direction is conserved separately 
in the bosons and fermions. As a corollary, we conclude that if we have a fermion 
in the (NS, R$-$) sector, there must be a one in the (R$+$, NS) sector and vice versa 
to obtain a nontrivial amplitude.

\subsection{(NS, NS)-(R$+$, R$-$)-(R$-$, R$+$)}
\label{sec:3pt_BBB}
We first compute the three-point amplitude among the fields of (NS, NS), (R$+$, R$-$) and (R$-$, R$+$): 
\bea
\hat{V}_1(z_1, \bar{z}_1) & = & \hat{T}_{k_1}(z_1)\,\hat{\bar{T}}_{-k_1}(\bar{z}_1)
\qquad \left(k_1\in {\bf Z}+\frac12\right), \nn \\
\hat{V}_2(z_2, \bar{z}_2) & = & \hat{V}_{k_2,\,+1}(z_2)\,\hat{\bar{V}}_{-k_2, \,-1}(\bar{z}_2)
\qquad \left(k_2=\frac12, \frac32, \cdots\right), \nn \\
\hat{V}_3(z_3,\bar{z}_3) & = & \hat{V}_{k_3,\,-1}(z_3)\,\hat{\bar{V}}_{-k_3,\,+1}(\bar{z}_3)
\qquad (k_3=0,-1,-2, \cdots), 
\eea
which is compared to the matrix model amplitude (\ref{vevPhi_MM}) in section~\ref{sec:vevPhi_MM}. 
{}From the conservation of $H$ and $\bar{H}$ charges (or equivalently, integrals over the zero-modes 
$h_0$ and $\bar{h}_0$ in (\ref{amp_CFT})), only the $s=0$ case is possibly nonvanishing. 
The $s=0$ amplitude reads 
\bea
 & & \left.\vev{\prod_{i=1}^3\hat{V}_i(z_i,\bar{z}_i)}\right|_{s=0} = \left(2\ln \frac{1}{\mu_1}\right)
\frac{1}{V_L} \nn \\
 & & \hspace{7mm} \times \bra 0|\hat{T}_{k_1}(z_1)\,\hat{\bar{T}}_{-k_1}(\bar{z}_1)\,
\hat{V}_{k_2,\,+1}(z_2)\,\hat{\bar{V}}_{-k_2, \,-1}(\bar{z}_2)\,
\hat{V}_{k_3,\,-1}(z_3)\,\hat{\bar{V}}_{-k_3,\,+1}(\bar{z}_3)|0\ket .
\label{3pt_BBB}
\eea
Here, the bra vacuum $\bra 0|$ has the background charge $(+2, +2)$ for the bosonized superconformal ghost 
$(\phi, \bar{\phi})$ and the background charge $-\frac{Q}{2}\,\chi(S^2)=-2$ for the Liouville field, 
while the ket vacuum $|0\ket$ is neutral for both of these charges. 
By using (\ref{Tbar_V_hat}) and (\ref{Vbar_V_hat}), the last line in (\ref{3pt_BBB}) becomes 
\bea
e^{i2\pi\beta(-\frac32+\sum_{i<j}k_ik_j)} & & 
\hspace{-5mm}\bra 0|\hat{T}_{k_1}(z_1)\,\hat{V}_{k_2,\,+1}(z_2)\,\hat{V}_{k_3,\,-1}(z_3) \nn \\
& & \times \hat{\bar{T}}_{-k_1}(\bar{z}_1)\,\hat{\bar{V}}_{-k_2, \,-1}(\bar{z}_2)\,\hat{\bar{V}}_{-k_3,\,+1}(\bar{z}_3)|0\ket. 
\label{3pt_BBB_lastline}
\eea
We move the three cocycle factors in the last line to act on $|0\ket$.  
Since $|0\ket$ is annihilated by $p_\phi$, $p_h$ and $p_x$, the cocycle factors do not work anymore: 
\be
\hat{\bar{T}}_{-k_1}(\bar{z}_1)\,\hat{\bar{V}}_{-k_2, \,-1}(\bar{z}_2)\,\hat{\bar{V}}_{-k_3,\,+1}(\bar{z}_3)|0\ket
=\bar{T}_{-k_1}(\bar{z}_1)\,\bar{V}_{-k_2, \,-1}(\bar{z}_2)\,\bar{V}_{-k_3,\,+1}(\bar{z}_3)|0\ket .
\ee
Similarly, by moving the three cocycle factors in the first line to act on $\bra 0|$, 
the phase factor $e^{i4\pi\beta}$ arises picking up the background charge for $\bar{\phi}$. 
However, because of $\beta\in {\bf Z}+\frac12$ the phase is trivial. Thus,  
\be
\bra 0|\hat{T}_{k_1}(z_1)\,\hat{V}_{k_2,\,+1}(z_2)\,\hat{V}_{k_3,\,-1}(z_3)
=\bra 0|T_{k_1}(z_1)\,V_{k_2,\,+1}(z_2)\,V_{k_3,\,-1}(z_3). 
\ee
Now, the amplitude is factorized into the holomorphic and the anti-holomorphic part as 
\bea
& & \left.\vev{\prod_{i=1}^3\hat{V}_i(z_i,\bar{z}_i)}\right|_{s=0} = \left(2\ln \frac{1}{\mu_1}\right)
\frac{1}{V_L} \,e^{i2\pi\beta(-\frac32+\sum_{i<j}k_ik_j)} \nn \\
& & \hspace{7mm} \times \bra 0|T_{k_1}(z_1)\,V_{k_2,\,+1}(z_2)\,V_{k_3,\,-1}(z_3)|0\ket \, 
 \bra 0| \bar{T}_{-k_1}(\bar{z}_1)\,\bar{V}_{-k_2,\,-1}(\bar{z}_2)\,\bar{V}_{-k_3,\,+1}(\bar{z}_3)|0\ket . \nn \\
& & \label{3pt_BBB_1} 
\eea
The last line in (\ref{3pt_BBB_1}) is computed by the Wick contraction. We end up with 
\bea
 & & \left.\vev{\prod_{i=1}^3\hat{V}_i(z_i,\bar{z}_i)}\right|_{s=0} = \delta_{\sum_ik_i, \,0}\, 
 \delta_{\sum_ip_{\ell_i}, \,2}\,
\left(2\ln \frac{1}{\mu_1}\right) \,e^{i2\pi\beta(-\frac32-\frac12\sum_ik_i^2)} \nn \\
& & \hspace{7mm} \times |z_1-z_2|^{-1}|z_1-z_3|^{-1}|z_2-z_3|^{-1}
\prod_{i<j}|z_i-z_j|^{2(k_ik_j-p_{\ell_i}p_{\ell_j})}, 
\label{3pt_BBB_2}
\eea
where the factor $\frac{1}{V_L}$ in (\ref{3pt_BBB_1}) is canceled with $V_L$ 
from the delta-function of the conservation of the Liouville momentum: 
\be
\delta\left(\sum_ip_{\ell_i}-2\right)=V_L\,\delta_{\sum_ip_{\ell_i}, \,2},
\label{Liouvillevol}
\ee
and the phase factor in (\ref{3pt_BBB_1}) was recast by using the $x$-winding conservation $\sum_ik_i=0$. 
We fix three positions as $(z_1,z_2,z_3)=(\infty, 1, 0)$ 
with inserting $c\bar{c}$ at each of them. The result is 
\be
\left.\vev{\prod_{i=1}^3c\bar{c}\hat{V}_i(z_i,\bar{z}_i)}\right|_{s=0, \,(z_1,z_2,z_3)=(\infty, 1, 0)} = 
\delta_{\sum_ik_i, \,0}\,\delta_{\sum_ip_{\ell_i}, \,2}\,
\left(2\ln \frac{1}{\mu_1}\right) \,e^{i2\pi\beta(-\frac32-\frac12\sum_ik_i^2)}. 
\label{3pt_BBB_f}
\ee
The kinematical constraints ($\sum_ik_i=0$ and $\sum_ip_{\ell_i}=2$) are met by 
\be
(k_1, \,k_2, \,k_3) = \left(-\frac12, \,\frac12, \,0\right), \qquad 
(p_{\ell_1}, \,p_{\ell_2}, \,p_{\ell_3}) = \left(\frac12, \,\frac12, \,1\right)
\ee
for the local branch of $\hat{V}_3$, and by 
\be
(k_1, \,k_2, \,k_3) =\left(-\frac12, \,k+\frac12, \,-k\right), \qquad  
(p_{\ell_1}, \,p_{\ell_2}, \,p_{\ell_3}) = \left(\frac12, \,-k+\frac12, \,k+1\right) 
\label{3pt_BBB_kin_NL}
\ee
with $k=1,2,\cdots$ 
for the nonlocal branch. 
Notice that if we did not allow the nonlocal branch, $k_2$ could not take values in ${\bf N}+\frac12$.  

\subsection{2(R$+$, R$-$)-2(R$-$, R$+$)}
\label{sec:4pt_BBBB}
Let us compute the four-point amplitude of two (R$+$, R$-$) and two (R$-$, R$+$) fields: 
\bea
\hat{V}_a(z_a,\bar{z}_a) & = & \hat{V}_{k_a, \,+1}(z_a)\,\hat{\bar{V}}_{-k_a, \,-1}(\bar{z}_a) \qquad 
\left(k_a=\frac12, \frac32, \cdots\right), \nn \\
\hat{V}_b(z_b,\bar{z}_b) & = & \hat{V}_{k_b,\,-1}(z_b)\,\hat{\bar{V}}_{-k_b,\,+1}(\bar{z}_b) \qquad
\left(k_b=0,-1,-2,\cdots\right)
\eea
with $a=1,2$ and $b=3,4$. 
It gives the counterpart of the matrix model result (\ref{vevPhiPhi_MM}) or (\ref{vevPhiPhi_MM2})
as we see in section~\ref{sec:vevPhiPhi_MM}. 
Only the $s=0$ case in the amplitude (\ref{amp_CFT}) can be nontrivial 
from the conservation of $H$ and $\bar{H}$ charges. 
A parallel argument to (\ref{3pt_BBB})-(\ref{3pt_BBB_2}) leads to  
\bea
& & \left.\vev{\prod_{i=1}^4\hat{V}_i(z_i,\bar{z}_i)}\right|_{s=0} = \delta_{\sum_ik_i,\, 0}
\,\delta_{\sum_ip_{\ell_i},\,2}
\,\left(2\ln \frac{1}{\mu_1}\right) \,e^{-i\pi\beta\sum_ik_i^2} \nn \\
& & \hspace{7mm} \times |z_1-z_3|^{-1}|z_1-z_4|^{-1}|z_2-z_3|^{-1}|z_2-z_4|^{-1}
\prod_{i<j}|z_i-z_j|^{2(k_ik_j-p_{\ell_i}p_{\ell_j})} .
\label{4pt_BBBB}
\eea 
The corresponding string amplitude on the trivial background is obtained from (\ref{4pt_BBBB}) by fixing 
the first three positions as 
$(z_1,z_2,z_3)=(\infty, 1,0)$ and integrating the rest ($z_4$). Then, we have 
\bea
& & \left.\vev{\prod_{i=1}^3c\bar{c}\hat{V}_i(z_i,\bar{z}_i)\,
\int d^2z_4\hat{V}_4(z_4,\bar{z}_4)}\right|_{s=0,\,(z_1,z_2,z_3)=(\infty, 1,0)} \nn \\
& & \hspace{7mm} = \delta_{\sum_ik_i, \,0}\,\delta_{\sum_ip_{\ell_i},\,2} \, 
\left(2\ln \frac{1}{\mu_1}\right) \, e^{-i\pi\beta\sum_ik_i^2}  \, \cI_{(1,0)},  
\label{4pt_BBBB_2}
\eea
where $\cI_{(1,0)}$ is the integral $I_{(1,0)}$ defined by (\ref{I(1,0)}) with 
\be
\alpha=\bar{\alpha}=k_3k_4-p_{\ell_3}p_{\ell_4}, \qquad 
\beta=\bar{\beta}=k_2k_4-p_{\ell_2}p_{\ell_4}-\frac12.  
\label{4pt_alpha_beta}
\ee 
The kinematics restricts $k_i$ and $p_{\ell_i}$ as follows. For both of $\hat{V}_b$ belonging to 
the local branch, 
\be
(k_1,\,k_2,\,k_3,\,k_4)=\left(\frac12,\,\frac12,\,0,\,-1\right) \quad \mbox{or} \quad 
\left(\frac12,\,\frac12,\,-1,\,0\right)
\label{4pt_kin_L_k}
\ee
with the corresponding Liouville momenta  
\be
(p_{\ell_1},\,p_{\ell_2},\,p_{\ell_3},\,p_{\ell_4})=\left(\frac12,\,\frac12,\,1,\,0\right) \quad
 \mbox{or} \quad \left(\frac12,\,\frac12,\,0,\,1\right), 
 \label{4pt_kin_L_pl}
\ee
respectively. For one of $\hat{V}_b$ (say, $\hat{V}_3$) local and the rest ($\hat{V}_4$) nonlocal, 
\bea
(k_1,\,k_2,\,k_3,\,k_4)& = & \left(n_1+\frac12,\,n_2+\frac12,\,-1,\,-n_1-n_2\right), \nn \\
(p_{\ell_1},\,p_{\ell_2},\,p_{\ell_3},\,p_{\ell_4}) & = &
\left(-n_1+\frac12,\,-n_2+\frac12,\,0,\,n_1+n_2+1\right)
\label{4pt_kin_NL}
\eea
with $n_1, n_2=0,1,2,\cdots$ but $(n_1, n_2)\neq (0,0)$.   
The case of both of $\hat{V}_b$ nonlocal is not allowed. 

If we try to plug these on-shell values into $I_{(1,0)}$ in (\ref{I(1,0)_f_II}) (or (\ref{I(1,0)_f_I})) directly 
to get $\cI_{(1,0)}$, it becomes indefinite or divergent. 
Thus we adopt the following prescription as a regularization. 
All the powers of the integrand in $I_{(1,0)}$ are uniformly shifted, i.e. 
$\alpha$, $\bar{\alpha}$, $\beta$, $\bar{\beta}$ are shifted by the same small quantity $\varepsilon$. 
Note that the uniform shift preserves the mutual locality of vertex operators and thus 
the equality between (\ref{I(1,0)_f_II}) and (\ref{I(1,0)_f_I}) (see (\ref{cond_I=II})). 
In the regularized result, we take $\frac{1}{\varepsilon}$ proportional to the volume of 
the Liouville direction: 
\be
\frac{1}{\varepsilon}=c_L\,\left(2\ln \frac{1}{\mu_1}\right) 
\label{varepsilon_VL}
\ee
with $c_L$ being a proportional constant. 
Since the divergence can be interpreted as a resonance in string theory, 
it seems plausible to regard $\frac{1}{\varepsilon}$ as the Liouville 
volume. Namely, it essentially has the same origin as in \eqref{Liouville_V}. 
Similar treatment is found in $c=1$ noncritical bosonic string theory~\cite{Klebanov:1991qa}.

\subsubsection{Case of both of $\hat{V}_b$ local} 
For the case of both of $\hat{V}_b$ local, $\alpha=\bar{\alpha}=0$ and $\beta=\bar{\beta}=-1$ at 
(\ref{4pt_kin_L_k}) and (\ref{4pt_kin_L_pl}). Then, 
\bea
\cI_{(1,0)} & = & \pi\,\frac{\Gamma(1+\varepsilon)\,\Gamma(\varepsilon)}{\Gamma(1+2\varepsilon)}\,
\frac{\Gamma(-2\varepsilon)}{\Gamma(-\varepsilon)\,\Gamma(1-\varepsilon)}
 = \frac{\pi}{2}\,\frac{1}{\varepsilon} +\cO(1) \nn \\
 & = & \frac{\pi}{2}\,c_L\,\left(2\ln\frac{1}{\mu_1}\right) + \cO(1).
 \label{4pt_I(1,0)_L}
\eea
Plugging (\ref{4pt_I(1,0)_L}) into (\ref{4pt_BBBB_2}), we have 
\bea 
 & & \left.\vev{\prod_{i=1}^3c\bar{c}\hat{V}_i(z_i,\bar{z}_i)\,
\int d^2z_4\hat{V}_4(z_4,\bar{z}_4)}\right|_{s=0,\,(z_1,z_2,z_3)=(\infty, 1,0)} \nn \\
& & \hspace{7mm}= \delta_{\sum_ik_i, \,0}\,\delta_{\sum_ip_{\ell_i},\,2} \,
\left(2\ln \frac{1}{\mu_1}\right)^2 \, e^{-i\pi\beta\sum_ik_i^2} \, 
\frac{\pi}{2}\,c_L. 
\label{4pt_BBBB_L_f}
\eea

\subsubsection{Case of one of $\hat{V}_b$ local} 
For the case of one of $\hat{V}_b$ (say, $\hat{V}_3$) local and the rest ($\hat{V}_4$) nonlocal, 
$\alpha=\bar{\alpha}=n_1+n_2$ and $\beta=\bar{\beta}=-n_1-1$ at (\ref{4pt_kin_NL}). 
As a result of the regularization with (\ref{varepsilon_VL}), 
\bea
\cI_{(1,0)} & = & \pi\,
\frac{\Gamma(n_1+n_2+1+\varepsilon)\,\Gamma(-n_1+\varepsilon)}{\Gamma(n_2+1+2\varepsilon)}\,
\frac{\Gamma(-n_2-2\varepsilon)}{\Gamma(-n_1-n_2-\varepsilon)\,\Gamma(n_1+1-\varepsilon)} \nn \\
 & = & \frac{\pi}{2}\,
\left(\frac{(n_1+n_2)!}{n_1!n_2!}\right)^2\,c_L\,\left(2\ln\frac{1}{\mu_1}\right) + \cO(1).
 \label{4pt_I(1,0)_NL}
\eea
Then, the amplitude finally becomes 
\bea 
 & & \left.\vev{\prod_{i=1}^3c\bar{c}\hat{V}_i(z_i,\bar{z}_i)\,
\int d^2z_4\hat{V}_4(z_4,\bar{z}_4)}\right|_{s=0,\,(z_1,z_2,z_3)=(\infty, 1,0)} \nn \\
& & \hspace{7mm}= \delta_{\sum_ik_i, \,0}\,\delta_{\sum_ip_{\ell_i},\,2} \,
\left(2\ln \frac{1}{\mu_1}\right)^2 \, e^{-i\pi\beta\sum_ik_i^2} \, 
\frac{\pi}{2}\,\left(\frac{(n_1+n_2)!}{n_1!n_2!}\right)^2\,c_L. 
\label{4pt_BBBB_NL_f}
\eea

As a consistency check, we can see that the case of $\hat{V}_4$ local and $\hat{V}_3$ nonlocal 
gives the identical result. 
It should be so, since $\hat{V}_b$ are target-space bosons.

\subsection{(NS, R$-$)-(R$+$, NS)-(R$-$, R$+$)}
Next we turn to the amplitude including the target-space fermions. 
We consider the three-point amplitude of (NS, R$-$) and (R$+$, NS) fermions and an (R$-$, R$+$) field: 
\bea
\hat{V}_1(z_1,\bar{z}_1) & = & \hat{T}_{k_1}(z_1)\,\hat{\bar{V}}_{k_1,\,-1}(\bar{z}_1) 
\qquad \left(k_1=-\frac12, -\frac32, \cdots\right), \nn\\
\hat{V}_2(z_2,\bar{z}_2) & = & \hat{V}_{k_2,\,+1}(z_2)\,\hat{\bar{T}}_{k_2}(\bar{z}_2)
\qquad \left(k_2=\frac12, \frac32, \cdots\right), \nn\\
\hat{V}_3(z_3,\bar{z}_3) & = & \hat{V}_{k_3,\,-1}(z_3)\,\hat{\bar{V}}_{-k_3,\,+1}(\bar{z}_3) 
\qquad (k_3=0,-1,-2,\cdots), 
\eea
which corresponds to the matrix-model amplitude (\ref{vevPsiPsi_MM}) with $k=\ell=0$ as is seen in 
section~\ref{sec:vevPsiPsi_MM1}.  
The $s=0$ case alone satisfies the conservation of $H$ and $\bar{H}$ charges in (\ref{amp_CFT}). 
After a similar calculation as in the previous one, we have 
\bea
 & & \left.\vev{\prod_{i=1}^3\hat{V}_i(z_i,\bar{z}_i)}\right|_{s=0} = \delta_{k_1+k_2, \,0}\,\delta_{k_3,\,0}\, 
 \delta_{\sum_ip_{\ell_i}, \,2}\,
\left(2\ln \frac{1}{\mu_1}\right) \, e^{i2\pi\beta(-1+k_1^2)} \nn \\ 
 & & \hspace{7mm} \times |z_1-z_2|^{-1}|z_1-z_3|^{-1}|z_2-z_3|^{-1}
\prod_{i<j}(z_i-z_j)^{k_ik_j-p_{\ell_i}p_{\ell_j}}\,
(\bar{z}_i-\bar{z}_j)^{\tilde{k}_i\tilde{k}_j-p_{\ell_i}p_{\ell_j}} \nn \\
& & 
\eea
with $\tilde{k}_1=k_1$, $\tilde{k}_2=k_2$, $\tilde{k}_3=-k_3$. 
Here as we mentioned at the beginning of this section, $\delta_{k_1+k_2, \,0}$ and $\delta_{k_3,\,0}$ represent 
the conservations of $x$-momentum and of $x$-winding, respectively. The string amplitude is obtained as 
\be
\left.\vev{\prod_{i=1}^3c\bar{c}\hat{V}_i(z_i,\bar{z}_i)}\right|_{s=0, \,(z_1,z_2,z_3)=(\infty, 1, 0)} = 
\delta_{k_1+k_2, \,0}\,\delta_{k_3,\,0}\,\delta_{\sum_ip_{\ell_i}, \,2}\,
\left(2\ln \frac{1}{\mu_1}\right) \,e^{i2\pi\beta(-1+k_1^2)}. 
\label{3pt_FFB_f}
\ee
The kinematics allows only the possibility 
\be
(k_1,\,k_2,\,k_3)=\left(-\frac12,\,\frac12,\,0\right), \qquad 
(p_{\ell_1},\,p_{\ell_2},\,p_{\ell_3})=\left(\frac12,\,\frac12,\,1\right)
\label{3pt_FFB_kin}
\ee
with the local branch of $\hat{V}_3$.  Its nonlocal branch is not allowed.

\subsection{(NS, R$-$)-(R$+$, NS)-3(R$-$, R$+$)}
In order to obtain the counterpart of (\ref{vevPsiPsi_MM}) with $k=\ell=1$, 
we consider the five-point amplitude of (NS, R$-$) and (R$+$, NS) fermions and three (R$-$, R$+$) 
fields: 
\bea
\hat{V}_1(z_1,\bar{z}_1) & = & \hat{T}^{(0)}_{k_1}(z_1)\,\hat{\bar{V}}_{k_1,\,-1}(\bar{z}_1) 
\qquad \left(k_1=-\frac12, -\frac32, \cdots\right), \nn\\
\hat{V}_2(z_2,\bar{z}_2) & = & \hat{V}_{k_2,\,+1}(z_2)\,\hat{\bar{T}}^{(0)}_{k_2}(\bar{z}_2)
\qquad \left(k_2=\frac12, \frac32, \cdots\right), \nn\\
\hat{V}_a(z_a,\bar{z}_a) & = & \hat{V}_{k_a,\,-1}(z_a)\,\hat{\bar{V}}_{-k_a,\,+1}(\bar{z}_a) 
\qquad (k_a=0,-1,-2,\cdots) 
\eea
with $a=3,4,5$. From the conservation of $H$ and $\bar{H}$ charges, 
the case other than $s=0, 2$ in 
(\ref{amp_CFT}) vanishes. Furthermore, the conservation of $x$-winding ($\sum_ak_a-\frac{s}{2}=0$) 
singles out the $s=0$ case as the nontrivial one and then $k_a=0$ for all $a$~\footnote{ 
The fact that $s=0$ and $k_a=0$ for all $a$ is also the case with general $k=\ell\neq 0,1$, 
where we have a $(2k+3)$-point amplitude with (NS, R$-$) and (R$+$, NS) fermions and $2k+1$ (R$-$, R$+$) fields.
}. 

From the Wick contraction,  
\bea
 & & \left.\vev{\prod_{i=1}^5\hat{V}_i(z_i,\bar{z}_i)}\right|_{s=0} = 
\frac{-1}{2}\,(p_{\ell_1}-k_1)(p_{\ell_2}+k_2)\,\delta_{k_1+k_2, \,0}\,\delta_{\sum_ak_a,\,0}\, 
 \delta_{\sum_ip_{\ell_i}, \,2}\,\nn \\
&  & \hspace{7mm} \times \left(2\ln \frac{1}{\mu_1}\right) \, 
e^{i2\pi\beta(-3+\frac12\sum_{i=1}^5k_i\tilde{k}_i)}\,|z_1-z_2| \nn \\ 
 & & \hspace{7mm} \times \left(\prod_{i=1}^2\prod_{a=3}^5|z_i-z_a|^{-1}\right)\,
\prod_{i<j}(z_i-z_j)^{k_ik_j-p_{\ell_i}p_{\ell_j}}\,(\bar{z}_i-\bar{z}_j)^{\tilde{k}_i\tilde{k}_j-p_{\ell_i}p_{\ell_j}}, 
\nn \\
& & 
\eea
where $\tilde{k}_1= k_1$, $\tilde{k}_2= k_2$, but $\tilde{k}_a= -k_a$. 
Note that the $\epsilon=-1$ part of $\hat{T}^{(0)}_{k_1}$ or 
the $\bar{\epsilon}=+1$ part of $\hat{\bar{T}}_{k_2}$ does not contribute to the amplitude, which ensures 
the correct target-space statistics as discussed in appendix~\ref{app:cocycle_T0}.
 
The kinematics restricts $k_i$ and $p_{\ell_i}$ as 
\be
(k_1,\,k_2,\,k_a)=\left(-\frac32, \,\frac32, \,0\right), \qquad 
(p_{\ell_1}, \,p_{\ell_2}, \,p_{\ell_a})= \left(-\frac12, \,-\frac12,\,1\right)
\label{5pt_kin}
\ee
and all of $\hat{V}_a$ local.  
It leads to the string amplitude 
\bea
& & \left.\vev{\prod_{i=1}^3c\bar{c}\hat{V}_i(z_i,\bar{z}_i)\,
\prod_{j=4,5}\int d^2z_j\,\hat{V}_j(z_j,\bar{z}_j)}\right|_{s=0, \,(z_1,z_2,z_3)=(\infty, 1, 0)} \nn \\
& & \hspace{7mm} = \frac{-1}{2}\,
\delta_{k_1+k_2, \,0}\,\delta_{k_3+k_4+k_5,\,0}\,\delta_{\sum_ip_{\ell_i}, \,2} \, 
\left(2\ln \frac{1}{\mu_1}\right) \,e^{i2\pi\beta(-1+k_1^2)} \,\cI_{(1,1)}. 
\label{5pt_1}
\eea
Here, $\cI_{(1,1)}$ is the integral $I_{(1,1)}$ given by (\ref{I(1,1)}) with 
\bea
\alpha=\bar{\alpha}=k_3k_4-p_{\ell_3}p_{\ell_4}, & & \alpha'=\bar{\alpha}'=k_3k_5-p_{\ell_3}p_{\ell_5}, \nn \\
\beta = k_2k_4-p_{\ell_2}p_{\ell_4}-\frac12, & & \beta'=k_2k_5-p_{\ell_2}p_{\ell_5}-\frac12, \nn \\
\bar{\beta} = -k_2k_4-p_{\ell_2}p_{\ell_4}-\frac12, & & \bar{\beta}'=-k_2k_5-p_{\ell_2}p_{\ell_5}-\frac12, 
\nn \\
2\sigma=k_4k_5-p_{\ell_4}p_{\ell_5}. & & 
\label{5pt_alpha_sigma}
\eea
To evaluate $\cI_{(1,1)}$ at the on-shell momenta (\ref{5pt_kin}) by (\ref{I(1,1)_f_II}), 
we use the regularization mentioned above. 
Namely, all the powers of the integrand in $I_{(1,1)}$ 
($\alpha$, $\beta$, $\alpha'$, $\beta'$, $\bar{\alpha}$, $\bar{\beta}$, $\bar{\alpha}'$, $\bar{\beta}'$, 
$2\sigma$) are uniformly shifted by $\varepsilon$. Then, $\gamma$ and $\gamma'$ given by (\ref{gamma_gamma'}) 
become 
\be
\gamma\to \gamma -3\varepsilon, \qquad \gamma'\to \gamma'-3\varepsilon. 
\ee 
Under the shift, we have 
\bea
& & C^{12}[\bar{\alpha}_i,\,\bar{\alpha}_i'] = C^{12}[\bar{\alpha}'_i,\,\bar{\alpha}_i] = 
-\frac{3}{\varepsilon}+\cO(1), \nn \\
& & C^{23}[\alpha_i,\,\alpha_i'] = C^{23}[\alpha_i', \,\alpha_i]=\frac{1}{\varepsilon} + \cO(1)
\eea
from the expressions (\ref{C12}) and (\ref{C23}). Here, the formula
\be
_3F_2(x,y,z\,;x-y+1,x-z+1;\,1) 
=\frac{\Gamma(\frac{x}{2}+1)\,\Gamma(x-y+1)\,\Gamma(x-z+1)\,\Gamma(\frac{x}{2}-y-z+1)}{\Gamma(x+1)\,
\Gamma(\frac{x}{2}-y+1)\,\Gamma(\frac{x}{2}-z+1)\,\Gamma(x-y-z+1)} 
\ee
is useful. The factors $s(\beta)$, $s(\beta')$, $s(\beta+2\sigma)$ and $s(\beta'+2\sigma)$ 
in (\ref{I(1,1)_f_II}) become $\cO(\varepsilon)$ quantities, 
which absorb the divergent factors from $C^{12}$'s and $C^{23}$'s. Thus, 
the final result of $\cI_{(1,1)}$ is finite: 
\be
\cI_{(1,1)}=6\pi^2.
\label{5pt_I(1,1)_f}
\ee
{}From (\ref{5pt_1}) and (\ref{5pt_I(1,1)_f}), we end up with 
\bea
 & & \left.\vev{\prod_{i=1}^3c\bar{c}\hat{V}_i(z_i,\bar{z}_i)\,\prod_{j=4,5}\hat{V}_j(z_j,\bar{z}_j)}\right|_{s=0, \,(z_1,z_2,z_3)=(\infty, 1, 0)} \nn \\
& & \hspace{7mm} = 
\delta_{k_1+k_2, \,0}\,\delta_{k_3+k_4+k_5,\,0}\,\delta_{\sum_ip_{\ell_i}, \,2} \, 
\left(2\ln \frac{1}{\mu_1}\right) \,e^{i2\pi\beta(-1+k_1^2)} \,(-3\pi^2). 
\label{5pt_f}
\eea

\subsection{(NS, NS)-(NS, R$-$)-(R$+$, NS)-(R$+$, R$-$)}
\label{sec:4pt_BFFB}
Let us compute the four-point amplitude of (NS, NS), (R$+$, R$-$) bosons and (NS, R$-$), (R$+$, NS) fermions: 
\bea
\hat{V}_1(z_1, \bar{z}_1) & = & \hat{T}^{(0)}_{k_1}(z_1)\,\hat{\bar{T}}^{(0)}_{-k_1}(\bar{z}_1) \qquad 
\left(k_1\in {\bf Z}+\frac12\right), \nn \\
\hat{V}_2(z_2,\bar{z}_2) & = & \hat{T}_{k_2}(z_2)\,\hat{\bar{V}}_{k_2,\,-1}(\bar{z}_2) \qquad
\left(k_2=-\frac12,-\frac32, \cdots\right), \nn \\
\hat{V}_2(z_2,\bar{z}_2) & = & \hat{V}_{k_3,\,+1}(z_3)\,\hat{\bar{T}}_{k_3}(\bar{z}_3) \qquad 
\left(k_3=\frac12,\frac32,\cdots\right), \nn \\
\hat{V}_4(z_4,\bar{z}_4) & = & \hat{V}_{k_4,\,+1}(z_4)\,\hat{\bar{V}}_{-k_4,\,-1}(\bar{z}_4) \qquad 
\left(k_4=\frac12,\frac32,\cdots\right).
\eea
The conservation of $H$ and $\bar{H}$ charges shows that the amplitude (\ref{amp_CFT}) 
for $s=0, -2$ can be nontrivial. 
Although the $s=-2$ case is not calculable by the standard CFT technique, the kinematical constraint from 
the Liouville momentum leads to $\sum_i|k_i|=1$, which however is not met because of $|k_i|\geq \frac12$ 
for all $i$. Thus, the $s=-2$ amplitude should vanish, and the $s=0$ case alone remains to be considered~\footnote{
Note that arguments based on the charge or momentum/winding conservation can be applied even in the case of 
negative $s$, since it concerns solely the corresponding zero-modes.}. 
The Wick contraction leads to 
\bea
& & \left. \vev{\prod_{i=1}^4\hat{V}_i(z_i,\bar{z}_i)}\right|_{s=0} =
\frac{-1}{2}\,(p_{\ell_1}+k_1)^2\,\delta_{k_1+k_4,\,0}\delta_{k_2+k_3,\,0}\,
\delta_{\sum_ip_{\ell_i},\,2}\,\left(2\ln\frac{1}{\mu_1}\right)\,e^{i2\pi\beta(-\frac32-k_1^2+k_2^2)}\nn \\
& & \hspace{7mm}\times 
|z_1-z_4|^{-1}|z_2-z_3|^{-1}(z_1-z_3)^{-1/2}(z_2-z_4)^{-1/2}
(\bar{z}_1-\bar{z}_2)^{-1/2}(\bar{z}_3-\bar{z}_4)^{-1/2}\nn\\
& & \hspace{7mm}\times \prod_{i<j}(z_i-z_j)^{k_ik_j-p_{\ell_i}p_{\ell_j}}\,
(\bar{z}_i-\bar{z}_j)^{\tilde{k}_i\tilde{k}_j-p_{\ell_i}p_{\ell_j}} 
\eea
with $\tilde{k}_1=-k_1$, $\tilde{k}_2=k_2$, $\tilde{k}_3=k_3$, $\tilde{k}_4=-k_4$. 
The kinematical constraints allow 
\be
k_1=k_2=-\frac12, \quad k_3=k_4=\frac12, \qquad p_{\ell_i}=\frac12 \quad \mbox{for all }i.
\label{4pt_BFFB_kin}
\ee

Then, the corresponding string amplitude is expressed as 
\bea
& & \left.\vev{\prod_{i=1}^3c\bar{c}\hat{V}_i(z_i,\bar{z}_i)\,
\int d^2z_4\,\hat{V}_4(z_4,\bar{z}_4)}\right|_{s=0, \,(z_1,z_2,z_3)=(\infty, 1, 0)} \nn \\
& & \hspace{7mm} = \frac{1}{2}\,(p_{\ell_1}+k_1)^2\,
\delta_{k_1+k_4, \,0}\,\delta_{k_2+k_3,\,0}\,\delta_{\sum_ip_{\ell_i}, \,2} \, 
\left(2\ln \frac{1}{\mu_1}\right) \,e^{-i3\pi\beta}\,\cI_{(1,0)}, 
\label{4pt_BFFB_1}
\eea
where $\cI_{(1,0)}$ is given by $I_{(1,0)}$ in (\ref{I(1,0)}) with 
\bea
\alpha = k_3k_4-p_{\ell_3}p_{\ell_4}, & & \bar{\alpha} = -k_3k_4-p_{\ell_3}p_{\ell_4}-\frac12, \nn \\
\beta = k_2k_4-p_{\ell_2}p_{\ell_4}-\frac12, & & \bar{\beta} = -k_2k_4-p_{\ell_2}p_{\ell_4}.
\eea
We calculate $\cI_{(1,0)}$ at the on-shell value (\ref{4pt_BFFB_kin}) by the regularization method to obtain 
\be
\cI_{(1,0)}=\frac{\pi}{2}\,\frac{1}{\varepsilon}=\frac{\pi}{2}\,c_L\left(2\ln\frac{1}{\mu_1}\right). 
\ee
However, since the factor $(p_{\ell_1}+k_1)^2$ vanishes for (\ref{4pt_BFFB_kin}),  
we conclude that the amplitude is trivial:  
\be
\left.\vev{\prod_{i=1}^3c\bar{c}\hat{V}_i(z_i,\bar{z}_i)\,
\int d^2z_4\,\hat{V}_4(z_4,\bar{z}_4)}\right|_{s=0, \,(z_1,z_2,z_3)=(\infty, 1, 0)} =0.
\label{4pt_BFFB_f}
\ee

The result (\ref{4pt_BFFB_f}) can be intuitively understood. From the $H$ and $\bar{H}$ charge conservation, 
only the $(\epsilon, \bar{\epsilon})=(-1, +1)$ part of 
\be
\hat{V}_1= \hat{T}^{(0)}_{k_1}\,\hat{\bar{T}}^{(0)}_{-k_1}=\sum_{\epsilon, \bar{\epsilon}=\pm 1} 
\hat{T}^{(0)}_{k_1,\,\epsilon}\,\hat{\bar{T}}^{(0)}_{-k_1,\,\bar{\epsilon}}
\ee 
is allowed to contribute to the amplitude. 
But, that part disappears at the on-shell \eqref{4pt_BFFB_kin} from \eqref{T0}, 
and the amplitude must vanish. It supports the validity of the regularization method.   

\subsection{2(NS, NS)-2(R$+$, R$-$)}
The four-point amplitude of
\bea
\hat{V}_1(z_1,\bar{z}_1) & = & \hat{T}_{k_1}(z_1)\,\hat{\bar{T}}_{-k_1}(\bar{z}_1), \nn \\
\hat{V}_2(z_2,\bar{z}_2) & = & \hat{T}^{(0)}_{k_2}(z_2)\,\hat{\bar{T}}^{(0)}_{-k_2}(\bar{z}_2) \qquad 
\left(k_1,k_2\in {\bf Z}+\frac12\right), \nn \\
\hat{V}_b(z_b,\bar{z}_b) & = & \hat{V}_{k_b,\,+1}(z_b)\,\hat{\bar{V}}_{-k_b,\,-1}(\bar{z}_b) \qquad 
\left(k_b=\frac12,\frac32,\cdots\right)
\eea
with $b=3,4$ can be computed by following the same lines as in section~\ref{sec:4pt_BFFB}. 
$H$ and $\bar{H}$ charges conserve only for $s=0, -2$, and the kinematical constraint for the Liouville momentum 
allows only the $s=0$ case. From the result  
\bea
& & \left. \vev{\prod_{i=1}^4\hat{V}_i(z_i,\bar{z}_i)}\right|_{s=0} =
\frac{-1}{2}\,(p_{\ell_2}+k_2)^2\,\delta_{\sum_ik_i,\,0}\,
\delta_{\sum_ip_{\ell_i},\,2}\,\left(2\ln\frac{1}{\mu_1}\right)\,e^{-i\pi\beta\sum_ik_i^2}\nn \\
& & \hspace{42mm}\times \left(\prod_{a=1,2}\prod_{b=3,4}|z_a-z_b|^{-1}\right)\,
\prod_{i<j}|z_i-z_j|^{2(k_ik_j-p_{\ell_i}p_{\ell_j})},
\eea
we have the expression of the string amplitude
\bea
& & \left.\vev{\prod_{i=1}^3c\bar{c}\hat{V}_i(z_i,\bar{z}_i)\,
\int d^2z_4\,\hat{V}_4(z_4,\bar{z}_4)}\right|_{s=0, \,(z_1,z_2,z_3)=(\infty, 1, 0)} \nn \\
& & \hspace{7mm} = \frac{-1}{2}\,(p_{\ell_2}+k_2)^2\,
\delta_{\sum_ik_i, \,0}\,\delta_{\sum_ip_{\ell_i}, \,2} \, 
\left(2\ln \frac{1}{\mu_1}\right) \,e^{-i\pi\beta\sum_ik_i^2}\,\cI_{(1,0)}.  
\label{4pt_BBBB3}
\eea
Here, $\cI_{(1,0)}$ is (\ref{I(1,0)}) with 
\be
\alpha=\bar{\alpha}=k_3k_4-p_{\ell_3}p_{\ell_4}, 
\qquad
\beta=\bar{\beta}=k_2k_4-p_{\ell_2}p_{\ell_4}-\frac12.
\ee
$\cI_{(1,0)}$ evaluated at the on-shell value 
\be
k_1=k_2=-\frac12, \quad k_3=k_4=\frac12, \qquad p_{\ell_i}=\frac12 \quad \mbox{for all }i
\label{4pt_BBBB3_kin}
\ee
by the regularization becomes
\be
\cI_{(1,0)}=\frac{\pi}{2}\,c_L\left(2\ln\frac{1}{\mu_1}\right). 
\ee
Since the factor $(p_{\ell_2}+k_2)^2$ vanishes at the on-shell, we find 
\be
\left.\vev{\prod_{i=1}^3c\bar{c}\hat{V}_i(z_i,\bar{z}_i)\,
\int d^2z_4\,\hat{V}_4(z_4,\bar{z}_4)}\right|_{s=0, \,(z_1,z_2,z_3)=(\infty, 1, 0)}=0.
\label{4pt_BBBB3_f}
\ee

\subsection{4(NS, NS)}
\label{sec:tachyons}
For readers who are interested in ``tachyon'' amplitudes, we present the $s=0$ amplitude of four 
(NS, NS) ``tachyons''~\footnote{
Since it is not used to check the correspondence with the matrix model in this paper, 
readers who want to see the correspondence quickly can skip this subsection.}   
\bea
\hat{V}_a(z_a, \bar{z}_a) & = & \hat{T}_{k_a}(z_a)\,\hat{\bar{T}}_{-k_a}(\bar{z}_a)
\qquad \left(k_a\in {\bf Z}+\frac12, \, a=1,2\right), \nn \\
\hat{V}_b(z_b,\bar{z}_b) & = & \hat{T}^{(0)}_{k_b}(z_b)\,\hat{\bar{T}}^{(0)}_{-k_b}(\bar{z}_b)
\qquad \left(k_b\in {\bf Z}+\frac12, \, b=3,4\right)  
\eea
as 
\bea
& & \left.\vev{\prod_{i=1}^3c\bar{c}\hat{V}_i(z_i,\bar{z}_i)\,
\int d^2z_4\,\hat{V}_4(z_4,\bar{z}_4)}\right|_{s=0, \,(z_1,z_2,z_3)=(\infty, 1, 0)} \nn \\
& & \hspace{7mm} = (p_{\ell_3}p_{\ell_4} -k_3k_4)^2\,
\delta_{\sum_ik_i, \,0}\,\delta_{\sum_ip_{\ell_i}, \,2} \, 
\left(2\ln \frac{1}{\mu_1}\right) \,e^{-i\pi\beta\sum_ik_i^2}\,\cI_{(1,0)},  
\label{4pt_TTTT}
\eea
where $\cI_{(1,0)}$ is (\ref{I(1,0)}) with 
\be
\alpha=\bar{\alpha}=k_3k_4-p_{\ell_3}p_{\ell_4}-1, 
\qquad
\beta=\bar{\beta}=k_2k_4-p_{\ell_2}p_{\ell_4}.
\ee
{}From (\ref{I(1,0)_f_II}) or (\ref{I(1,0)_f_I}), we can formally rewrite (\ref{4pt_TTTT}) to the form:  
\bea
& & \left.\vev{\prod_{i=1}^3c\bar{c}\hat{V}_i(z_i,\bar{z}_i)\,
\int d^2z_4\,\hat{V}_4(z_4,\bar{z}_4)}\right|_{s=0, \,(z_1,z_2,z_3)=(\infty, 1, 0)} \nn \\
& & \hspace{7mm} = 
\delta_{\sum_ik_i, \,0}\,\delta_{\sum_ip_{\ell_i}, \,2} \, 
\left(2\ln \frac{1}{\mu_1}\right) \,e^{-i\pi\beta\sum_ik_i^2}\,(-\pi)\prod_{i=1}^3
\frac{\Gamma(k_ik_4-p_{\ell_i}p_{\ell_4}+1)}{\Gamma(-k_ik_4+p_{\ell_i}p_{\ell_4})}.  
\label{4pt_TTTT2}
\eea
The factors of the gamma functions are common in ``tachyon'' amplitudes in two-dimensional 
(super)string theory, 
for example  eq. (3.14) in~\cite{Di Francesco:1991ud}.  

However, at the on-shell momenta 
\be
k_1=k_2=\frac12, \quad k_3=k_4=-\frac12, \qquad p_{\ell_i}=\frac12 \quad \mbox{for all }i, 
\label{4pt_TTTT_kin}
\ee
our regularization scheme gives 
\be
\cI_{(1,0)}=\pi c_L\left(2\ln\frac{1}{\mu_1}\right).  
\ee
Then we obtain the vanishing amplitude~\footnote
{Although in \eqref{4pt_TTTT2} we have included contributions from the 0-picture tachyons 
with $\epsilon=\bar\epsilon$ which have incorrect target-space statistics as shown in appendix \ref{app:cocycle_T0}, 
it is easy to check that  they themselves vanish at the on-shell value \eqref{4pt_TTTT_kin}.} 
\be
 \left.\vev{\prod_{i=1}^3c\bar{c}\hat{V}_i(z_i,\bar{z}_i)\,
\int d^2z_4\,\hat{V}_4(z_4,\bar{z}_4)}\right|_{s=0, \,(z_1,z_2,z_3)=(\infty, 1, 0)} =0.
\label{4pt_TTTT_f}
\ee

The result (\ref{4pt_TTTT_f}) immediately implies that the three-point function of ``tachyons'' 
for $s=1$ also vanishes from \eqref{amp_CFT} and \eqref{Liouville_V}. 
Together with the conservation of $H$, $\bar{H}$ charges and the Liouville momentum, 
we conclude that it vanishes for any $s$.

\section{Check of the correspondence}
\label{sec:check}
\setcounter{equation}{0}
In this section, we compute various amplitudes in the IIA theory on the RR background based on the result 
in the previous section, and show that the correspondence (\ref{mat_IIA_op1}) and 
(\ref{mat_IIA_op2}) hold at the level of the amplitudes. 

Here we consider leading nontrivial contributions in the perturbation by $W_{{\rm RR}}$. 
Then, the RR charge $q_{{\rm RR}}$ will be identified with the quantity $(\nu_+-\nu_-)$ in the matrix model:~\footnote
{A similar observation is made in a matrix model for noncritical type 0B string theory~\cite{Douglas:2003up}, 
where eigenvalues asymmetrically filled in two potential wells are interpreted as an RR field.} 
\be
\nu_+-\nu_-=q_{{\rm RR}},
\label{qRR_leading} 
\ee
where a proportional constant is absorbed into $a_k$ in (\ref{WRR0}). 
By a constant shift of the Liouville field $\varphi_{{\rm tot}}=\varphi+\bar{\varphi}$, 
we can adjust the value of $\mu_1$ 
to be equal to $\omega$, where $\omega$ is the parameter in the matrix model (\ref{omega})~\footnote{
Here we have implicitly assumed that $\mu_1>0$. 
}.
Then, 
the $\omega$-dependence of the matrix model action (\ref{S}) and the $\mu_1$-dependence of the Liouville-like 
interaction term (\ref{Sint}) suggest the identification~\footnote{
The additional term $-\frac14\,\frac{\partial}{\partial\mu_1}W_{{\rm RR}}$ 
could appear in the r.h.s. of (\ref{identify_B}) due to the $\mu_1$-dependence of the RR background. 
It gives small corrections of the order $(\nu_+-\nu_-)$ compared to the leading $\frac14\cV_B^{(0,0)}(0)$. 
In this section, we focus on contributions from the leading and neglect these corrections.} 
\be
N\,\tr(-iB) \cong \frac14\cV_B^{(0,0)}(0).
\label{identify_B}
\ee
It is consistent with the last line in (\ref{mat_IIA_op1}) up to the choice of 
the picture with 
\be
\frac{1}{N}\cong g_s.
\label{1/N_gs}
\ee 
To make (\ref{mat_IIA_op1}) and (\ref{mat_IIA_op2}) more precise, 
we introduce numerical coefficients $c_k$, $d_k$ and $\bar{d}_k$ and write   
\be
\Phi_{2k+1}\cong c_k\,\cV_{\phi}(k), \qquad 
\Psi_{2k+1}\cong d_k\,\cV_{\psi}(k), \qquad
\bar{\Psi}_{2k+1}\cong \bar{d}_k\,\cV_{\bar{\psi}}(k). 
\label{identify_fields}
\ee
As we will see later, these coefficients appear to contain no divergence. 
It is contrast to the correspondence of two-dimensional bosonic string theory to the $c=1$ matrix model 
or the Penner model, where momentum-dependent divergent factors, the so-called leg factors, should be put 
to connect quantities in the string theory with those in the matrix model~\cite{Klebanov:1991qa,Mukhi:2003sz}.  
   
\subsection{$\vev{N\,\tr(-iB)\,\Phi_{2k+1}}_{C,0}$}
\label{sec:vevPhi_MM}
The matrix-model amplitude $\vev{N\,\tr(-iB)\,\Phi_{2k+1}}_{C,0}$ is obtained by differentiating 
(\ref{vevPhi_MM}) with respect to $\omega$: 
\bea
& & \left.\vev{N\,\tr(-iB)\,\Phi_{2k+1}}_{C,0}\right|_{\rm sing.}
= -\frac14\frac{\partial}{\partial\omega}\,\left.\vev{\Phi_{2k+1}}_0\right|_{\rm sing.}
\nn \\
& & \hspace{7mm} = -\frac14\,(\nu_+-\nu_-)\,\frac{2^{k+2}}{\pi}\,\frac{(2k+1)!!}{(k+1)!}\,\omega^{k+1}\,\ln\omega +(\mbox{less singular)}. 
\label{vevBPhi_MM}
\eea
Note that this is equal to the leading contribution to $\vev{N\tr(-iB)\,\Phi_{2k+1}}_C$ at large $N$ 
as seen from (\ref{vev_MM}). 
The corresponding IIA amplitude is 
$\cN\,g_s^{-2}\left\bra\!\!\vev{\frac14\cV_B^{(0,0)}\,c_k\cV_{\phi}(k)}\!\!\right\ket$, 
where we put an overall normalization constant $\cN$ independent of fields, and the bare string coupling $g_s^{-2}$ 
due to the sphere topology (a string tree amplitude)~\footnote{
In general, $g_s^{2h-2}$ is put for a string $h$-loop amplitude.}. 
The leading nontrivial contribution in the small $(\nu_+-\nu_-)$-expansion comes from the linear order 
of $W_{\rm RR}$ in (\ref{vev_IIA_RR}). 
Under an appropriate choice of the picture, it reads 
\bea
\cN\,g_s^{-2}\left\bra\!\!\!\vev{\frac14\cV_B^{(0,0)}(0)\,c_k\cV_{\phi}(k)}\!\!\!\right\ket 
& = & \frac14\,\cN\,g_s^{-4}\,c_k\,(\nu_+-\nu_-)\sum_{\ell\in {\bf Z}}a_\ell\,\omega^{\ell+1}
 \vev{\cV_B(0)\,\cV_{\phi}(k)\,\cV^{\rm RR}_{\ell}}
 \nn \\
&  = & -\frac14\,(\nu_+-\nu_-)\,2\cN\,c_k a_k\,\omega^{k+1}(\ln\omega) \,e^{i2\pi\beta(-k^2-\frac12k+\frac14)}.
\nn \\
& & \label{vevBPhi_IIA}
\eea
(\ref{3pt_BBB_f})-(\ref{3pt_BBB_kin_NL}) were used in the last equality. 
We find that dependence on $\omega$ as well as $\nu_{\pm}$ completely coincides 
in \eqref{vevBPhi_MM} and \eqref{vevBPhi_IIA} for any $k$. 
In particular, as we have noticed at the end of section \ref{sec:3pt_BBB}, 
the existence of the nonlocal branch enables this agreement to hold for any $k\in{\bf N}$.  
Furthermore by identifying their coefficients, 
we have a relation 
\be 
\cN\,\hat{c}_k\hat{a}_k\,e^{i\pi\beta\frac34}=\frac{2}{\pi}\frac{(2k+1)!}{k!(k+1)!}
\label{rel_vevBPhi} 
\ee
with 
\be
\hat{c}_k\equiv c_k \,e^{-i\pi\beta(k+\frac12)^2}, \qquad 
\hat{a}_k \equiv a_k\,e^{-i\pi\beta k^2}. 
\label{chat_ahat}
\ee

\subsection{$\vev{\Phi_{2k_1+1}\,\Phi_{2k_2+1}}_{C,0}$}
\label{sec:vevPhiPhi_MM}
The large-$N$ leading part of the two-point function $\vev{\Phi_{2k_1+1}\,\Phi_{2k_2+1}}_C$ 
in the matrix model reads from (\ref{vevPhiPhi_MM2}) as
\bea
& & \left.\frac{1}{N^2}\vev{\Phi_{2k_1+1}\,\Phi_{2k_2+1}}_{C,0}\right|_{\rm sing.} \nn \\
&  & \hspace{7mm}=\frac{1}{N^2}\left\{
-\frac{(\nu_+-\nu_-)^2}{2\pi^2}\frac{1}{k_1+k_2+1}\frac{(2k_1+1)!}{(k_1!)^2}
\frac{(2k_2+1)!}{(k_2!)^2}\,\omega^{k_1+k_2+1}(\ln \omega)^2 \right.\nn \\
 & & \left.\frac{}{}\hspace{18mm} +(\mbox{less singular})\right\}.
\label{vevPhiPhi_MM3}
\eea
The corresponding IIA amplitude is  
$
\cN\,g_s^{-2}\bra\!\vev{c_{k_1}\cV_{\phi}(k_1)\,c_{k_2}\cV_{\phi}(k_2)}\!\ket,   
$
whose leading nontrivial contribution comes from the quadratic order of $W_{\rm RR}$ as 
\bea
 & & \cN\,g_s^{-2}\bra\!\vev{c_{k_1}\cV_{\phi}(k_1)\,c_{k_2}\cV_{\phi}(k_2)}\!\ket = 
\frac12\,\cN\,g_s^{-2}\,c_{k_1}c_{k_2}\,(\nu_+-\nu_-)^2\nn \\
& & \hspace{24mm}\times \sum_{\ell_1,\ell_2\in {\bf Z}}a_{\ell_1}a_{\ell_2}\,
\omega^{\ell_1+\ell_2+2}\,
\vev{\cV_{\phi}(k_1)\,\cV_{\phi}(k_2)\,\cV_{\ell_1}^{{\rm RR}}\,\cV_{\ell_2}^{{\rm RR}}}. 
\eea
{}From the result in section~\ref{sec:4pt_BBBB}, we have 
\bea
& & \vev{\cV_{\phi}(k_1)\,\cV_{\phi}(k_2)\,\cV_{\ell_1}^{{\rm RR}}\,\cV_{\ell_2}^{{\rm RR}}} 
=g_s^4\left(\delta_{\ell_1,\,k_1+k_2}\,\delta_{\ell_2,\,-1}+(\ell_1\leftrightarrow \ell_2)\right) \nn \\
& & \hspace{7mm}\times (2\ln\omega)^2\,e^{-i\pi\beta\{\sum_{i=1}^2(k_i+\frac12)^2+\sum_{i=1}^2\ell_i^2\}}\, 
\frac{\pi}{2}\,\left(\frac{(k_1+k_2)!}{k_1!k_2!}\right)^2\,c_L.
\eea
There appears the square of the Liouville volume $(2\ln\omega)^2$. 
One of them is from 
the integral over the Liouville constant mode as in \eqref{Liouville_V} as usual, 
while the other from the resonance of on-shell particles and the background as mentioned in \eqref{varepsilon_VL}. 
Thus we obtain  
\bea
 & & \cN\,g_s^{-2}\bra\!\vev{c_{k_1}\cV_{\phi}(k_1)\,c_{k_2}\cV_{\phi}(k_2)}\!\ket \nn \\
& & \hspace{3mm} = (\nu_+-\nu_-)^2\,\cN\,g_s^2\,c_L\,\hat{c}_{k_1}\hat{c}_{k_2}\hat{a}_{k_1+k_2}\hat{a}_{-1}\,
2\pi\,\left(\frac{(k_1+k_2)!}{k_1!k_2!}\right)^2\,\omega^{k_1+k_2+1}\,(\ln\omega)^2. 
\label{vevPhiPhi_IIA}
\eea
\eqref{vevPhiPhi_MM3} and \eqref{vevPhiPhi_IIA} indeed have the same dependence on $\nu_{\pm}$ and $\omega$ 
for any $k_1$ and $k_2$. 
Moreover, the dependence on $k_1$ and $k_2$ of the coefficient in \eqref{vevPhiPhi_IIA} is 
written in a factorized form as $f(k_1)\,f(k_2)\,g(k_1+k_2)$, where $f$ and $g$ are some functions.  
It serves as a nontrivial check to see that the matrix model result exhibits the same 
factorization as well. It is not manifest at all in the original expression (\ref{vevPhiPhi_MM}), 
but (\ref{vevPhiPhi_MM2}) and therefore 
(\ref{vevPhiPhi_MM3}) are indeed so. 
Identifying (\ref{vevPhiPhi_MM3}) with (\ref{vevPhiPhi_IIA}) leads to 
\be
\left(\frac{\hat{c}_{k_1}}{(2k_1+1)!}\right)\,\left(\frac{\hat{c}_{k_2}}{(2k_2+1)!}\right)\, 
\left(\hat{a}_{k_1+k_2}\,(k_1+k_2)!\,(k_1+k_2+1)!\right) = 
-\frac{1}{4\pi^3}\,\frac{1}{\cN\,c_L\,\hat{a}_{-1}}.   
\ee
Notice that the r.h.s. is independent of $k_1$ and $k_2$, and thus that the product of the first two factors 
in the l.h.s. must give a function of $k_1+k_2$. 
It determines the $k$-dependence of $\hat{c}_k$ and $\hat{a}_k$ as 
\be
\hat{c}_k = \hat{c}_0\,e^{\gamma k}\,(2k+1)!, \qquad 
\hat{a}_k = \frac{\hat{a}_0\,e^{-\gamma k}}{k!(k+1)!} \qquad (k=0,1,2,\cdots) 
\label{chat_ahat_2}
\ee
with $\gamma$ being a numerical constant and 
\be
\hat{c}_0^2\,\hat{a}_0=- \frac{1}{4\pi^3}\,\frac{1}{\cN\,c_L\,\hat{a}_{-1}}.
\label{chat0_ahat0}
\ee

As another nontrivial check, we can see that 
(\ref{chat_ahat_2}) correctly reproduces the $k$-dependence of (\ref{rel_vevBPhi}) which is obtained 
from a separate amplitude. Then, we have 
\be
\cN\,\hat{c}_0\,\hat{a}_0\,e^{i\pi\beta\frac34}=\frac{2}{\pi}. 
\label{N_chat0_ahat0}
\ee 
{}From (\ref{chat0_ahat0}) and (\ref{N_chat0_ahat0}), $\hat{c}_0$ and $\hat{a}_0$ are expressed as 
\bea
 & & \hat{c}_0 = -\frac{1}{8\pi^2}\,\frac{1}{c_L\,\hat{a}_{-1}}\,e^{i\pi\beta\frac34}, \qquad 
c_0=\frac{1}{8\pi^2}\,\frac{1}{c_L\,a_{-1}}, \nn \\
& & a_0=\hat{a}_0=\frac{16\pi}{\cN}\,c_L\,\hat{a}_{-1}\,e^{i\pi\beta\frac12}
=\frac{16\pi}{\cN}\,c_L\,a_{-1}\,e^{-i\pi\beta\frac12}. 
\eea
(Note that $e^{i2\pi\beta}=-1$ due to (\ref{beta}).)

\subsection{$\vev{\Psi_{1}\,\bar{\Psi}_{1}}_{C,0}$}
\label{sec:vevPsiPsi_MM1}
The large-$N$ leading part of $\vev{\Psi_{1}\,\bar{\Psi}_{1}}_C$, which is given by 
the $k=\ell=0$ case of the matrix-model amplitude (\ref{vevPsiPsi_MM}),  
\be
\left.\frac{1}{N^2}\vev{\Psi_1\,\bar{\Psi}_1}_{C, 0}\right|_{\rm sing.} 
=\frac{1}{N^2}\left\{(\nu_+-\nu_-)\,\frac{1}{\pi}\,\omega\ln\omega +(\mbox{less singular})\right\},  
\label{vevPsiPsi_MM1}
\ee
is compared with $\cN\,g_s^{-2}\bra\!\vev{d_0\cV_{\psi}(0)\,\bar{d}_0\cV_{\bar{\psi}}(0)}\!\ket$.  
Its leading nontrivial contribution with respect to small $(\nu_+-\nu_-)$ comes from the linear order 
of $W_{\rm RR}$ in (\ref{vev_IIA_RR}):  
\bea
\cN\,g_s^{-2}\bra\!\vev{d_0\cV_{\psi}(0)\,\bar{d}_0\cV_{\bar{\psi}}(0)}\!\ket & = & 
\cN\,g_s^{-2}\,d_0\,\bar{d}_0\,(\nu_+-\nu_-)\sum_{\ell\in {\bf Z}}a_{\ell}\,\omega^{\ell+1}
\,\vev{\cV_{\psi}(0)\,\cV_{\bar{\psi}}(0)\,\cV^{{\rm RR}}_{\ell}} \nn \\
& = & \cN\,g_s^2\,d_0\,\bar{d}_0\,(-2a_0)\,(\nu_+-\nu_-)\,\omega(\ln\omega)\,e^{-i\pi\beta\frac32}. 
\label{vevPsiPsi_IIA1}
\eea
In the last line, we used (\ref{3pt_FFB_f}) and (\ref{3pt_FFB_kin}). 
This takes the same form as in \eqref{vevPsiPsi_MM1} as a function of $\nu_{\pm}$ and $\omega$. 
{}From the further comparison of their coefficients, we have 
\be
\cN\,d_0\,\bar{d}_0\,a_0\,e^{i\pi\beta\frac12} = \frac{1}{2\pi}. 
\label{rel_vevPsiPsi1}
\ee
This and (\ref{N_chat0_ahat0}) give a relation of $d_0, \bar{d}_0$ to $c_0$:  
\be
d_0\,\bar{d}_0=\frac14\,c_0,  
\label{d0dbar0_c0}
\ee
which is relevant to the target-space supersymmetry as is seen later.

\subsection{$\vev{\Psi_{3}\,\bar{\Psi}_{3}}_{C,0}$}
\label{sec:vevPsiPsi_MM3}
The large-$N$ leading of $\vev{\Psi_{3}\,\bar{\Psi}_{3}}_C$ reads from the $k=\ell=1$ case of (\ref{vevPsiPsi_MM}) as 
\be
\left.\frac{1}{N^2}\vev{\Psi_3\,\bar{\Psi}_3}_{C, 0}\right|_{\rm sing.} 
=\frac{1}{N^2}\left\{(\nu_+-\nu_-)^3\,\frac{6}{\pi}\,\omega^3\ln\omega +(\mbox{less singular}) \right\} .
\label{vevPsiPsi_MM3}
\ee
It corresponds to the IIA amplitude 
$\cN\,g_s^{-2} \,\bra\!\vev{d_1\cV_{\psi}(1)\,\bar{d}_1\cV_{\bar{\psi}}(1)}\!\ket$, 
whose leading nontrivial contribution in the $(\nu_+-\nu_-)$-expansion arises from the cubic order of $W_{\rm RR}$: 
\bea
& & \cN \,g_s^{-2}\bra\!\vev{d_1\cV_{\psi}(1)\,\bar{d}_1\cV_{\bar{\psi}}(1)}\!\ket 
= \frac{1}{3!}\,\cN\,g_s^{-2}\,d_1\,\bar{d}_1\,(\nu_+-\nu_-)^3 \nn \\
& & \hspace{14mm}\times \sum_{\ell_1, \ell_2,\ell_3\in {\bf Z}} 
a_{\ell_1}\,a_{\ell_2}\,a_{\ell_3}\,\omega^{\ell_1+\ell_2+\ell_3+3}\,
\vev{\cV_{\psi}(1)\,\cV_{\bar{\psi}}(1)\,\cV^{{\rm RR}}_{\ell_1}\,\cV^{{\rm RR}}_{\ell_2}\,\cV^{{\rm RR}}_{\ell_3}}. 
\eea
By making use of (\ref{5pt_kin}) and (\ref{5pt_f}), the final expression becomes 
\be
\cN \,g_s^{-2}\bra\!\vev{d_1\cV_{\psi}(1)\,\bar{d}_1\cV_{\bar{\psi}}(1)}\!\ket 
= (\nu_+-\nu_-)^3 \,\cN\,g_s^2\,d_1\,\bar{d}_1\,a_0^3\,\pi^2\,\omega^3(\ln\omega)\,
e^{-i\pi\beta\frac32}. 
\label{vevPsiPsi_IIA3}
\ee
We again find that \eqref{vevPsiPsi_MM3} and \eqref{vevPsiPsi_IIA3} have exactly the same dependence 
on $\nu_{\pm}$ and $\omega$. 
Comparing their coefficients gives 
\be
\cN\,d_1\,\bar{d}_1\,a_0^3\,e^{i\pi\beta\frac12}=-\frac{6}{\pi^3}. 
\label{rel_vevPsiPsi3}
\ee
Together with (\ref{rel_vevPsiPsi1}) and (\ref{d0dbar0_c0}), the relation (\ref{rel_vevPsiPsi3}) leads to 
\be
d_1\,\bar{d}_1= \left(-\frac{12}{\pi^2}\,\frac{1}{a_0^2}\right)\,d_0\,\bar{d}_0
=-\frac{3}{\pi^2}\,\frac{c_0}{a_0^2}. 
\label{d1dbar1_d0dbar0}
\ee 
 
As we have seen so far, it is remarkable that the single choice of the RR background 
\eqref{WRR0} and \eqref{WRR} realizes the agreement between 
several kinds of IIA amplitudes (\eqref{vevBPhi_IIA}, \eqref{vevPhiPhi_IIA}, 
\eqref{vevPsiPsi_IIA1} and \eqref{vevPsiPsi_IIA3}) and the corresponding matrix-model correlators, 
with respect not only to the dependence of $\nu_{\pm}$ and $\omega$ but also to prefactors (depending on 
$x$, $\bar{x}$-momenta/powers of matrices).

\subsection{Target-space supersymmetry}
Corresponding to (\ref{identify_fields}), 
let us identify the supercharges in the matrix model $Q$ and $\bar{Q}$ with those in the IIA theory as 
\be
Q\cong \alpha\,\hat{Q}_+,\qquad \bar{Q} \cong \bar{\alpha}\,\hat{\bar{Q}}_-
\ee
by putting the coefficients $\alpha$, $\bar{\alpha}$. 

{}From (\ref{identify_B}) and (\ref{identify_fields}), each of 
the quartet $(\Phi_1, \,\Psi_1, \,\bar{\Psi}_1, \,\frac{1}{N}\tr(-iB))$ with respect to $Q$, $\bar{Q}$ 
should be precisely mapped to each of  
$(c_0\cV_{\phi}(0),\,d_0\cV_{\psi}(0), \,\bar{d}_0\cV_{\bar{\psi}}(0), \,\frac14\cV_B(0))$ 
with respect to $\hat{Q}_+$, $\hat{\bar{Q}}_-$. 
This assertion implies 
\be
d_0=\alpha\,c_0, \qquad \bar{d}_0=\bar{\alpha}\,c_0, \qquad \alpha\,\bar{\alpha}\,c_0=\frac14, 
\ee
from which we obtain (\ref{d0dbar0_c0}) again. 
Note that the argument here does not refer to any amplitudes. Nevertheless, it reproduces the relation 
(\ref{d0dbar0_c0}) which was derived from amplitudes in the previous subsection. 
This also shows consistency of the correspondence, in particular, the identification of the supercharges 
in both sides.

\subsection{$\vev{N\tr(-iB)\,\Psi_{1}\,\bar{\Psi}_{1}\,\Phi_1}_{C,0}$ and 
$\vev{(\tr(-iB))^2\,\Phi_1^2}_{C,0}$}
The matrix model correlator 
$\frac{1}{N^4}\vev{N\tr(-iB)\,\Psi_{1}\,\bar{\Psi}_{1}\,\Phi_1}_{C,0}$ 
is computed as 
\be
\frac{1}{N^4}\vev{N\tr(-iB)\,\Psi_{1}\,\bar{\Psi}_{1}\,\Phi_1}_{C,0}
=-\frac18\frac{\partial}{\partial\omega}\,\frac{1}{N^4}\vev{\frac{1}{N}\tr\,\phi^{-1}\,\frac{1}{N}\tr\,\phi}_{C,0}, 
\ee
which is proportional to $(\nu_+-\nu_-)^2$ as discussed in appendix B.2 in~\cite{Kuroki:2012nt}. 
Here $\frac1N\tr\,\phi^{-1}$ arises by the contraction of $\Psi_1$ and $\bar\Psi_1$ \cite{Kuroki:2012nt}. 
Thus, there is no contribution of the order $(\nu_+-\nu_-)^0$, which corresponds to the IIA amplitude
\be
g_s^{-2}\vev{\frac14\cV_B^{(0,0)}(0)\,d_0\cV_{\psi}(0)\,\bar{d}_0\cV_{\bar{\psi}}(0)\,c_0\cV_{\phi}(0)}
\ee
without insertions of $W_{\rm RR}$. According to (\ref{4pt_BFFB_kin}) and (\ref{4pt_BFFB_f}), it vanishes, 
and we see that the correspondence holds at the order $(\nu_+-\nu_-)^0$. 
To check the correspondence 
up to the order of $(\nu_+-\nu_-)^2$, we have to compute a six-point CFT amplitude. 
We leave it as a future subject. 

Similarly, $\frac{1}{N^4}\vev{(\tr(-iB))^2\,\Phi_1^2}_{C,0}$ in the matrix model vanishes at 
the order $(\nu_+-\nu_-)^0$. It corresponds to the IIA amplitude 
\be
g_s^{-2}\vev{\frac14\cV_B(0)\,\frac14\cV_B^{(0,0)}(0)\,\left\{c_0\cV_{\phi}(0)\right\}^2} 
\ee
that is proportional to (\ref{4pt_BBBB3_f}) with (\ref{4pt_BBBB3_kin}). 
The result is also zero, showing the validity of the correspondence.

\subsection{Torus partition function}
\label{sec::torusZ}
In this subsection we confirm that the IIA theory on the RR background is a consistent superstring theory 
by checking that it has a modular invariant torus partition function. We also see that the result is consistent 
with the torus free energy of the matrix model. 

The genus-one amplitude among vertex operators $\cV_i$ is 
\be
\int_{\cF}\frac{d^2\tau}{\mbox{Vol(CKG($T^2$))}}\,\vev{c\bar{c}(0)\,B\bar{B}\,\prod_i\cV_i}. 
\ee 
Here, the volume of the conformal Killing group of $T^2$ is nothing but the area of the torus of 
the worldsheet. The corresponding $c$, $\bar{c}$-ghost zero-modes are fixed by the insertion $c\bar{c}(0)$. 
$B$, $\bar{B}$ are $b$, $\bar{b}$-ghost insertions associated with the integration with respect to 
the torus moduli $\tau$. The integration is over the fundamental region $\cF$. 

In a similar manner as in section~\ref{sec:IIA_RR}, 
the torus partition function under the RR background is expressed as 
\bea
\bra\!\vev{1}\!\ket & = & \vev{e^{W_{\rm RR}}} \nn \\
& = & \int_{\cF}\frac{d^2\tau}{\mbox{Vol(CKG($T^2$))}}\sum_{n=0}^\infty\frac{1}{n!}\,
2\Gamma(-s)\,\mu_1^s\,\frac{1}{V_L}
\vev{c\bar{c}(0)\,B\bar{B}\,(W_{\rm RR})^n\,\cV_B^{(0,0)}(0)^s}_{\rm CFT} \nn \\
& & \label{torusZ}
\eea
with $s=-2\sum_ip_{\ell_i}$ due to $\chi(T^2)=0$. 
Note that every vertex operator in $W_{\rm RR}$ has nonpositive $x$-winding from (\ref{WRR}). 
For $s\geq 0$, the $s=0$ case alone possibly give nonvanishing contribution to $\bra\!\vev{1}\!\ket$, 
and only $\cV^{{\rm RR}}_{k=0}$ remains in $W_{\rm RR}$. 
Furthermore, 
the conservation of the Liouville momentum $\sum_ip_{\ell_i}=0$ tells us that even the remaining 
$\cV_{k=0}^{{\rm RR}}$ gives no effect. Thus, just the first term of $n=0$ in the sum 
$\sum_{n=0}^{\infty}\frac{1}{n!}\,(W_{\rm RR})^n$ 
can contribute to the partition function. Because this argument relies only on the conservation of the 
$x$-winding and the Liouville momentum, it holds irrespective of a way to distribute the picture charges.  

Next, let us consider the $s<0$ case. Although the computation cannot be carried out by the standard CFT technique, 
we can argue that the partition function should be nil from the conservation of the picture and $H$ charges. 
Since the $x$-winding of $\cV^{{\rm RR}}_k$ is $k\in {\bf Z}$, its conservation law means 
\be
s=-2,-4,-6, \cdots.
\label{s<0}
\ee
The total picture must be $(0,0)$ to provide a nontrivial amplitude of the torus topology. 
We consider contribution from  $(W_{\rm RR})^{n}$ ($n=0,1,2,\cdots$) 
to \eqref{torusZ}. 
Any RR field there has the $(-\frac12,\,-\frac12)$-picture originally as in \eqref{Rvertex}, 
\eqref{WRR0} and \eqref{WRR}. Hence $n$ should be even and there must be insertion 
of $\frac{n}{2}$ picture raising operators 
($Q_{{\rm BRST}}(2\xi\,\cdot \,)$) in the holomorphic sector so that the total picture will be zero: 
$\left(-\frac12\right)\times n+1\times \frac n2=0$. 
(Similarly in the anti-holomorphic sector.) 
Then let us see what happens to its $H$ charge.  
Every RR field in $(W_{\rm RR})^{n}$ has the $H$ charge $\frac{\epsilon}{2}=-\frac12$ as in \eqref{WRR}, and 
each of 
the picture raising operators increases the $H$ charge at most by one as seen from (\ref{TmF})-(\ref{TmFm}) 
and (\ref{QBRST})~\footnote{
For example, the explicit form of the $(+\frac12)$-picture R vertex operator ($p_{\ell}=1-|k|$, $k=\epsilon |k|$) reads 
\bea
& & Q_{\rm BRST}(2\xi(z)V_{k,\,\epsilon}(z))  =  \partial (2c\xi V_{k,\,\epsilon})(z) 
-\frac{i}{\sqrt{2}}(p_{\ell}-\epsilon k+1)\,e^{\frac12\phi+i\epsilon\frac32 H+ikx+p_{\ell}\varphi}(z) \nn \\
& &\hspace{7mm} +\frac{i}{\sqrt{2}}\partial(\varphi-i\epsilon x+2i\epsilon H)\,
e^{\frac12\phi-i\epsilon\frac12 H+ikx+p_{\ell}\varphi}(z) 
+\frac12b\eta\, e^{\frac32\phi+i\epsilon\frac12 H+ikx+p_{\ell}\varphi}(z).
\eea
}. 
Thus the total $H$ charge of $(W_{\rm RR})^{n}$ with the $\frac{n}{2}$ picture raising operators will be 
at most $\left(-\frac12\right)\times n+1\times \frac n2=0$, while that of $\cV_B^{(0,0)}(0)^s$ is $s$. 
Recalling (\ref{s<0}), we see that the total $H$ charge of the amplitude cannot be conserved, 
and the contribution from $(W_{\rm RR})^{n}$ to the partition function should be zero.

Now we conclude that the torus partition function becomes 
\be
\bra\!\vev{1}\!\ket = \left.\vev{1}\right|_{s=0}= \left(2\ln \frac{1}{\mu_1}\right)\,
\frac{1}{V_L} \int_{\cF}\frac{d^2\tau}{\mbox{Vol(CKG($T^2$))}}\,
\vev{c\bar{c}(0)\,B\bar{B}}_{\rm CFT} . 
\label{torusZ_2}
\ee
As expected from the fact that the two-dimensional string has no dynamical degrees of freedom of 
oscillator modes, contributions from oscillators cancel leaving 
those from $x$-winding/momentum~\cite{Bershadsky:1991zs}: 
\bea
\frac{1}{V_L}\int_{\cF}\frac{d^2\tau}{\mbox{Vol(CKG($T^2$))}}\,
\vev{c\bar{c}(0)\,B\bar{B}}_{\rm CFT} & = & 
\int_{\cF}(d\tau)\,\left[Z_{(\rNS, \rNS)}(\tau,\bar{\tau}) 
+ Z_{(\rR +, \rR -)}(\tau,\bar{\tau}) \right. \nn \\
& & \hspace{12mm} 
\left. + Z_{(\rNS, \rR -)}(\tau,\bar{\tau}) +Z_{(\rR +, \rNS)}(\tau,\bar{\tau}) \right]  
\nn \\
& & 
\label{partition_sum}
\eea
with $(d\tau)\propto d^2\tau/({\rm Im}\,\tau)^2$ modular invariant. 
On the integrand in the r.h.s., neither the level matching condition~\footnote{
The level matching condition is taken into account by performing the integral of ${\rm Re} \,\tau$.} 
nor the Dirac equation constraint is imposed~\cite{Murthy:2003es,Ita:2005ne}.  
Holomorphic and anti-holomorphic $x$-momenta $(k, \bar{k})$ in the (NS, NS) state sum 
are restricted as follows~\footnote{A similar argument is presented in~\cite{Kutasov:1991pv}.}. 
$k, \bar{k}\in {\bf Z}+\frac12$ 
for the corresponding vertex operator 
($e^{-\phi+ikx+p_{\ell}\varphi}\,e^{-\bar{\phi}+i\bar{k}\bar{x}+p_{\ell}\bar{\varphi}}$) to be local with 
the target-space supercurrents $q_+$ and $\bar{q}_-$. 
Also, from the locality between two (NS, NS) vertex operators with the momenta $(k, \bar{k})$ and $(k',\bar{k}')$, 
we have $kk'-\bar{k}\bar{k}'\in {\bf Z}$. 
These two conditions lead to 
\be
k-\bar{k}\in 2{\bf Z} \qquad \mbox{or}\qquad k+\bar{k}\in 2{\bf Z}. 
\label{NSNS_condition}
\ee
(Then $\frac12k^2-\frac12\bar{k}^2\in {\bf Z}$ is satisfied and 
$Z_{(\rNS, \rNS)}(\tau+1,\bar{\tau}+1)=Z_{(\rNS, \rNS)}(\tau,\bar{\tau})$.) 
The former contains the ``momentum background'', while the latter does the ``winding background'' 
(\ref{spectrum_before}) that we are considering. 
The restriction is an analog of the GSO projection in critical string theory~\cite{Kutasov:1990ua}.  
Thus, 
\be
Z_{(\rNS, \rNS)}(\tau,\bar{\tau}) = \sum_{n,m\in {\bf Z}}q^{\frac12(n+\frac12)^2}\,
\bar{q}^{\frac12(2m-n-\frac12)^2} 
=\frac12|\theta_2(\tau)|^2, 
\label{ZNSNS}
\ee
where $q=e^{i2\pi\tau}$, and $\theta_2(\tau)$ is one of Jacobi's theta functions: 
$\theta_2(\tau)=\sum_{r\in {\bf Z}+\frac12}q^{\frac12 r^2}$. The identical condition 
(\ref{NSNS_condition}) also arises for the other three sectors. 
The latter of (\ref{NSNS_condition}) is taken in the (R$+$, R$-$) sector, while the former is 
in the (NS, R$-$) and (R$+$, NS) sectors, in such a way to contain the ``winding background" (\ref{spectrum_before}). 
The result is 
\bea
& & Z_{(\rR +, \rR -)}(\tau,\bar{\tau})=Z_{(\rNS, \rNS)}(\tau,\bar{\tau}), \nn \\
& & Z_{(\rR +, \rNS)}(\tau,\bar{\tau})= Z_{(\rNS, \rR -)}(\tau,\bar{\tau})=
-\sum_{n,m\in {\bf Z}}q^{\frac12(n+\frac12)^2}\,\bar{q}^{\frac12(n-2m+\frac12)^2} \nn \\
& & \hspace{55mm}=-Z_{(\rNS, \rNS)}(\tau,\bar{\tau}). 
\label{ZRR}
\eea
(\ref{ZNSNS}) and (\ref{ZRR}) give the vanishing torus partition function~\footnote{
The result of the torus partition sum is different from that  
on the trivial background computed in~\cite{Kutasov:1991pv,Murthy:2003es}. 
The partition sums in the (NS, R$-$) and (R$+$, NS) sectors obtained there 
are the half of our result, and cancellation with the (NS, NS) sector is observed 
(See eqs. (3.14), (3.15) in~\cite{Kutasov:1991pv} and appendix B.1 in~\cite{Murthy:2003es}.). 
We guess that the Dirac equation constraint 
is imposed in the (NS, R) and (R, NS) sectors but not in the (R, R) sector 
in~\cite{Kutasov:1991pv,Murthy:2003es}. In contrast, we do not consider the Dirac equation constraint 
for any R sectors in (\ref{partition_sum}). 
Their result is modular invariant as well as ours.}   
: 
\be
\bra\!\vev{1}\!\ket=0. 
\ee
This is consistent with the matrix model results (\ref{torus_vevB_MM}) and (\ref{Z_MM}) 
which mean the zero torus free energy.  

We obtain the same result 
even if the integrand is assumed to be slightly generalized to linear combinations as   
\be
Z_{(\rNS, \rNS)}(\tau,\bar{\tau}) 
+ a\,Z_{(\rR +, \rR -)}(\tau,\bar{\tau}) + b\,Z_{(\rNS, \rR -)}(\tau,\bar{\tau})
+c\,Z_{(\rR +, \rNS)}(\tau,\bar{\tau}), 
\label{general_integrand}
\ee
where coefficients $a$, $b$ and $c$ are fixed so that (\ref{general_integrand}) is modular invariant. 
{}From (\ref{ZNSNS}) and (\ref{ZRR}), 
\be
(\ref{general_integrand}) = \frac{1+a-b-c}{2}\,|\theta_2(\tau)|^2,
\ee
for which to be modular invariant there is no other possibility than the prefactor $1+a-b-c$ being null. 
Then the torus partition function vanishes. 

Note that our conclusion of the vanishing torus partition function is not usually expected from 
the supercurrents (\ref{supercurrents}) 
carrying $x$, $\bar{x}$-momenta~\cite{Kutasov:1990ua,Kutasov:1991pv,Murthy:2003es}. 
However, in our case where the (R$-$, R$+$) vertex operators represent the fixed RR background 
and does not take part in the torus partition sum, 
contributions from the remaining three sectors are balanced as seen in the above. 
Furthermore, the RR background itself does not spoil the supersymmetry, which is consistent 
with the fact that the (R$-$, R$+$) fields are singlets under the supersymmetries, 
as shown in section \ref{sec:correspondence to MM}.
As a result, we obtain what is naively expected from a supersymmetric theory.


\section{Remarks on nonlocal RR vertex operators}
\label{sec:nonlocalRR}
\setcounter{equation}{0}
The nonlocal vertex operators appearing in the background $W_{{\rm RR}}$ do not satisfy the Dirac equation 
constraint as we have pointed out below \eqref{WRR}. 
It would be acceptable from the point of view of representing a background and not on-shell particles. 
As a consequence, however they are not BRST-closed: 
\bea
Q_{\rm BRST} \left(V_{k,\,-1}^{({\rm nonlocal})}(z)\right) & = & \partial \left(cV_{k,\,-1}^{{\rm (nonlocal)}}\right)(z) 
-\frac{i}{2\sqrt{2}}(p_{\ell}-k-1)\,\eta\,e^{\frac12\phi+i\frac12H+ikx+p_{\ell}\varphi}(z), 
\nn \\
Q_{\rm BRST} \left(cV_{k,\,-1}^{({\rm nonlocal})}(z)\right) & = & 
\frac{i}{2\sqrt{2}}(p_{\ell}-k-1)\,c\eta\,e^{\frac12\phi+i\frac12H+ikx+p_{\ell}\varphi}(z).
\label{nonBRSTclosed}
\eea

Nevertheless, we can formally see that this violation of the BRST invariance does not contribute to amplitudes 
among BRST-closed physical vertex operators ($Q_{\rm BRST}(cV_{i,\,{\rm phys}}(z))=0$). For example, 
\be
\vev{\prod_{i=1}^3c(z_i)V_{i,\,{\rm phys}}(z_i)\,\prod_{j\geq 4}\int dz_j\,V_{j,\,{\rm phys}}(z_j)\,
\int dz\,Q_{\rm BRST} \left(V_{k,\,-1}^{({\rm nonlocal})}(z)\right)}=0
\ee
by deforming the contour of the BRST current~\footnote{ 
We might also give somewhat similar but more formal argument by assuming the existence of 
the CFT state even corresponding to the nonlocal operator. 
As mentioned below \eqref{WRR},  
$G_0^+\left|V_{k,\,-1}^{({\rm nonlocal})}\right\ket\neq 0$,
but we could see 
$G_0^-G_0^+\left|V_{k,\,-1}^{({\rm nonlocal})}\right\ket=0$,
which implies that the breaking of the BRST invariance would have zero norm 
$\left\| G_0^+\left|V_{k,\,-1}^{({\rm nonlocal})}\right\ket \right\|=0$ in the matter CFT sector 
and would be decoupled in our case where the matter CFT is unitary. 
However, that is not always true as we see that the integrand in $\cB$ of (\ref{cB_nonzero}) is not zero 
at $z_4=0$.}.   
Actually in relevant amplitudes in the previous sections, no more than one nonlocal operator is inserted. 

We emphasize that (\ref{nonBRSTclosed}) does not immediately mean inconsistency of the theory. 
Indeed, the nonlocal 
vertex operators have the conformal weight $(1,1)$ preserving the worldsheet conformal symmetry. 
Namely, they are marginal perturbations around the flat background given by $S_{\rm CFT}$ 
in (\ref{SCFT})~\footnote{
Note that we treat the RR background as a perturbation around the flat background 
(recall $S_{{\rm int}}$ in \eqref{Sint} is also regarded as a perturbation) as shown in 
\eqref{vev_IIA_RR} and \eqref{torusZ}.}, 
under which the theory should make sense as a string theory. 
We have also seen in section \ref{sec::torusZ} that the theory is modular invariant. 
Furthermore, the breaking (\ref{nonBRSTclosed}) solely comes from the breaking of the global worldsheet supersymmetry $G^+_0$. 
Thus we can construct a BRST-like charge $Q_0+Q_1^-$, where 
\be
Q_1^{\pm}\equiv - \oint \frac{dz}{2\pi i} \,\gamma(z)T^{\pm}_{{\rm m},\,F}(z)
= -\oint \frac{dz}{2\pi i} \,e^{\phi}\,\eta\,T^{\pm}_{{\rm m},\,F}(z)
\ee
with (\ref{TmF}) and (\ref{QBRST2}). 
This charge is nilpotent and annihilates the physical vertex operators (\ref{spectrum_after}) 
as well as the nonlocal vertex operators. 
This situation is reminiscent of boundary states as mentioned in section~\ref{sec:IIA_RR}. 
For amplitudes containing boundary states in that case, physical vertex operators inserted in the bulk 
are invariant under each of the holomorphic and anti-holomorphic BRST operators, while the boundary states 
break the invariance under a half of them and preserve the other half (a certain linear combination of them). 
As discussed in \cite{Callan:1988wz,Becker:2011bw}, 
it requires intricate examinations to confirm the BRST invariance for such amplitudes 
due to $b$-ghost insertions associated with the boundary states and to 
continuation to off-shell momenta as a regularization.  
Although it is not easy to prove decoupling of the BRST-exact operators 
and independence of the way to distribute the picture charges in a generic amplitude,  
the literature investigates this issue by taking some concrete amplitudes 
and gets affirmative consequences. 
Similarly, let us see evidences supporting that (\ref{nonBRSTclosed}) does not ruin consistency 
of the theory at least for amplitudes discussed in the previous sections. 

\subsection{Target-space gauge symmetry}
(\ref{nonBRSTclosed}) may imply that BRST-exact operators potentially do not 
decouple from amplitudes in the presence of the 
nonlocal operators. In general, decoupling of BRST-exact operators guarantees gauge symmetry 
in target space, and its breaking would run into serious inconsistency of theory.  
 
Note that in the two-dimensional superstring we are considering, 
only the RR vertex operators concern gauge particles, and they do not couple to its gauge potential, 
but directly to gauge invariant $U(1)$ RR field strength. 
Thus there must be no gauge transformation in the target space expressed by BRST-exact operators. 
Actually, as is shown in the following, we cannot construct any BRST-exact shift to the $(-1)$-picture 
NS field $cT_k(z)=c\,e^{-\phi+ikx+p_{\ell}\varphi}(z)$: 
\be
cT_k(z) \to cT_k(z) + Q_{{\rm BRST}}(U(z)) 
\ee
in a consistent manner. 
$U(z)$ should have the ghost number zero, the picture $(-1)$ and the weight zero. 
Moreover, it should have the term $e^{ikx+p_{\ell}\varphi}$ of the weight $\frac12$ which is common to 
$cT_k$. Hence the prefactor of $e^{ikx+p_{\ell}\varphi}$ appearing in $U(z)$ 
has the weight $(-\frac12)$.   
As discussed in~\cite{Becker:2011bw}, 
let us consider the 
operators consisting of $b$, $c$, $\xi$, $\eta$ and $\phi$ 
with the ghost number zero and the picture $P$ given by~\footnote{Note that $\xi$ and $\eta$ have 
the picture charges $(+1)$ and $(-1)$, respectively.}  
\bea
A_+& \equiv & e^{(n+P)\phi}\,\eta\,(\partial\eta)\cdots(\partial^{n-1}\eta)\,
b \,(\partial b)\cdots (\partial^{n-1}b) \qquad (n= 0, 1,2 \cdots), \nn \\
A_-& \equiv & e^{(n+P)\phi}\,(\partial\xi)\,(\partial^2\xi)\cdots(\partial^{|n|}\xi)\,
c\,(\partial c)\cdots (\partial^{|n|-1}c) \qquad (n=0,-1,-2,\cdots)  
\eea   
with their weights 
\be
[A_+]=\frac12\left\{-P^2-2(n+1)P+n(n+2)\right\}, \qquad
[A_-]=\frac12\left\{-P^2+2(|n|-1)P+n^2\right\}.
\ee
The prefactor of $e^{ikx+p_{\ell}\varphi}$ in $U(z)$ has the form of $A_+$ or $A_-$ with $P=-1$ 
multiplied by polynomials of derivatives of $x$, $\varphi$, 
$H$ and $\phi$. It may also be multiplied by factors of a form $:\partial^kb\partial^{\ell}c:$ or 
$:(\partial^{p+1}\xi)\partial^q\eta:$ ($k,\ell, p, q\geq 0$) that do not change the ghost number or 
the picture number. 
Note that such multiplicative fields increase or keep the weight, but never decrease it.  
Thus, the weight must be 
\be
[A_+]\le -\frac12 \qquad \mbox{or}\qquad
[A_-]\le -\frac12 
\ee 
for $P=-1$. However, any integer $n$ does not satisfy this condition, meaning that such $U(z)$ does not exist. 
Similar argument is possible for other vertex operators~\footnote{
We have explicitly seen that no BRST-exact shift is allowed for the 0-picture NS field 
and R fields with the pictures $(-\frac12)$, $(+\frac12)$ and $(-\frac32)$.}.

\subsection{Picture changing operation}
\label{sec:pic}
Since the picture changing operation discussed in~\cite{Friedan:1985ge} relies on the BRST invariance 
of vertex operators, one may wonder if \eqref{nonBRSTclosed} prevents it. 
Here we concretely demonstrate the picture changing operation in the presence of 
the nonlocal operators. Relevant amplitudes investigated in this paper are 
in sections~\ref{sec:3pt_BBB} and \ref{sec:4pt_BBBB}. 
  
Let us consider the holomorphic part of the amplitude in section~\ref{sec:3pt_BBB}:~\footnote{
A parallel argument can be applied to the anti-holomorphic part.}   
\be
\bra 0|cT_{k_1}(z_1)\,cV_{k_2,\,+1}(z_2)\,cV^{({\rm nonlocal})}_{k_3,\,-1}(z_3) |0\ket.
\label{3pt_BBB_hol}
\ee
We change the picture assignment of the NS and R$+$ fields from 
the $(-1)$ and $(-\frac12)$ pictures to the 0 and $(-\frac32)$ pictures by use of  
\bea
\left(cT_{k_1}\right)^{(0)}(z_1) & = & Q_{{\rm BRST}}\left(2\xi(z_1) \,cT_{k_1}(z_1)\right) \nn \\
 & = & cT^{(0)}_{k_1}(z_1) -\frac12\,\eta\,e^{\phi+ik_1x+p_{\ell_1}\varphi}(z_1),  
\label{cT_0}\\
cV_{k_2,\,+1}(z_2) & = & Q_{{\rm BRST}}\left(2\xi(z_2)\,\left(cV_{k_2,\,+1}\right)^{(-3/2)}(z_2)\right), 
\label{cV+_-3/2}
\eea
where $T^{(0)}_k(z)$ is given by (\ref{T0}). 
Note that the picture changing operation does not commute with the multiplication of the $c$ ghost. 
The $(-\frac32)$-picture field is obtained by the inverse picture changing operator 
$Y(z)=2c\,(\partial\xi)\,e^{-2\phi}(z)$ as 
\be
\left(cV_{k,\,+1}\right)^{(-3/2)}(z)= \lim_{w\to z}Y(w)\,cV_{k,\,+1}(z)=2c(\partial c)(\partial \xi)\, 
e^{-\frac52\phi+i\frac12 H+ikx+p_{\ell}\varphi}(z).
\ee
We introduce the $\xi$ zero-mode to move to the large Hilbert space: 
\be
\left.\bra 0|\xi_0cT_{k_1}(z_1)\,cV_{k_2,\,+1}(z_2)\,cV^{({\rm nonlocal})}_{k_3,\,-1}(z_3) |0\ket\right|_{{\rm large}}.
\label{3pt_BBB_hol_large}
\ee
Here, $\xi_0$ can be replaced with $\xi(z_1)=\sum_{n\in {\bf Z}}\xi_n\,z_1^{-n}$ 
because $\left.\bra 0|\cdots|0\ket\right|_{{\rm large}}=0$ when $\cdots$ 
does not contain $\xi_0$. Plugging (\ref{cV+_-3/2}) into (\ref{3pt_BBB_hol_large}) and manipulating the contour 
of the BRST current leads to 
\bea 
(\ref{3pt_BBB_hol_large}) & = & \left.-\bra 0|\left(cT_{k_1}\right)^{(0)}(z_1)\,\xi(z_2)
\left(cV_{k_2,\,+1}\right)^{(-3/2)}(z_2)\,cV^{({\rm nonlocal})}_{k_3,\,-1}(z_3)|0\ket\right|_{{\rm large}} 
\nn \\
&  & \left.-2\bra 0|\xi(z_1)cT_{k_1}(z_1)\,\xi(z_2)\left(cV_{k_2,\,+1}\right)^{(-3/2)}(z_2)\,
Q_{{\rm BRST}}\left(cV_{k_3,\,-1}^{({\rm nonlocal})}(z_3)\right)|0\ket\right|_{{\rm large}}. \nn \\
& & 
\eea
The first and second terms come from the BRST currents encircling the points $z_1$ and $z_3$, respectively. 
We move $\xi(z_2)$ to the bra vacuum in the first term, 
and go back to the expression on the small Hilbert space: 
\be
 \bra 0|\left(cT_{k_1}\right)^{(0)}(z_1)\,
\left(cV_{k_2,\,+1}\right)^{(-3/2)}(z_2)\,cV^{({\rm nonlocal})}_{k_3,\,-1}(z_3)|0\ket , 
\ee
that is the realization of the picture changing operation. 
Although the second term would seem to remain due to (\ref{nonBRSTclosed}), 
a closer look shows that this is not the case. 
Actually, it reads 
\bea
& & -i\sqrt{2}(p_{\ell_3}-k_3-1)\,\bra 0|\xi c\,e^{-\phi+ik_1x+p_{\ell_1}\varphi}(z_1) \,
\xi(\partial\xi)c(\partial c)\,e^{-\frac52\phi+i\frac12H+ik_2x+p_{\ell_2}\varphi}(z_2) \nn \\
& & \hspace{37mm}\left.\times
c\eta\, e^{\frac12\phi+i\frac12H+ik_3x+p_{\ell_3}\varphi}(z_3)|0\ket\right|_{{\rm large}}. 
\label{3pt_BBB_hol_large2}
\eea
Notice (\ref{3pt_BBB_hol_large2}) does not conserve 
any of the ($\xi, \eta$) fermion number, the $c$ ghost number, the $\phi$ charge and the $H$ charge. 
Thus, we see that (\ref{3pt_BBB_hol_large2}) vanishes, and that the usual result of picture 
changing  
\bea
& & \bra 0|cT_{k_1}(z_1)\,cV_{k_2,\,+1}(z_2)\,cV^{({\rm nonlocal})}_{k_3,\,-1}(z_3) |0\ket \nn \\
& & \hspace{7mm} =\bra 0|\left(cT_{k_1}\right)^{(0)}(z_1)\,
\left(cV_{k_2,\,+1}\right)^{(-3/2)}(z_2)\,cV^{({\rm nonlocal})}_{k_3,\,-1}(z_3)|0\ket
\label{picture_change1}
\eea
is obtained. 

A similar argument for the amplitude in section~\ref{sec:4pt_BBBB} again leads to the usual result 
\bea
& & \bra 0|cV_{k_1,\,+1}(z_1)\,cV_{k_2,\,+1}(z_2)\,cV_{k_3,\,-1}(z_3)\,
\int dz_4\,V^{({\rm nonlocal})}_{k_4,\,-1}(z_4) |0\ket 
\nn \\
& & \hspace{7mm}  = \bra 0|\left(cV_{k_1,\,+1}\right)^{(+1/2)}(z_1)\,\left(cV_{k_2,\,+1}\right)^{(-3/2)}(z_2)\,
cV_{k_3,\,-1}(z_3)\,\int dz_4\,V^{({\rm nonlocal})}_{k_4,\,-1}(z_4) |0\ket \nn \\
& &  \hspace{7mm} =\bra 0|\left(cV_{k_1,\,+1}\right)^{(+1/2)}(z_1)\,cV_{k_2,\,+1}(z_2)\,
\left(cV_{k_3,\,-1}\right)^{(-3/2)}(z_3)\,\int dz_4\,V^{({\rm nonlocal})}_{k_4,\,-1}(z_4) |0\ket  
\nn \\
& & \label{picture_change2}
\eea
in spite of the presence of the nonlocal operator. 
(\ref{picture_change1}) and the first equality of (\ref{picture_change2}) can be regarded as evidence 
that the correspondence 
(\ref{mat_IIA_op1}) and (\ref{mat_IIA_op2}) holds 
independently of the choice of the picture in the IIA theory. 

We should notice that the usual result of picture changing does not hold for every amplitude 
in the presence of the nonlocal operators. 
Appendix~\ref{app:4pt_BBBB2_pic} presents an amplitude where the picture changing operation 
induces a nonvanishing term containing the BRST transformation of nonlocal operators. 
Since we have not identified the matrix-model counterpart to the positive-winding ``tachyons'' 
($\hat{V}_1, \hat{V}_2$ with $k_1, k_2=\frac12,\frac32,\cdots$ in~(\ref{4pt_BBBB2_V})), 
it is not clear which amplitude in the matrix model corresponds to the amplitude discussed there. 
In general, 
the BRST charge acting on $b$ ghosts in higher-genus amplitudes 
amounts to picking up contribution from boundaries of the 
moduli space~\cite{Atick:1987wt}. 
It could give a hint to understand the nonvanishing effect from the nonlocal operators in a geometrical 
manner.   

As far as IIA string amplitudes 
whose correspondence to matrix-model amplitudes is given in this paper, 
we have seen in the above that the picture changing manipulation works as usual around the natural pictures.

\section{Discussions}
\label{sec:discussions}
\setcounter{equation}{0}
In this paper, we computed various amplitudes in two-dimensional type IIA superstring theory 
on a nontrivial (R$-$, R$+$) background, where the background is expressed by vertex operators 
as a small perturbation. 
By comparing the results with 
correlators in the matrix model calculated in~\cite{Kuroki:2012nt}, we 
checked the correspondence of the type IIA theory 
to the supersymmetric double-well matrix model, which was previously discussed 
from the viewpoint of symmetries and spectrum in~\cite{Kuroki:2012nt}. 

To enable the comparison at the quantitative level, 
we explicitly constructed cocycle factors to vertex operators 
in such a way that the target-space statistics is respected.     
The evaluation of IIA string amplitudes at the on-shell momenta often needs to be regularized. 
We found a certain regularization scheme which seems reasonable from the viewpoint of resonance structure 
in the amplitudes. 
As a result of the comparison, 
there arise various relations among coefficients that connect quantities of the 
matrix model to those of the type IIA theory. Remarkably, all of such relations obtained so far 
are consistent with each other, which convinces us of the validity of the correspondence. 

We mainly investigated two-point amplitudes of the IIA theory on the nontrivial 
(R$-$, R$+$)  background. By taking into account the background in the perturbation theory, however they 
amount to the computation of three-, four- and five-point functions on the trivial background. 
Since the analysis of spectrum in the previous paper~\cite{Kuroki:2012nt} is based on the 
vertex operators (\ref{spectrum_after}) on the trivial background, 
the computation here is important to see the effect of the RR background.      
Some amplitudes have the factor of the square of the Liouville volume. Its physical interpretation 
is as follows. One of the volume factors is from the integral over the Liouville constant mode as usual, 
while the other factor is due to the resonance of external particles and the background. 
Although it increases technical complexity, it will be meaningful to examine the correspondence for 
higher-point or higher-genus amplitudes.  
Also, we considered leading nontrivial contributions to the amplitudes in the perturbation of 
the background $W_{{\rm RR}}$. It is interesting to analyze subleading contributions. Then, 
we would have to take into account deformation of the BRST charge as backreaction from the background, and 
the relation (\ref{qRR_leading}) would be renormalized as 
\be
\nu_+-\nu_-=q_{{\rm RR}} \left( 1+\sum_{n=1}^\infty u_n q_{{\rm RR}}^{2n}\right)
\ee
with $u_n$ coefficients.    

So far we have not yet clarified the matrix-model counterpart to the positive winding ``tachyons'' 
$g_s^2\int d^2z\,\hat{T}_{k+\frac12}(z)\,\hat{\bar{T}}_{-k-\frac12}(\bar{z})$ ($k=0,1,2\cdots$), 
while the negative winding ones ($k=-1,-2,\cdots$) would correspond to 
$\{\frac{1}{N}\tr(-iB)^{k+1}\}$ up to some mixing terms. 
We expect to make it clear by introducing source terms of an external matrix to the matrix model 
as discussed in the case of the Penner model in~\cite{Mukhi:2003sz,Imbimbo:1995yv}. 
If it succeeds, it will be interesting to investigate 
the correspondence for amplitudes concerning the ``tachyons'' 
(for example, what we presented in section~\ref{sec:tachyons} and appendix~\ref{app:4pt_BBBB2}). 
Then it would become clear 
how the picture changing issue of the type IIA theory on the (R$-$, R$+$) background 
in appendix~\ref{app:4pt_BBBB2_pic} is understood in the matrix model.  

The investigation here and in the previous paper~\cite{Kuroki:2012nt} focuses on massless degrees of freedom 
in the type IIA string theory. As discussed in the two-dimensional 
NSR string~\cite{Itoh:1991qb,Bouwknegt:1991am}, the type IIA theory also has massive states at fixed momenta 
called ``discrete states''. In the matrix model, it seems natural that such massive states correspond to 
single-trace operators involving several kinds of matrices like $\frac{1}{N}\tr\,(\phi^k\,\psi^{2\ell+1})$, 
$\frac{1}{N}\tr\,(\phi^k\,\psi^\ell\,B^m\, \bar{\psi}^n)$ and so on. 
It is intriguing to extend the correspondence to include the massive excitations. 

The correspondence we have discussed concerns fundamental string degrees of freedom in the type IIA theory. 
If we push forward this interpretation, 
each element of the matrix variables in the matrix model could be regarded as a 
sort of short string or string bit carrying a unit of winding or momentum along the $S^1$ direction, 
which is somewhat similar to the matrix string theory~\cite{Dijkgraaf:1997vv}. 
We could also consider the correspondence from another direction based on 
a relation of solitonic objects (D-branes) in the IIA theory with the matrix model 
as in~\cite{Douglas:2003up,McGreevy:2003kb,Klebanov:2003km,McGreevy:2003ep,Takayanagi:2003sm}, 
where matrix elements are interpreted as open string degrees of freedom on a bunch of D-branes, and 
the matrix models describe closed string dynamics via open-closed string duality. 
The single-trace operators could be regarded as sources of closed strings rather than the strings themselves. 
It would be worth considering both of the interpretations in a complementary manner 
to obtain deeper correspondence between the matrix model and the type IIA superstring theory.  

\section*{Acknowledgements}
We would like to thank Rajesh~Gopakumar, Yasuaki~Hikida, Kazuo~Hosomichi, Hirotaka~Irie, Satoshi~Iso, Hiroshi~Itoyama, 
Shoichi~Kawamoto, Sanefumi~Moriyama, Koichi~Murakami, Hidehiko~Shimada, 
Yuji~Sugawara, Hiroshi~Suzuki, Tadashi Takayanagi and Tamiaki~Yoneya for useful discussions and comments. 
F.~S. thanks KITP Santa Barbara for hospitality during his stay (February, 2012), 
where a part of this work was done. 
The authors thank Osaka City University and the Yukawa Institute for Theoretical Physics at Kyoto University. 
Discussions during the conference ``Progress in Quantum Field Theory and String Theory'' (April, 2012) 
and the YITP workshop ``Gauge/Gravity duality'' (October, 2012) 
were useful to complete this work. 
The work of T.~K. is supported in part by Rikkyo University Special Fund for Research and 
a Grant-in-Aid for Scientific Research (C), 25400274. 
The work of F.~S. is supported in part by a Grant-in-Aid for Scientific Research (C), 
21540290 and 25400289.

\appendix
\section{Summation of $m$ in (\ref{vevPhiPhi_MM})}
\label{app:sum_m}
\setcounter{equation}{0}
In this appendix, 
we show the equality between the r.h.s. of (\ref{vevPhiPhi_MM}) and of (\ref{vevPhiPhi_MM2}). 
Note that it is equivalent to prove that 
$
\tilde{S}(k,m)=\tilde{\cI}(k,m)
$
with 
\bea
\tilde{S}(k,m)& \equiv & \sum_{p=1}^m
B\left(p+k+\frac12,\,\frac12\right)\,B\left(k+m-p+\frac12,\,\frac32\right), 
\label{Stildekm} \\
\tilde{\cI}(k,m)& \equiv & \frac{m}{2k+m+1}\,B\left(k+\frac12,\,\frac12\right)\,B\left(k+m+\frac12,\,\frac12\right). 
\label{Itildekm}   
\eea

We see that both of $\tilde{S}(k,m)$ and $\tilde{\cI}(k,m)$ obey the same recursion relation.  
For (\ref{Stildekm})  
\bea
\tilde{S}(k+1, m-2) & = & \tilde{S}(k,m) -B\left(k+\frac32,\,\frac12\right)\,B\left(k+m-\frac12,\,\frac32\right)
\nn \\
 & & \hspace{15mm}-B\left(k+\frac12,\,\frac32\right)\,B\left(k+m+\frac12,\,\frac12\right) \nn \\
 & = & \tilde{S}(k,m)
-\frac{2k+m}{2(k+1)(k+m+\frac12)}\,B\left(k+\frac12,\,\frac12\right)\,B\left(k+m+\frac12,\,\frac12\right),
\nn \\
& & 
\eea
and for (\ref{Itildekm})
\bea
\tilde{\cI}(k+1, m-2) & = & \tilde{\cI}(k,m)
-\frac{2k+m}{2(k+1)(k+m+\frac12)}\,B\left(k+\frac12,\,\frac12\right)\,B\left(k+m+\frac12,\,\frac12\right). 
\nn \\
& & 
\eea
Since  $\tilde{S}$ and $\tilde{\cI}$ satisfy the same initial conditions: 
\bea
& & \tilde{S}(k,0)=\tilde{\cI}(k,0)=0, \nn \\
& & \tilde{S}(k,1)=\tilde{\cI}(k,1)
=\frac{1}{2(k+1)}\,B\left(k+\frac12,\,\frac12\right)\,B\left(k+\frac32,\,\frac12\right)
\eea
for $k=0,1,\cdots$, we can conclude that 
$
\tilde{S}(k,m)=\tilde{\cI}(k,m). 
$

\section{Worldsheet superconformal symmetry}
\label{app:SCFT}
\setcounter{equation}{0}

Superconformal generators 
\bea
T_F & = &  T_{{\rm m},\,F}+T_{{\rm gh}, \,F}, \\
T_{{\rm m},\,F} & = & \frac{i}{2}\psi_x\partial x +\frac{i}{2} \psi_\ell\partial\varphi -\frac{i}{2}Q\partial\psi_\ell, \\
T_{{\rm gh},\,F} & = & \frac12 b\gamma -\partial \beta c -\frac32\beta\partial c
\eea 
and the energy-momentum tensors (\ref{EMtensor}) have the OPEs: 
\bea
T_{\bA}(z)T_{\bB}(w) & \sim & 
\delta_{\bA,\bB}\left[\frac{c_{\bB}/2}{(z-w)^4}+\frac{2}{(z-w)^2}T_{\bB}(w)+\frac{1}{z-w}\partial T_{\bB}(w)\right], \nn \\
T_{\bA}(z)T_{\bB,\,F}(w) & \sim & 
\delta_{\bA,\bB}\left[\frac{3/2}{(z-w)^2}T_{\bB, \,F}(w)+\frac{1}{z-w}\partial T_{\bB,\,F}(w)\right], \nn\\
T_{\bA,\,F}(z)T_{\bB,\,F}(w) & \sim & 
\delta_{\bA,\bB}\left[\frac{c_{\bB}/6}{(z-w)^3} +\frac{1}{z-w}\frac12 T_{\bB}(w)\right]
\label{OPE_N=1SCA}
\eea
with $\bA, \bB = {\rm m}, {\rm gh}$. The central charges for the matter sector ($(x,\psi_x)$ and $(\varphi, \psi_\ell)$) 
and the ghost sector ($(b,c)$ and $(\beta, \gamma)$) are 
\be
c_{\rm m} =  \frac32 + \left(\frac32 + 3Q^2\right)= 3+3Q^2 , \qquad
c_{\rm gh}  =  -26 + 11 = -15. 
\ee
Thus the total central charge  $c=c_{\rm m}+c_{\rm gh}$ vanishes by using $Q=2$. 
In terms of modes 
\be
T_{\bA}(z)= \sum_{n\in {\bf Z}}L_n^{\bA} z^{-n-2}, \qquad T_{\bA,\,F}(z)= \frac12 \sum_r G_r^{\bA}z^{-r-3/2} 
\ee
($r\in {\bf Z}+\frac12$ for the NS sector, $r\in{\bf Z}$ for the R sector),  
the OPEs (\ref{OPE_N=1SCA}) represent the $\cN=1$ superconformal algebra
\bea
 & & [L^{\bA}_n,L^{\bB}_m] = \delta_{\bA,\bB}\left[(n-m)L_{n+m} +\frac{c_{\bB}}{12}(n^3-n) \delta_{n+m,0}\right], \nn \\
& & [L^{\bA}_n, G^{\bB}_r] = \delta_{\bA,\bB}\left[\left(\frac{n}{2}-r\right) G^{\bB}_{n+r}\right], \nn \\
& & \{G^{\bA}_r, G^{\bB}_s\} = \delta_{\bA,\bB}\left[2L^{\bB}_{r+s}+\frac{c_{\bB}}{3}\left(r^2-\frac14\right)\delta_{r+s,0}\right]. 
\eea

The matter part has an enlarged symmetry, namely the $\cN=2$ superconformal symmetry. By dividing the superconformal generator as 
\bea
T_{{\rm m},\,F} & = & T_{{\rm m},\,F}^+ + T_{{\rm m},\,F}^-, \label{TmF}\\
T_{{\rm m},\,F}^+& \equiv & \frac{i}{4}\left[\Psi^\dagger \partial(\varphi+ix)-Q\partial\Psi^\dagger\right]
=  \frac{i}{2\sqrt{2}}\left[e^{iH}\partial(\varphi+ix)-Q\partial e^{iH}\right], \label{TmFp}\\
T_{{\rm m},\,F}^- & \equiv &  \frac{i}{4}\left[\Psi \partial(\varphi-ix)-Q\partial\Psi\right] 
= \frac{i}{2\sqrt{2}}\left[e^{-iH}\partial(\varphi-ix)-Q\partial e^{-iH}\right], \label{TmFm}
\eea
and defining the $U(1)$ current as 
\be
J\equiv -\frac12\Psi\Psi^\dagger+iQ\partial x=i\partial H+iQ\partial x, 
\ee
one can see the OPEs for the $\cN=2$ superconformal algebra: 
\bea
T_{\rm m}(z) T_{\rm m}(w) & \sim & \frac{c_{\rm m}/2}{(z-w)^4} +\frac{2}{(z-w)^2}\,T_m(w) +\frac{1}{z-w}\,\partial T_m(w), \nn \\
T_{\rm m}(z) T^{\pm}_{{\rm m},\,F}(w) & \sim & \frac{3/2}{(z-w)^2}\,T^{\pm}_{m,\,F}(w) +\frac{1}{z-w}\,\partial T^{\pm}_{m,\,F}(w), 
\nn \\
T_{\rm m}(z) J(w) & \sim & \frac{1}{(z-w)^2}\,J(w) + \frac{1}{z-w}\,\partial J(w), \nn \\
T^{+}_{{\rm m},\,F}(z)T^{-}_{{\rm m},\,F}(w) & \sim & \frac{c_m/12}{(z-w)^3} + \frac{1}{(z-w)^2}\frac14\,J(w) \nn \\
 & & +\frac{1}{z-w}\left(\frac14\,T_m(w)+\frac18\,\partial J(w)\right), \nn \\
J(z)T^{\pm}_{{\rm m},\,F}(w) & \sim & \frac{\pm1}{z-w}\,T^{\pm}_{m,\,F}(w) , \nn \\
J(z)J(w) & \sim & \frac{c_{\rm m}/3}{(z-w)^2} , \nn \\
T^{\pm}_{{\rm m},\,F}(z)T^{\pm}_{{\rm m},\,F}(w) & \sim & 0. 
\label{OPE_N=2SCA}
\eea
In terms of modes 
\be
T_{{\rm m},\,F}^{\pm}(z) = \frac{1}{2\sqrt{2}}\sum_r G^\pm_r z^{r-3/2}, \qquad J(z) = \sum_{n\in {\bf Z}} J_n z^{-n-1}, 
\ee
(\ref{OPE_N=2SCA}) are recast as 
\bea
 & & [L^{\rm m}_n,L^{\rm m}_m] = (n-m)L^{\rm m}_{n+m} +\frac{c_{\rm m}}{12}(n^3-n) \delta_{n+m,0}, \nn \\
& & [L^{\rm m}_n, G^{\pm}_r] = \left(\frac{n}{2}-r\right) G^{\pm}_{n+r}, \nn \\
& & [L^{\rm m}_n, J_m] = -m J_{n+m}, \nn \\
& & \{G^+_r, G^-_s\} = 2L^{\rm m}_{r+s} +(r-s)J_{r+s} +\frac{c_{\rm m}}{3}\left(r^2-\frac14\right)\delta_{r+s,0}, \nn \\
& & [J_n, G^\pm_r] = \pm G^\pm_{n+r}, \nn \\
& & [J_n,J_m] = \frac{c_{\rm m}}{3}\,n\delta_{n+m,0}, \nn \\
& & \{G^\pm_r,G^\pm_s\} = 0. 
\eea

The BRST charge 
\bea
Q_{{\rm BRST}} & = & \oint\frac{dz}{2\pi i}\,j_{{\rm BRST}}(z), 
\nn \\ 
j_{{\rm BRST}}(z) & \equiv & c(z)\left(T_{\rm m}(z)+\frac12T_{\rm gh}(z)\right)
-\gamma(z)\left(T_{{\rm m},\,F}(z)+\frac12T_{{\rm gh},\,F}(z)\right) 
\label{QBRST}
\eea
is decomposed into the three pieces 
\bea
Q_{{\rm BRST}} & = & Q_0 + Q_1 +Q_2, \\
Q_0 & \equiv & \oint \frac{dz}{2\pi i}\left[c(z)\left(T_{{\rm m}}(z) +T_{\beta\gamma}(z)\right)+bc\partial c(z)\right], \nn \\
Q_1 & \equiv & - \oint \frac{dz}{2\pi i} \,\gamma(z)T_{{\rm m},\,F}(z) = -\oint \frac{dz}{2\pi i} \,e^{\phi}\,\eta\,T_{{\rm m},\,F}(z), \nn \\
Q_2 & \equiv & -\frac14  \oint \frac{dz}{2\pi i} \,b\gamma^2(z) = 
\frac14\oint \frac{dz}{2\pi i} \,e^{2\phi}(\partial\eta)\eta b(z)
\label{QBRST2}
\eea
according to the superconformal ghost charges. 
Here, $T_{\beta\gamma}=-\frac32\beta\partial\gamma-\frac12\partial \beta\gamma$ is the $\beta\gamma$-part of $T_{\rm gh}$. 
One can check the nilpotency of the BRST charge ($Q_{{\rm BRST}}^2=0$). 

Although we discussed only the holomorphic part, we can repeat a parallel argument for the anti-holomorphic part.

\section{Cocycle factors for 0-picture NS fields}
\label{app:cocycle_T0}
\setcounter{equation}{0}
The 0-picture holomorphic NS field $T^{(0)}_k$ is obtained by the picture changing operation 
\be
Q_{{\rm BRST}}\left(2\xi(z)T_k(z)\right) 
\equiv \oint_z\frac{dw}{2\pi i}\,j_{{\rm BRST}}(w)\,2\xi(z)T_k(z)= \partial (2c\xi T_k)(z) + T_k^{(0)}(z), 
\ee
where the first term does not contribute upon the $z$ integration and can be neglected. 
The second term represents 
\be
T^{(0)}_k(z) = T^{(0)}_{k,\,+1}(z) + T^{(0)}_{k, \,-1}(z), \qquad
T^{(0)}_{k,\,\epsilon}(z)  \equiv 
\frac{i}{\sqrt{2}}\,(p_\ell-\epsilon k)\,e^{i\epsilon H+ikx+p_\ell \varphi}(z). 
\label{T0}
\ee
The anti-holomorphic field is similarly given by 
\be  
\bar{T}^{(0)}_{\bar{k}}(\bar{z}) = \bar{T}^{(0)}_{\bar{k},\,+1}(\bar{z}) + \bar{T}^{(0)}_{\bar{k}, \,-1}(\bar{z}), \qquad 
\bar{T}^{(0)}_{\bar{k},\,\bar{\epsilon}}(\bar{z}) \equiv 
\frac{i}{\sqrt{2}}\,(p_\ell-\bar{\epsilon} \bar{k})\,e^{i\bar{\epsilon} \bar{H}+i\bar{k}\bar{x}+p_\ell \bar{\varphi}}(\bar{z}).
\label{Tbar0}
\ee

The OPEs 
\bea
q_+(z)\,T^{(0)}_{k,\,\epsilon}(w) & = & \frac{1}{(z-w)^{\frac{\epsilon}{2}+k}}\,:q_+(z)\,T^{(0)}_{k,\,\epsilon}(w): \,, 
\\
\bar{q}_-(\bar{z})\,\bar{T}^{(0)}_{\bar{k},\,\bar{\epsilon}}(\bar{w}) & = & \frac{1}{(\bar{z}-\bar{w})^{-\frac{\bar{\epsilon}}{2}-\bar{k}}}\,:\bar{q}_-(\bar{z})\,\bar{T}^{(0)}_{\bar{k},\,\bar{\epsilon}}(\bar{w}): 
\eea
show that the degree of the poles differs by one depending on the values of $\epsilon, \bar{\epsilon}$. 
It suggests that the $\epsilon=\pm 1$ parts of $T^{(0)}_k$ have different target-space statistics. 
Similar is the case of $\bar{T}^{(0)}_{\bar{k}}$. 

In the same manner as in section~\ref{sec:cocycle}, we introduce the cocycle factor to the 0-picture NS 
fields~\footnote{
Note that the cocycle factor for $T^{(0)}_{k,\,\epsilon}(z)$ is different from the one for $T_k(z)$ in \eqref{cocycle_TV}. 
If we put appropriate cocycle factors to the BRST charge (\ref{QBRST2}) corresponding to 
exponential operators appearing there and similarly to the anti-holomorphic BRST charge, 
the result (\ref{cocycle_T0}) will be directly obtained from the picture changing operation.}: 
\bea
\hat{T}^{(0)}_k(z) & = & \hat{T}^{(0)}_{k,\,+1}(z) + \hat{T}^{(0)}_{k, \,-1}(z), \qquad
\hat{T}^{(0)}_{k,\,\epsilon}(z)  \equiv e^{\pi \beta (i\epsilon p_{\bar{h}}+ikp_{\bar{x}})}\, T^{(0)}_{k,\,\epsilon}(z), \nn \\
\hat{\bar{T}}^{(0)}_{\bar{k}}(\bar{z}) & = & \hat{\bar{T}}^{(0)}_{\bar{k},\,+1}(\bar{z}) + \hat{\bar{T}}^{(0)}_{\bar{k}, \,-1}(\bar{z}), \qquad 
\hat{\bar{T}}^{(0)}_{\bar{k},\,\bar{\epsilon}}(\bar{z}) \equiv 
e^{-\pi\beta (i\bar{\epsilon}p_h+i\bar{k}p_x)}\,\bar{T}^{(0)}_{\bar{k},\,\bar{\epsilon}}(\bar{z}). 
\label{cocycle_T0}
\eea 
Then, we find 
\bea
\hat{\bar{T}}^{(0)}_{\bar{k},\,\bar{\epsilon}}(\bar{z})\,\hat{q}_+(w) & = & 
e^{i2\pi\beta (\frac{\bar{\epsilon}}{2}+\bar{k})}\,\hat{q}_+(w)\,\hat{\bar{T}}^{(0)}_{\bar{k},\,\bar{\epsilon}}(\bar{z}), \\
\hat{\bar{q}}_-(\bar{z})\,\hat{T}_{k,\,\epsilon}(w) & = & 
e^{i2\pi\beta(-\frac{\epsilon}{2}-k)}\,\hat{T}_{k,\,\epsilon}(w)\,\hat{\bar{q}}_-(\bar{z}), 
\eea
and 
\bea
\hat{\bar{T}}^{(0)}_{\bar{k},\,\bar{\epsilon}}(\bar{z})\,\hat{T}_{k'}(w) & = & e^{i2\pi\beta(-\bar{k}k')}\,
\hat{T}_{k'}(w)\,\hat{\bar{T}}^{(0)}_{\bar{k},\,\bar{\epsilon}}(\bar{z}), \\
\hat{\bar{T}}^{(0)}_{\bar{k},\,\bar{\epsilon}}(\bar{z})\,\hat{V}_{k',\,\epsilon'}(w) & = & 
e^{2\pi\beta(-\frac12\bar{\epsilon}\epsilon'-\bar{k}k')}\,\hat{V}_{k',\,\epsilon'}(w)\,
\hat{\bar{T}}^{(0)}_{\bar{k},\,\bar{\epsilon}}(\bar{z}), \\
\hat{\bar{T}}_{\bar{k}}(\bar{z})\,\hat{T}^{(0)}_{k',\,\epsilon}(w) & = & e^{i2\pi\beta(-\bar{k}k')}\,
\hat{T}^{(0)}_{k',\,\epsilon}(w)\, \hat{\bar{T}}_{\bar{k}}(\bar{z}), \\
\hat{\bar{V}}_{\bar{k},\,\bar{\epsilon}}(\bar{z})\,\hat{T}^{(0)}_{k',\,\epsilon'}(w) & = & 
e^{i2\pi\beta(-\frac12\bar{\epsilon}\epsilon'-\bar{k}k')}\, \hat{T}^{(0)}_{k',\,\epsilon'}(w)\,
\hat{\bar{V}}_{\bar{k},\,\bar{\epsilon}}(\bar{z}), \\
\hat{\bar{T}}^{(0)}_{\bar{k},\,\bar{\epsilon}}(\bar{z})\,\hat{T}^{(0)}_{k',\,\epsilon'}(w) & = & 
e^{i2\pi\beta(-\bar{\epsilon}\epsilon'-\bar{k}k')}\, \hat{T}^{(0)}_{k',\,\epsilon'}(w)\,
\hat{\bar{T}}^{(0)}_{\bar{k},\,\bar{\epsilon}}(\bar{z}). 
\eea

\paragraph{(NS$^{(0)}$, NS$^{(0)}$) sector} 
{}From the above result, we see that the target-space supercharges $\hat{Q}_+$, $\hat{\bar{Q}}_-$ act on 
the $(\epsilon,\bar{\epsilon})=(+1,-1), (-1,+1)$ parts of 
the (NS$^{(0)}$, NS$^{(0)}$) field 
\be
\hat{T}^{(0)}_k(z)\,\hat{\bar{T}}^{(0)}_{-k}(\bar{z}) = \sum_{\epsilon, \bar{\epsilon}=\pm 1} 
\hat{T}^{(0)}_{k,\,\epsilon}(z)\,\hat{\bar{T}}^{(0)}_{-k,\,\bar{\epsilon}}(\bar{z})
\label{cocycle_T0T0bar}
\ee 
in the form of a commutator, but on the $(\epsilon,\bar{\epsilon})=(+1,+1), (-1,-1)$ parts in the form of 
an anti-commutator, so that they can be expressed as the contour integral of the radial ordering. 
Moreover, among fields in the natural picture ($(-1)$-picture for NS, $(-1/2)$-picture for R), 
target-space fermions ((NS, R$-$), (R$+$, NS)) commute with the $(\epsilon,\bar{\epsilon})=(+1,-1), (-1,+1)$ 
parts, but anti-commute with the $(\epsilon,\bar{\epsilon})=(+1,+1), (-1,-1)$ parts. 
Target-space bosons ((NS, NS), (R$+$, R$-$), (R$-$, R$+$)) commute with the all parts of $(\epsilon,\bar{\epsilon})$. 
Thus, we conclude that the $(\epsilon,\bar{\epsilon})=(+1,-1), (-1,+1)$ parts have the correct target-space 
statistics, while the $(\epsilon,\bar{\epsilon})=(+1,+1), (-1,-1)$ parts do not. 

Interestingly, the $(\epsilon,\bar{\epsilon})=(+1,+1), (-1,-1)$ parts do not contribute to any amplitudes 
computed in this paper. It is likely that the target-space statistics is correctly realized at the level of 
the amplitudes. 

\paragraph{(NS$^{(0)}$, R$-$) sector} 
Similarly, we see that the $\epsilon=+1$ part of the (NS$^{(0)}$, R$-$) field 
\be
\hat{T}_k^{(0)}(z)\,\hat{\bar{V}}_{k,\,-1}(\bar{z})=\sum_{\epsilon=\pm 1}
\hat{T}_{k,\,\epsilon}^{(0)}(z)\,\hat{\bar{V}}_{k,\,-1}(\bar{z})
\label{cocycle_T0Vbar}
\ee
has the correct target-space statistics, but the $\epsilon=-1$ part does not. 

The $\epsilon=-1$ part does not contribute to the amplitudes in this paper, 
and the target-space statistics is correct in the amplitudes. 

\paragraph{(R$+$, NS$^{(0)}$) sector}
In the (R$+$, NS$^{(0)}$) field
\be
\hat{V}_{k,\,+1}(z)\,\hat{\bar{T}}^{(0)}_k(\bar{z}) = \sum_{\bar{\epsilon}=\pm 1} 
\hat{V}_{k,\,+1}(z)\,\hat{\bar{T}}^{(0)}_{k,\,\bar{\epsilon}}(\bar{z}),
\label{cocycle_VT0bar}
\ee
the $\bar{\epsilon}=-1$ part obeys the correct statistics, while the $\bar{\epsilon}=+1$ part does not. 

We see that the $\bar{\epsilon}=+1$ part gives no contribution to the computed amplitudes.    
The correct statistics is realized in the amplitudes.

\section{Integral formulas}
\label{app:integrals}
\setcounter{equation}{0}
In this appendix, we present formulas for the two integrals  
\bea
I_{(1,0)} & \equiv & \int d^2z\,z^\alpha \,\bar{z}^{\bar{\alpha}} \,(1-z)^\beta\,(1-\bar{z})^{\bar{\beta}}, 
\label{I(1,0)}
\\
I_{(1,1)} & \equiv & \int d^2z\,d^2w\,z^\alpha \,\bar{z}^{\bar{\alpha}} \,(1-z)^\beta\,
(1-\bar{z})^{\bar{\beta}}\,w^{\alpha'}\,\bar{w}^{\bar{\alpha}'}\,(1-w)^{\beta'}\,(1-\bar{\beta})^{\bar{\beta}'}
\,|z-w|^{4\sigma},\nn \\
& & \label{I(1,1)}
\eea
where $z=x+iy$, $\bar{z}=x-iy$, $w=u+iv$, $\bar{w}=u-iv$, $d^2z=dx\,dy$ and $d^2w=du\,dv$. 
The powers appearing in the integrands $\alpha, \bar{\alpha},\cdots, \beta', \bar{\beta}', \sigma$   
are independent. 

More general integrals including (\ref{I(1,0)}), (\ref{I(1,1)}) are computed in~\cite{Fukuda:2001jd}. 
However, in order to make this paper reasonably self-contained and 
to remark on the conventions for complex phases in integrands that is not mentioned in~\cite{Fukuda:2001jd}, 
we give computational details.

\subsection{$I_{(1,0)}$}
\label{app:I(1,0)}
As discussed in \cite{Fukuda:2001jd,Kawai:1985xq}, 
we rotate the integration contour of $y$ by almost $-90$ degrees as 
\be
y\to e^{-i(\frac{\pi}{2}-\epsilon)}\,y=-i(1+i\epsilon)\,y,   
\label{y_contour_deform}
\ee
and the real axis by $-\epsilon$, so $x\to (1-i\epsilon)x$, $1-x\to(1-i\epsilon)(1-x)$. 
Then, in terms of $\eta\equiv x+y, \chi\equiv x-y$, (\ref{I(1,0)}) becomes 
\be
I_{(1,0)}= \frac{-i}{2}\int^\infty_{-\infty}d\eta\,d\chi\,(\eta-i\epsilon\chi)^{\alpha}\,
(\chi-i\epsilon\eta)^{\bar{\alpha}}\,(1-\eta-i\epsilon(1-\chi))^{\beta}\,(1-\chi-i\epsilon(1-\eta))^{\bar{\beta}}.
\label{I(1,0)_1}
\ee
We assumed no contribution from the infinity in the contour deformation (\ref{y_contour_deform}), 
which is justified when 
\be
{\rm Re}\,(\alpha+\bar{\alpha}+\beta+\bar{\beta})<-1. 
\label{int_cond1}
\ee
The complex phase in the integrand should be carefully treated. We consider the following 
two phase conventions. The one (I) is 
\bea
 & & \left\{(\eta-i\epsilon\chi)(\chi-i\epsilon\eta)\right\}^{\alpha}\,
(\chi-i\epsilon\eta)^{\bar{\alpha}-\alpha} \nn \\
& & \times \left\{(1-\eta-i\epsilon(1-\chi))(1-\chi-i\epsilon(1-\eta))\right\}^{\beta}\,
 (1-\chi-i\epsilon(1-\eta))^{\bar{\beta}-\beta},
\eea
and the other (II) is 
\bea
 & & \left\{(\eta-i\epsilon\chi)(\chi-i\epsilon\eta)\right\}^{\bar{\alpha}}\,
(\eta-i\epsilon\chi)^{\alpha-\bar{\alpha}} \nn \\
& & \times \left\{(1-\eta-i\epsilon(1-\chi))(1-\chi-i\epsilon(1-\eta))\right\}^{\bar{\beta}}\,
 (1-\eta-i\epsilon(1-\chi))^{\beta-\bar{\beta}}. 
\eea 
For example, $\left\{(\eta-i\epsilon\chi)(\chi-i\epsilon\eta)\right\}^{\alpha}$ means 
\be
 \left\{(\eta-i\epsilon\chi)(\chi-i\epsilon\eta)\right\}^{\alpha}=(\eta\chi-i\epsilon)^\alpha
=\begin{cases} |\eta\chi|^\alpha & (\eta\chi>0) \\ e^{-i\pi\alpha}|\eta\chi|^\alpha & (\eta\chi<0).
\end{cases}
\ee
By noting the phase in the case (II), 
we see that (\ref{I(1,0)_1}) can be divided into the three parts according to integration regions of $\chi$: 
\bea
I_{(1,0)} & = & \frac{-i}{2}\int^0_{-\infty}d\chi\, \int^\infty_{-\infty}d\eta\,
(\eta+i\epsilon)^\alpha\,(\chi-i\epsilon)^{\bar{\alpha}}\,(1-\eta-i\epsilon)^\beta\,(1-\chi)^{\bar{\beta}}
\nn \\
& & + \frac{-i}{2}\int^\infty_1d\chi\, \int^\infty_{-\infty}d\eta\,
(\eta-i\epsilon)^\alpha\,\chi^{\bar{\alpha}}\,(1-\eta+i\epsilon)^\beta\,(1-\chi-i\epsilon)^{\bar{\beta}}
\nn \\
& & +\frac{-i}{2}\int^1_0d\chi\, \int^\infty_{-\infty}d\eta\,
(\eta-i\epsilon)^\alpha\,\chi^{\bar{\alpha}}\,(1-\eta-i\epsilon)^\beta\,(1-\chi)^{\bar{\beta}}.
\label{I(1,0)_2}
\eea
Since the integrand of the first line (the second line) in (\ref{I(1,0)_2}) is regular with respect 
to the upper (lower) half plane of $\eta$, 
the $\eta$-integral vanishes by closing the contour with a large semi-circle there. 
Contribution from the large semi-circle can be neglected when 
\be
{\rm Re}\,(\alpha+\beta)<-1. 
\label{int_cond2}
\ee
The $\chi$-integral in the last line gives $B(\bar{\alpha}+1, \bar{\beta}+1)$, and 
the $\eta$-integral can be computed as 
\bea 
\int^\infty_{-\infty}d\eta\,(\eta-i\epsilon)^\alpha\,(1-\eta-i\epsilon)^\beta
 & = & -2i\sin(\pi\beta)\int^\infty_1d\eta\,\eta^\alpha(\eta-1)^\beta \nn \\
& = & -2i\sin(\pi\beta)\,B(\beta+1,-\alpha-\beta-1),
\eea
where the first equality holds in the region (\ref{int_cond2}). 
Thus, we end up with 
\bea
I_{(1,0)} & = & -\sin(\pi\beta)\,B(\bar{\alpha}+1, \bar{\beta}+1)\, B(\beta+1,-\alpha-\beta-1)
\nn \\
 & = & \pi\frac{\Gamma(\bar{\alpha}+1)\,\Gamma(\bar{\beta}+1)}{\Gamma(\bar{\alpha}+\bar{\beta}+2)}\,
\frac{\Gamma(-\alpha-\beta-1)}{\Gamma(-\alpha)\,\Gamma(-\beta)}
\label{I(1,0)_f_II}
\eea
for the case (II). 
The derivation of (\ref{I(1,0)_f_II}) is valid for
\be
-1<{\rm Re}\,\alpha, \,{\rm Re}\, \bar{\alpha}, \,{\rm Re}\,\beta, \,{\rm Re}\,\bar{\beta} <-\frac12. 
\label{I(1,0)_cond}
\ee
We define $I_{(1,0)}$ for generic $\alpha, \bar{\alpha}, \beta, \bar{\beta}$ by analytic continuation 
from (\ref{I(1,0)_cond}).  

On the other hand, the result for the case (I) is obtained from (\ref{I(1,0)_f_II}) by the replacement 
$\chi\leftrightarrow \eta$, $\alpha\leftrightarrow \bar{\alpha}$ and $\beta\leftrightarrow \bar{\beta}$: 
\be
I_{(1,0)} = \pi\frac{\Gamma(\alpha+1)\,\Gamma(\beta+1)}{\Gamma(\alpha+\beta+2)}\,
\frac{\Gamma(-\bar{\alpha}-\bar{\beta}-1)}{\Gamma(-\bar{\alpha})\,\Gamma(-\bar{\beta})},
\label{I(1,0)_f_I}
\ee
which is obtained in ref.~\cite{Fukuda:2001jd}. 

(\ref{I(1,0)_f_II}) and (\ref{I(1,0)_f_I}) are different in general, but become coincident when 
\be
\frac{s(\alpha)\,s(\beta)}{s(\alpha+\beta)}=\frac{s(\bar{\alpha})\,s(\bar{\beta})}{s(\bar{\alpha}+\bar{\beta})}
\qquad \mbox{with} \qquad s(x)\equiv \sin(\pi x). 
\label{cond_I=II}
\ee
For amplitudes among mutually local vertex operators as we discuss in the text, 
the parameters satisfy $\alpha-\bar{\alpha},\,\beta-\bar{\beta}\in {\bf Z}$, and thus (\ref{cond_I=II}).

\subsection{$I_{(1,1)}$}
\label{app:I(1,1)}
Similar contour deformation as in the previous subsection for $y$ as well as for $v$ leads to 
\bea
I_{(1,1)} & = & \left(\frac{-i}{2}\right)^2\int^\infty_{-\infty}d\eta_1\,d\chi_1\,d\eta_{\hat{1}}\,d\chi_{\hat{1}}\,
(\eta_1-i\epsilon\chi_1)^{\alpha}\,(\chi_1-i\epsilon\eta_1)^{\bar{\alpha}}\,
(1-\eta_1-i\epsilon(1-\chi_1))^{\beta}\nn \\
& & \hspace{21mm}\times (1-\chi_1-i\epsilon(1-\eta_1))^{\bar{\beta}}\, 
(\eta_{\hat{1}}-i\epsilon\chi_{\hat{1}})^{\alpha'}\,(\chi_{\hat{1}}-i\epsilon\eta_{\hat{1}})^{\bar{\alpha}'}\nn \\
& & \hspace{21mm}\times(1-\eta_{\hat{1}}-i\epsilon(1-\chi_{\hat{1}}))^{\beta'}\,(1-\chi_{\hat{1}}-i\epsilon(1-\eta_{\hat{1}}))^{\bar{\beta}'} \nn \\
& & \hspace{21mm}\times \left(\eta_1-\eta_{\hat{1}}-i\epsilon(\chi_1-\chi_{\hat{1}})\right)^{2\sigma}
\,\left(\chi_1-\chi_{\hat{1}}-i\epsilon(\eta_1-\eta_{\hat{1}})\right)^{2\sigma},
\label{I(1,1)_1}
\eea
where we put $\eta_1=x+y$, $\chi_1=x-y$, $\eta_{\hat{1}}=u+v$, $\chi_{\hat{1}}=u-v$.
The phase convention of (I) for the integrand is 
\bea
& & \left\{(\eta_1-i\epsilon\chi_1)(\chi_1-i\epsilon\eta_1)\right\}^{\alpha}\,
(\chi_1-i\epsilon\eta_1)^{\bar{\alpha}-\alpha}\,
\left\{(1-\eta_1-i\epsilon (1-\chi_1))(1-\chi_1-i\epsilon (1-\eta_1))\right\}^{\beta}\nn \\
& & \times (1-\chi_1-i\epsilon (1-\eta_1))^{\bar{\beta}-\beta}\,
\left\{(\eta_{\hat{1}}-i\epsilon\chi_{\hat{1}})(\chi_{\hat{1}}-i\epsilon\eta_{\hat{1}})\right\}^{\alpha'}\,
(\chi_{\hat{1}}-i\epsilon\eta_{\hat{1}})^{\bar{\alpha}'-\alpha'} \nn \\
& & \times \left\{(1-\eta_{\hat{1}}-i\epsilon (1-\chi_{\hat{1}}))
(1-\chi_{\hat{1}}-i\epsilon (1-\eta_{\hat{1}}))\right\}^{\beta'}\,
(1-\chi_{\hat{1}}-i\epsilon (1-\eta_{\hat{1}}))^{\bar{\beta}'-\beta'}\nn \\
& & \times \left\{(\eta_1-\eta_{\hat{1}}-i\epsilon (\chi_1-\chi_{\hat{1}}))
(\chi_1-\chi_{\hat{1}}-i\epsilon (\eta_1-\eta_{\hat{1}}))\right\}^{2\sigma}, 
\eea 
and that of (II) is 
\bea
& & \left\{(\eta_1-i\epsilon\chi_1)(\chi_1-i\epsilon\eta_1)\right\}^{\bar{\alpha}}\,
(\eta_1-i\epsilon\chi_1)^{\alpha-\bar{\alpha}}\,
\left\{(1-\eta_1-i\epsilon (1-\chi_1))(1-\chi_1-i\epsilon (1-\eta_1))\right\}^{\bar{\beta}}\nn \\
& & \times (1-\eta_1-i\epsilon (1-\chi_1))^{\beta-\bar{\beta}}\,
\left\{(\eta_{\hat{1}}-i\epsilon\chi_{\hat{1}})(\chi_{\hat{1}}-i\epsilon\eta_{\hat{1}})\right\}^{\bar{\alpha}'}\,
(\eta_{\hat{1}}-i\epsilon\chi_{\hat{1}})^{\alpha'-\bar{\alpha}'} \nn \\
& & \times \left\{(1-\eta_{\hat{1}}-i\epsilon (1-\chi_{\hat{1}}))
(1-\chi_{\hat{1}}-i\epsilon (1-\eta_{\hat{1}}))\right\}^{\bar{\beta}'}\,
(1-\eta_{\hat{1}}-i\epsilon (1-\chi_{\hat{1}}))^{\beta'-\bar{\beta}'}\nn \\
& & \times \left\{(\eta_1-\eta_{\hat{1}}-i\epsilon (\chi_1-\chi_{\hat{1}}))
(\chi_1-\chi_{\hat{1}}-i\epsilon (\eta_1-\eta_{\hat{1}}))\right\}^{2\sigma}. 
\eea 

Similarly to the previous subsection, (\ref{I(1,1)_1}) in the case (II) can be divided into twelve parts 
according to integration regions of $\chi_1, \chi_{\hat{1}}$. Among them, only two parts corresponding to 
$0<\chi_{\hat{1}}<\chi_1<1$ and to $0<\chi_1<\chi_{\hat{1}}<1$ are nonvanishing. Then, we have 
\be
I_{(1,1)}= \left(\frac{-i}{2}\right)^2\left\{C^{12}[\bar{\alpha}_i,\,\bar{\alpha}_i']\,
P^{12}[\alpha_i,\,\alpha_i']+
C^{21}[\bar{\alpha}_i,\,\bar{\alpha}_i']\,
P^{21}[\alpha_i,\,\alpha_i']\right\}, 
\label{I(1,1)_2}
\ee
where 
\bea
C^{12}[\bar{\alpha}_i,\,\bar{\alpha}_i'] & = & \int_0^1d\chi_1\int_0^{\chi_1}d\chi_{\hat{1}}\,
\chi_1^{\bar{\alpha}}\,(1-\chi_1)^{\bar{\beta}}\,\chi_{\hat{1}}^{\bar{\alpha}'}\,
(1-\chi_{\hat{1}})^{\bar{\beta}'}
\,(\chi_1-\chi_{\hat{1}})^{2\sigma}, \nn\\
C^{21}[\bar{\alpha}_i,\,\bar{\alpha}_i'] & = & \int_0^1d\chi_1\int^1_{\chi_1}d\chi_{\hat{1}}\,
\chi_1^{\bar{\alpha}}\,(1-\chi_1)^{\bar{\beta}}\,\chi_{\hat{1}}^{\bar{\alpha}'}\,
(1-\chi_{\hat{1}})^{\bar{\beta}'}
\,(\chi_{\hat{1}}-\chi_1)^{2\sigma}, \nn\\
P^{12}[\alpha_i,\,\alpha_i'] & = & \int^\infty_{-\infty}d\eta_1\,d\eta_{\hat{1}}\,
(\eta_1-i\epsilon)^{\alpha}\,(1-\eta_1-i\epsilon)^{\beta}\,(\eta_{\hat{1}}-i\epsilon)^{\alpha'}\,
(1-\eta_{\hat{1}}-i\epsilon)^{\beta'}\nn \\
& & \hspace{18mm} \times (\eta_1-\eta_{\hat{1}}-i\epsilon)^{2\sigma}, \nn\\
P^{21}[\alpha_i,\,\alpha_i'] & = & \int^\infty_{-\infty}d\eta_1\,d\eta_{\hat{1}}\,
(\eta_1-i\epsilon)^{\alpha}\,(1-\eta_1-i\epsilon)^{\beta}\,(\eta_{\hat{1}}-i\epsilon)^{\alpha'}\,
(1-\eta_{\hat{1}}-i\epsilon)^{\beta'}\nn \\
& & \hspace{18mm} \times (\eta_{\hat{1}}-\eta_1-i\epsilon)^{2\sigma}. 
\label{CCPP}
\eea
In $P^{12}$ ($P^{21}$), $-i\epsilon$ in the last factor indicates that the contour of $\eta_{\hat{1}}$ avoids 
that of $\eta_1$ upward (downward). 
We have used the notation in~\cite{Fukuda:2001jd}, where $\alpha_1=\alpha$, $\alpha_2=\beta$, 
$\alpha_1'=\alpha'$, $\alpha_2'=\beta'$, etc. Since integration variables are dummy, we see 
\be
C^{21}[\bar{\alpha}_i,\,\bar{\alpha}_i']= C^{12}[\bar{\alpha}_i',\,\bar{\alpha}_i], \qquad
P^{21}[\alpha_i,\,\alpha_i']=P^{12}[\alpha_i',\,\alpha_i].
\ee

Next, we obtain a relation between $C^{12}$ and $P^{12}$ by introducing
\be
Q^{12}[\bar{\alpha}_i,\,\bar{\alpha}_i'] \equiv \int_0^1d\chi_1\,d\chi_{\hat{1}}\,
\chi_1^{\bar{\alpha}}\,(1-\chi_1)^{\bar{\beta}}\,\chi_{\hat{1}}^{\bar{\alpha}'}\,
(1-\chi_{\hat{1}})^{\bar{\beta}'}\,(\chi_1-\chi_{\hat{1}}-i\epsilon)^{2\sigma}. 
\ee
Splitting the integration region $[0,1]\times [0,1]$ into the part of $\chi_1>\chi_{\hat{1}}$ and that of 
$\chi_1<\chi_{\hat{1}}$ yields 
\be
Q^{12}[\bar{\alpha}_i,\,\bar{\alpha}_i'] =C^{12}[\bar{\alpha}_i,\,\bar{\alpha}_i']
+e^{-i2\pi\sigma}\,C^{12}[\bar{\alpha}_i',\,\bar{\alpha}_i].
\label{Q12_C12}
\ee
On the other hand, we deform the integration contours of $P^{12}[\alpha_i,\,\alpha_i']$ to surround 
the cut $[1,\infty)$:
\bea
P^{12}[\alpha_i,\,\alpha_i'] & = & -2is(\beta)\,\left[-e^{i\pi\beta'}\,\int^1_\infty d\eta_1\,d\eta_{\hat{1}}\,
\eta_1^{\alpha}\,(\eta_1-1)^{\beta}\,\eta_{\hat{1}}^{\alpha'}\,(\eta_{\hat{1}}-1)^{\beta'}\,
(\eta_1-\eta_{\hat{1}}+i\epsilon)^{2\sigma}\right. \nn \\
& & \left.\hspace{17mm}+e^{-i\pi\beta'}\,\int^1_\infty d\eta_1\,d\eta_{\hat{1}}\,
 \eta_1^{\alpha}\,(\eta_1-1)^{\beta}\,\eta_{\hat{1}}^{\alpha'}\,(\eta_{\hat{1}}-1)^{\beta'}\,
(\eta_1-\eta_{\hat{1}}-i\epsilon)^{2\sigma}\right]. \nn \\
 & & 
\eea
Changing variables $\eta_1=1/\xi_1$ and $\eta_{\hat{1}}=1/\xi_{\hat{1}}$, we have 
\be
P^{12}[\alpha_i,\,\alpha_i'] = -2is(\beta)\left\{e^{-i\pi\beta'}\,Q^{32}[\alpha_i',\,\alpha_i]
-e^{i\pi\beta'}e^{i2\pi\sigma}\,Q^{32}[\alpha_i,\,\alpha_i']\right\}, 
\label{P12_Q32}
\ee
where 
\be
Q^{32} [\alpha_i,\,\alpha_i'] \equiv \int^1_0 d\xi_1\,d\xi_{\hat{1}}\,\xi_1^{\gamma}\,(1-\xi_1)^{\beta}\,
\xi_{\hat{1}}^{\gamma'}\,(1-\xi_{\hat{1}})^{\beta'}\,(\xi_1-\xi_{\hat{1}}-i\epsilon)^{2\sigma}, 
\label{Q32}
\ee
\be
\gamma \equiv -\alpha-\beta-2\sigma-2 (\equiv \alpha_3), \qquad 
\gamma'\equiv -\alpha'-\beta'-2\sigma-2 (\equiv \alpha'_3), 
\label{gamma_gamma'}
\ee
and $Q^{32}[\alpha_i',\,\alpha_i]$ is obtained by $\beta\leftrightarrow \beta'$, $\gamma\leftrightarrow \gamma'$ in (\ref{Q32}). 
Changing variables $\xi_1=1-\chi_{\hat{1}}$, $\xi_{\hat{1}}=1-\chi_1$ in (\ref{Q32}) means 
\be
Q^{32} [\alpha_i,\,\alpha_i']=Q^{23}[\alpha_i',\,\alpha_i], \qquad 
Q^{32} [\alpha_i',\,\alpha_i]=Q^{23}[\alpha_i,\,\alpha_i'].
\label{Q32_Q23}
\ee
Together with (\ref{Q12_C12}), (\ref{P12_Q32}) and (\ref{Q32_Q23}), we obtain 
\be
P^{12}[\alpha_i,\,\alpha_i'] = (-2i)^2s(\beta)\,\left\{s(\beta')\,C^{23}[\alpha_i,\,\alpha_i']
+s(\beta'+2\sigma)\,C^{23}[\alpha_i',\,\alpha_i]\right\} 
\label{P12_C23}
\ee
and 
\be
P^{21}[\alpha_i,\,\alpha_i']=P^{12}[\alpha_i',\,\alpha_i]
=(-2i)^2s(\beta')\,\left\{s(\beta)\,C^{23}[\alpha_i',\,\alpha_i]
+s(\beta+2\sigma)\,C^{23}[\alpha_i,\,\alpha_i']\right\}. 
\label{P21_C23}
\ee

Plugging (\ref{P12_C23}) and (\ref{P21_C23}) into (\ref{I(1,1)_2}) expresses $I_{(1,1)}$ in terms of 
$C^{12}$'s and $C^{23}$'s: 
\bea
I_{(1,1)} & = & s(\beta)\,s(\beta')
\left\{C^{12}[\bar{\alpha}_i,\,\bar{\alpha}_i']\,C^{23}[\alpha_i,\,\alpha_i']
+C^{12}[\bar{\alpha}_i',\,\bar{\alpha}_i]\,C^{23}[\alpha_i',\,\alpha_i]\right\} \nn \\
& & +s(\beta)\,s(\beta'+2\sigma)\,C^{12}[\bar{\alpha}_i,\,\bar{\alpha}_i']\,C^{23}[\alpha_i',\,\alpha_i] \nn \\
& & +s(\beta')\,s(\beta+2\sigma)\,C^{12}[\bar{\alpha}_i',\,\bar{\alpha}_i]\,C^{23}[\alpha_i,\,\alpha_i']. 
\label{I(1,1)_f_II}
\eea
Once we know $C^{12}[\alpha_i,\,\alpha_i']$, 
all of the $C^{12}$'s and $C^{23}$'s appearing in (\ref{I(1,1)_f_II}) are obtained by replacing parameters. 
For example, the change $(\alpha, \alpha',\beta, \beta')\rightarrow (\beta, \beta',\gamma,\gamma')$ in 
$C^{12}[\alpha_i,\,\alpha_i']$ gives $C^{23}[\alpha_i,\,\alpha_i']$. 
As a result of the direct computation, 
$C^{12}[\alpha_i,\,\alpha_i']$ is represented by hypergeometric functions:
\bea
C^{12}[\alpha_i,\,\alpha_i'] & = & \frac{\Gamma(\alpha'+1)\,\Gamma(2\sigma+1)}{\Gamma(\alpha'+2\sigma+2)}\,
\int^1_0 d\chi_1\,\chi_1^{\alpha+\alpha'+2\sigma+1}\,(1-\chi_1)^{\beta}\nn \\
& & \hspace{47mm} \times F(-\beta',\alpha'+1,\alpha'+2\sigma+2;\,\chi_1) \nn \\
& = & \frac{\Gamma(\alpha+\alpha'+2\sigma+2)\,\Gamma(\beta+1)\,\Gamma(\alpha'+1)\,\Gamma(2\sigma+1)}{
\Gamma(\alpha+\alpha'+\beta+2\sigma+3)\,\Gamma(\alpha'+2\sigma+2)}\nn \\
& & \times _3F_2(-\beta',\alpha'+1,\alpha+\alpha'+2\sigma+2
;\,\alpha'+2\sigma+2,\alpha+\alpha'+\beta+2\sigma+3;\,1), \nn \\
& & 
\label{C12}
\eea
from which 
\bea
C^{23}[\alpha_i,\,\alpha_i'] 
& = & \frac{\Gamma(\beta+\beta'+2\sigma+2)\,\Gamma(\gamma+1)\,\Gamma(\beta'+1)\,\Gamma(2\sigma+1)}{
\Gamma(\beta+\beta'+\gamma+2\sigma+3)\,\Gamma(\beta'+2\sigma+2)}\nn \\
& & \times _3F_2(-\gamma',\beta'+1,\beta+\beta'+2\sigma+2
;\,\beta'+2\sigma+2,\beta+\beta'+\gamma+2\sigma+3;\,1). \nn \\
& & 
\label{C23}
\eea

We get the result in the phase convention (I) from that in (II) by the replacement 
\be
(\alpha, \alpha', \beta, \beta', \gamma, \gamma')\leftrightarrow 
(\bar{\alpha}, \bar{\alpha}', \bar{\beta}, \bar{\beta}', \bar{\gamma}, \bar{\gamma}'). 
\ee
It can be checked that string amplitudes in the text that are expressed by $I_{(1,1)}$ give the same results 
irrespective of the conventions (I) and (II).

\section{2(NS, NS)-2(R$-$, R$+$) amplitude and its picture changing}
\label{app:4pt_BBBB2_pic}
\setcounter{equation}{0}
In this appendix, we compute the four-point genus-zero amplitude of two (NS, NS) and two (R$-$, R$+$) fields:
\bea
\hat{V}_1(z_1,\bar{z}_1) & = & \hat{T}_{k_1}(z_1)\,\hat{\bar{T}}_{-k_1}(\bar{z}_1), \nn \\
\hat{V}_2(z_2,\bar{z}_2) & = & \hat{T}^{(0)}_{k_2}(z_2)\,\hat{\bar{T}}^{(0)}_{-k_2}(\bar{z}_2)  
\qquad \left(k_1, k_2\in {\bf Z}+\frac12\right), \nn \\
\hat{V}_b(z_b,\bar{z}_b) & = & \hat{V}_{k_b,\,-1}(z_b)\,\hat{\bar{V}}_{-k_b,\,+1}(\bar{z}_b) \qquad 
(k_b=0,-1,-2,\cdots)
\label{4pt_BBBB2_V}
\eea
with $b=3,4$. 
Although the matrix-model counterpart of the (NS, NS) fields $\hat{V}_1$ and $\hat{V}_2$ has not been found, 
we consider this amplitude because it exhibits nontrivial behavior in the picture changing operation 
when $\hat{V}_3$ or $\hat{V}_4$ is nonlocal. 

\subsection{The amplitude}
\label{app:4pt_BBBB2}
{}From the conservation of $H$ and $\bar{H}$ charges, $s=0, 2$ cases can give a nontrivial result. 
We here consider the $s=0$ case. Following the same procedure as in section~\ref{sec:basic} yields 
\bea
\left.\vev{\prod_{i=1}^4\hat{V}_i(z_i,\bar{z}_i)}\right|_{s=0} & = & 
\frac{-1}{2}(p_{\ell_2}-k_2)^2\,\delta_{\sum_ik_i,\,0}\,
\delta_{\sum_ip_{\ell_i},\,2}\,\left(2\ln\frac{1}{\mu_1}\right)\,e^{-i\pi\beta\sum_ik_i^2}  \nn \\
& & \times \left(\prod_{a=1,2}\prod_{b=3,4}|z_a-z_b|^{-1}\right)
\prod_{i<j}|z_i-z_j|^{2(k_ik_j-p_{\ell_i}p_{\ell_j})}.
\label{4pt_BBBB2}
\eea
The corresponding string amplitude reads 
\bea
& & \left.\vev{\prod_{i=1}^3c\bar{c}\hat{V}_i(z_i,\bar{z}_i)\,
\int d^2z_4\hat{V}_4(z_4,\bar{z}_4)}\right|_{s=0,\,(z_1,z_2,z_3)=(\infty, 1,0)} \nn \\
& & \hspace{7mm} = \frac{-1}{2}(p_{\ell_2}-k_2)^2\,\delta_{\sum_ik_i, \,0}\,\delta_{\sum_ip_{\ell_i},\,2} \, 
\left(2\ln \frac{1}{\mu_1}\right) \, e^{-i\pi\beta\sum_ik_i^2}  \, \cI_{(1,0)},  
\label{4pt_BBBB2_2}
\eea
where $\cI_{(1,0)}$ is the same form as what appears in~(\ref{4pt_BBBB_2}), i.e. 
the integral $I_{(1,0)}$ in (\ref{I(1,0)}) with (\ref{4pt_alpha_beta}). 
The kinematical condition is satisfied by the same momenta as in (\ref{4pt_kin_L_k})-(\ref{4pt_kin_NL}). 
The case of both of $\hat{V}_b$ nonlocal is forbidden. 
Thus, we use the result of the regularization (\ref{4pt_I(1,0)_L}) and (\ref{4pt_I(1,0)_NL}), and end up with 
\be 
\left.\vev{\prod_{i=1}^3c\bar{c}\hat{V}_i(z_i,\bar{z}_i)\,
\int d^2z_4\hat{V}_4(z_4,\bar{z}_4)}\right|_{s=0,\,(z_1,z_2,z_3)=(\infty, 1,0)} =0
\label{4pt_BBBB2_L_f}
\ee
for both of $\hat{V}_b$ local, and 
\bea
 & & \left.\vev{\prod_{i=1}^3c\bar{c}\hat{V}_i(z_i,\bar{z}_i)\,
\int d^2z_4\hat{V}_4(z_4,\bar{z}_4)}\right|_{s=0,\,(z_1,z_2,z_3)=(\infty, 1,0)} \nn \\
& & \hspace{7mm}= \delta_{\sum_ik_i, \,0}\,\delta_{\sum_ip_{\ell_i},\,2} \,
\left(2\ln \frac{1}{\mu_1}\right)^2 \, e^{-i\pi\beta\sum_ik_i^2} \, 
(-\pi)\,n_2^2\left(\frac{(n_1+n_2)!}{n_1!n_2!}\right)^2\,c_L 
\label{4pt_BBBB2_NL_f}
\eea
for $\hat{V}_3$ local and $\hat{V}_4$ nonlocal. 
The result is also identical for $\hat{V}_4$ local and $\hat{V}_3$ nonlocal.

\subsection{Picture changing operation}
(\ref{4pt_BBBB2_NL_f}) is not symmetric under $n_1\leftrightarrow n_2$ corresponding to 
$\hat{V}_1\leftrightarrow \hat{V}_2$. 
Since $\hat{V}_1$ and $\hat{V}_2$ are (NS, NS) fields with the pictures $(-1, -1)$ and $(0,0)$ respectively, 
the amplitude would be symmetric if the picture changing operation worked as usual. 
Let us see the situation explicitly focusing on the holomorphic part:
\be
\bra 0|cT_{k_1}(z_1)\,cT^{(0)}_{k_2}(z_2)\,cV_{k_3,\,-1}(z_3)\int dz_4 V^{({\rm nonlocal})}_{k_4,\,-1}(z_4)
|0\ket.
\ee
Here, $cT^{(0)}_{k_2}(z_2)$ can be replaced with $(cT_{k_2})^{(0)}(z_2)$ defined in \eqref{cT_0}, 
because the difference of them (the second term in the r.h.s. of (\ref{cT_0}) with changing $(k_1, z_1)$ to $(k_2,z_2)$)  
does not contribute to the amplitude from the conservation of various charges. 
After the same procedure as in section~\ref{sec:pic}, we have 
\bea
& & \bra 0|cT_{k_1}(z_1)\,cT^{(0)}_{k_2}(z_2)\,cV_{k_3,\,-1}(z_3)\int dz_4 V^{({\rm nonlocal})}_{k_4,\,-1}(z_4)
|0\ket \nn \\
& & \hspace{7mm} = \bra 0|cT^{(0)}_{k_1}(z_1)\,cT_{k_2}(z_2)\,cV_{k_3,\,-1}(z_3)\int dz_4 V^{({\rm nonlocal})}_{k_4,\,-1}(z_4)
|0\ket + \cB,
\eea
where $\cB$ remains, because $\hat{V}_{k_4,\,-1}^{({\rm nonlocal})}(z_4)$ is not BRST-closed: 
\bea
\cB& \equiv & \frac{-i}{\sqrt{2}}(p_{\ell_4}-k_4-1)\,
\bra 0|\prod_{a=1,2}\left\{\xi c\,e^{-\phi+ik_ax+p_{\ell_a}\varphi}(z_a)\right\}\,
c\,e^{-\frac12\phi-i\frac12H+ik_3x+p_{\ell_3}\varphi}(z_3) \nn \\
& & \hspace{34mm}\times \left.\int dz_4\,\eta\,e^{\frac12\phi+i\frac12H+ik_4x+p_{\ell_4}\varphi}(z_4)|0\ket
\right|_{{\rm large}}. 
\eea 
This satisfies charge conservations, and the Wick contraction leads to a nonvanishing result
\bea
\cB & = & \frac{-i}{\sqrt{2}}(p_{\ell_4}-k_4-1)\,(z_1-z_2)(z_1-z_3)^{1/2}(z_2-z_3)^{1/2} \nn \\
& & \times \int \frac{dz_4}{(z_1-z_4)^{1/2}(z_2-z_4)^{1/2}}\,
\prod_{i<j}(z_i-z_j)^{k_ik_j-p_{\ell_i}p_{\ell_j}} \,
\bra 0|:\prod_{i=1}^4 e^{ik_ix+p_{\ell_i}\varphi}(z_i):|0\ket. \nn \\
& & 
\label{cB_nonzero}
\eea 
This should be nonzero because (\ref{4pt_BBBB2_NL_f}) is not symmetric under 
$\hat{V}_1\leftrightarrow \hat{V}_2$.


\end{document}